\documentclass[11pt]{article}
\usepackage{graphicx}
\usepackage{epstopdf}
\usepackage{multirow}
\usepackage{amsfonts}
\usepackage{extarrows}
\usepackage{amssymb}
\usepackage{geometry} 
\let\ssection=\section
\renewcommand{\section}{\setcounter{equation}{0}\ssection}

\newcommand\mathC{\mkern1mu\raise2.2pt\hbox{$\scriptscriptstyle|$}
        {\mkern-7mu\rm C}}              
\newcommand{\mathR}{{\rm I\! R}}         

\newcommand{\be}{\begin{equation}}
\newcommand{\ee}{\end{equation}}

\newcommand{\abs}{absolute}

\newcommand{\ind}{indiscernible}

\newcommand{\raw}{\rightarrow}
\newcommand{\Raw}{\Rightarrow}

\newcommand{\riota}{\mathrm{\rotatebox[origin=c]{180}{$\iota$}}}

\begin{document}
\begin{center}
{\large\bf On Kinds of Indiscernibility in Logic and Metaphysics}

Adam Caulton\footnote{Institute of Philosophy, University of London, Senate House, Malet Street, London, WC1E 7HU.  adam.caulton@sas.ac.uk.} and Jeremy Butterfield\footnote{Trinity College, Cambridge, CB2 1TQ. jb56@cam.ac.uk.}
\end{center}


\begin{center}  Forthcoming, after an Appendectomy, in the \emph{British Journal for the Philosophy of Science}. \end{center}

\begin{abstract}

Using the Hilbert-Bernays account as a spring-board, we first define four ways in which
two objects can be discerned from one another, using the non-logical vocabulary of the
language concerned.
(These definitions are based on definitions made by Quine and Saunders.)
Because
of our use of the Hilbert-Bernays account, these definitions are in terms of the
syntax of the language. But we also relate our definitions to the idea of permutations
on the domain of quantification, and their being symmetries. These relations turn out to be subtle---some natural conjectures about them are false. We will see in particular that the idea of symmetry meshes with a species of indiscernibility that we will call `absolute indiscernibility'. 
We then  report all the logical implications between our four kinds of discernibility.

We use these four kinds as a resource for stating four metaphysical
theses about identity.
Three of these theses articulate two traditional philosophical themes: viz. the principle of
the identity of indiscernibles (which will come in two versions), and haecceitism.
The fourth is recent. Its most notable feature is that it makes diversity (i.e.
non-identity) weaker than what we will call individuality (being an individual): two objects can be distinct
but not individuals. For this reason, it has been advocated both for quantum particles and for spacetime points.

Finally, we locate this fourth metaphysical thesis in a broader position, which we call structuralism.
We conclude with a discussion of the semantics suitable for a structuralist, with particular reference to physical theories as well as elementary model theory.

\end{abstract}
\newpage
\tableofcontents

\section{Introduction}
\subsection{Prospectus}\label{prosp}
This is one of several papers about identity. Their overall topic is quantum theory: where
there has been discussion over many decades about the treatment of indistinguishable (also
known as: identical) particles.\footnote{Castellani [1998] is a valuable collection of
classic and contemporary articles. French and Krause [2006] is a thorough recent monograph.} This discussion has been invigorated in the last few
years, principally by Saunders' revival of the Hilbert-Bernays account of identity (also
briefly discussed by Quine) and his application of it to quantum theory.

In short, Saunders saw that there is an error in the consensus of the previous philosophical
literature. That literature had shown that for any assembly of indistinguishable quantum
particles (fermions or bosons), and any state of the assembly (appropriately
(anti)-symmetrized), and any two particles in the assembly: the reduced density matrices
(reduced states) of the particles (and so all probabilities for single-particle measurements)
were equal; and so were appropriate corresponding two-particle conditional probabilities.
This result strongly suggests that quantum theory endemically violates the principle of the
identity of indiscernibles: for any two particles in the assembly are surely
indiscernible.\footnote{This consensus seems to have been first stated by Margenau ([1944], pp.~201-3) and it is endorsed and elaborated by e.g.~French and Redhead ([1988], p.~241), Butterfield ([1993], p.~464), and French ([2006], \S 4).  Massimi ([2001], pp.~326-7) questions these authors' emphasis on monadic properties, but agrees that quantum mechanics violates the identity of indiscernibles when the quantum state is taken to codify purely \emph{relational} properties.}

Saunders' basic insight ([2006], p.~7) is that the Hilbert-Bernays account
provides ways that two objects can be distinct, which are not captured by these orthodox
quantum probabilities: and yet which {\em are} instantiated by quantum theory. Thus on the
Hilbert-Bernays account, two objects can be distinct, even while sharing all their monadic
properties and their relations to all other objects. For they can be distinguished by
either:\\
\indent (1): a relation $R$ between them holding one way but not the other (which is called
`being relatively discernible'); \\
\indent (2): a relation $R$ between them holding in both directions but neither object having
$R$ to itself (`being weakly discernible').\\
Saunders then argues that the second case, (2), is instantiated by fermions: the prototype
example being the relation $R$ = `\ldots\ has opposite value for spin (in any spatial direction)
to \underline{\phantom{\ldots}}' for two spin-$\frac{1}{2}$ fermions in the singlet state.

Saunders' proposals have led to several developments, including exploring the parallel between quantum particles and spacetime points. But these will be the focus of
other papers. Here we will have enough to do, answering some (broadly taxonomic) questions that arise from the Hilbert-Bernays account, and relating our answers to some associated (even venerable!) metaphysical theses---without
regard to quantum theory. 

More specifically, our aim here will be to define four different ways in which
objects can be discerned from one another (two corresponding to (1) and (2) above), and to
relate these definitions: (i) to the idea of symmetry; (ii) to one another, and (iii) to some metaphysical themes. These four ways of being discerned
will be (our formulations of)  definitions made by Quine and Saunders in terms of
the Hilbert-Bernays account.

The Hilbert-Bernays account is a reductive account, reducing identity to a conjunction of statements of
indiscernibility. So this account is controversial: many philosophers hold that
identity is {\em irreducible} to any sort of qualitative facts, but nevertheless wholly unproblematic
because completely understood.\footnote{For example, Lewis ([1986], pp.~192-193).  More generally: in the philosophy of logic, the Fregean tradition that identity is indefinable, but understood, remains strong---including as a response to the Hilbert-Bernays account: cf.~e.g.~Ketland ([2006], p.~305 and Sections 5, 7).} This latter view is
certainly defensible, perhaps orthodox, once one sets aside issues about diachronic
identity---as we will in this paper.  But this is not a
problem for us: for we are not committed to the Hilbert-Bernays account (and nor is Saunders'
basic insight about weak discernibility in quantum theory). On the contrary, we will eventually (in Section \ref{4metaph}) unveil a metaphysical thesis which unequivocally denies the Hilbert-Bernays account, and which elsewhere we will apply to quantum theory.  But these controversies do not undercut the rationale for our investigations into the kinds of discernibility; essentially because anyone, whatever their philosophy of identity or their attitude to the identity of indiscernibles, will accept that discernibility is a sufficient condition for
diversity (the `non-identity of discernibles').

So to sum up, our aim is to use  the Hilbert-Bernays account as a spring-board so as to give a
precise ``logical geography'' of discernibility. This logical geography will be in terms of
the syntax of a formal first-order language. But we will also relate our definitions to the
idea of permutations on the domain of quantification, and to the idea of these permutations being symmetries. These relations seem not to have been much studied in the recent philosophical literature about the Hilbert-Bernays  account; and we will see that they turn out to be subtle---some natural conjectures are false.\footnote{We thank N.~da Costa and J.~Ketland for alerting us to their results in this area, which we (culpably!)~had missed.  Further references below. \label{fn:KetlandCosta}}

The plan of the paper is as follows. After stating some stipulations about jargon (Section
\ref{Convent}), we will state the Hilbert-Bernays account in terms of a first-order 
language (Section \ref{HBsec}) and relate it to permutations on a domain of quantification
(Section \ref{permnslogic}). With these preparations in hand, we will be ready to define four
ways in which objects can be discerned from each other, using the predicates of the language
(Section \ref{4kinds}). These definitions will include precise versions of (1) and (2) above. Then in Section \ref{symmrevisit}, we give several results relating  permutations and symmetries to a kind of indiscernibility which we (following Quine and Saunders) will call `absolute'.
We admit that one could instead define kinds of indiscernibility in terms of permutations and symmetries, rather than the instantiation of formulas.  So in the second Appendix (Section 8) we explore this approach, emphasising the consequences for Section 4's results.  (The first Appendix, Section 7, surveys the logical implications between our four kinds of indiscernibility.)
 Finally, in Section \ref{4metaph}, we will use our four kinds of discernibility to state four metaphysical theses.  (There will also be a fifth ``umbrella" thesis called `structuralism'.)
Three of these theses articulate two traditional philosophical themes: viz.~the principle of
the identity of indiscernibles (which will come in two versions), and haecceitism.
The fourth is recent. Its most notable feature is that it makes diversity (i.e.
non-identity) weaker than what we will call individuality (being an individual): two objects can be distinct
but not individuals. For this reason, it has been advocated both for quantum particles and for spacetime points. We will classify and compare these theses in two ways: in terms of two philosophical questions (Section \ref{the4}), and in terms of the conflicting ways in which they count possible structures (Section \ref{hypercube}).  We conclude (Section \ref{structuralism}) by relating these theses to the position we call `structuralism': which prompts a non-standard (though familiar) interpretation of, or even a revision of, traditional formalisms in both semantics and physical theories.

\subsection{Stipulations about jargon}\label{Convent}
We make some stipulations about philosophical terms. We think that all of them are natural and innocuous, though the last one, about `individual', is a bit idiosyncratic.

{\em `Object', `identity', `discernibility'}:---
We will use `object' for the broad idea, in the tradition of Frege and Quine,
of a potential referent of a singular term, or value of a variable. The negation of the identity relation on objects we will call (indifferently): `non-identity', `distinctness', `diversity'; (and hence use cognate words like `distinct', `diverse').
When we have in mind that a formula applies to one of two objects but not the other, we will say that they are `discerned', or that the formula `discerns' them. We will also use `discernment' and `discernibility': these are synonyms; (though the former usefully avoids connotations of modality, the latter often sounds better). Their negation we will call  `indiscernibility'.

{\em `Individual'}:---
 Following a recent tradition started by Muller \& Saunders [2008], we will also use `individual' for a narrower notion than `object',
viz.~an
object that is discerned from others by a strong, and traditional, form of
difference---which we will call `absolute discernibility'. Anticipating the following section,  this usage may be illustrated by the fact a haecceitist (in our terms) would demand that all objects be individuals, in virtue of each possessing its own unique property.

We can see that it would be worth distinguishing cases according to whether the discernment is by an arbitrary language or an ``ideal" one. That is: since the Hilbert-Bernays account will be cast in a formal first-order language, and such languages can differ as to their non-logical vocabulary (primitive predicates), our discussion will sometimes be relative to a
choice of such vocabulary.  So for example an object that fails to be an individual by the lights of an impoverished language may yet be an individual in an ideal language adequate for expressing all facts (especially all facts about identity and diversity).  Therefore one might use the term `$L$-individual' for any object which is absolutely discerned from all others using the linguistic resources of the language $L$.  The un-prefixed term `individual' may then be reserved for the case of ideal language.  However, we will stick to the simple term `individual', since it will always be obvious which language is under consideration.  In Section \ref{4metaph} we will use `individual' in the strictly correct sense just proposed, since there we envisage a language which {\em is} taken to be adequate for expressing all facts.  

{\em `Haecceitism'}:--- Though the details will not be needed till Section \ref{4metaph}, we should say what we will mean by `haecceitism' (a venerable doctrine going back to Kaplan [1975] if not Duns Scotus!). The core meaning is advocacy of haecceities, i.e. non-qualitative thisness properties: almost always associated with the claim that every object has its own haecceity. But this core meaning is itself ambiguous, and authors differ about the implications and connotations of `haecceity'---about how ``thick'' a notion is advocated.  Some discussion is therefore in order.

\subsubsection{Haecceitism}\label{subhaec}
At first sight, there are (at least) three salient ways to construe haecceitism.\footnote{We owe much of the discussion here to our conversations with Fraser MacBride.}  For each we give a description and one or more proponents. All three will refer to possible worlds, and will use the language of Lewis's (especially [1986]) metaphysics (though it will not require a commitment his form of modal realism, or to any form of modal realism for that matter).  The first is the weakest and the second and third are equivalent; we favour the second and third (and prefer the formulation in the third).
\begin{enumerate}
\item  \emph{Haecceitistic differences.}  Following Lewis ([1986], p.~221), we may take haecceitism to be about the way possible worlds represent the modal properties of objects.  It is the denial of the following supervenience thesis: a world's representation of  \emph{de re} possibilities (that is, possibilities pertaining to particular objects)  supervenes on the qualitative mosaic, i.e.~the pattern of instantiation of qualitative properties and relations, within that world.\footnote{This prompts the question, `What is a \emph{qualitative} property or relation?'  That is not a question we will here attempt to answer.}  Thus a haecceitist allows that two worlds, exactly alike in their qualitative features, may still disagree as to which object partakes in which property or relation.\footnote{Note that this is a stronger position than one that just allows duplicate worlds, that is, several worlds exactly alike in their qualitative features.  The existence of duplicate worlds does not entail Haecceitism in Lewis's sense, since, for each individual, every world in a class of mutual duplicates may represent the same dossier of \emph{de re} possibilities.  Lewis ([1986], p.~224) himself refrains from committing either to duplicate worlds or a ``principle of identity of indiscernibles'' applied to worlds; though the question seems moot for anyone who does not take possible worlds to be concretely existing entities.}

This version of haecceitism makes no further claims as to what exactly is left out of the purely qualitative representations, and so it is the weakest of the three haecceitisms.  But since two qualitatively identical worlds may disagree about what they represent \emph{de re}, they must differ in some non-qualitative ways: ways which somehow represent (actual) objects in a way that does not rely on how things are qualitatively (whether accidentally or essentially), with those objects.  There are two natural candidates for these representatives: the objects themselves (or some abstract surrogate for them), divorced from their qualitative clothing; or some non-qualitative properties which suitably track the objects across worlds.  These two candidates prompt the second and third kinds of haecceitism (which, we argue below, are in fact equivalent).  We see no sensible alternative to these two candidates for the missing representatives---though perhaps the difference could be taken as a primitive relation between worlds.  But haecceitism in our first sense does not entail either of the haecceitisms below, though they each entail haecceitism in our first sense of the acceptance of haecceitistic differences.

\item \emph{Combinatorial independence.}  Lewis's definition was inspired by a definition by Kaplan ([1975], pp.~722-3); but Kaplan's is stronger.  It is phrased explicitly in terms of trans-world identification.\footnote{Kaplan:  [haecceitism is] ``the doctrine that holds that it does make sense to ask---without reference to common attributes and behavior---whether this is the same individual in another possible world, that individuals can be extended in logical space (i.e.,~through possible worlds) in much the way we commonly regard them as being extended in physical space and time, and that a common ``thisness" may underlie extreme dissimilarity or distinct thisnesses may underlie great resemblance." \label{fn:haecKap}} But we believe there is a version of haecceitism, clearer than Kaplan's, defined in terms of combinatorial possibility; a version which is still stronger than Lewis's, but does not commit one to claims about non-spatiotemporal overlap between worlds or trans-world mereological sums.

According to this version of haecceitism, objects partake independently---that is, independently of each other and of the qualitative properties and relations---in the exhaustive recombinations which generate the full space of possible worlds.  For example: with a domain of $N$ objects there are: $2^N$-many possible property extensions (each of them distinct); $2^{N^2}$-many distinct binary relation extensions (each of them distinct); so the number of distinct worlds\footnote{Or distinct equivalence classes of world-duplicates!} containing $N$-objects, $k$ monadic properties and $l$ binary relations (and no other relations), is $2^{kN + lN^2}$.  

Combinatorial independence would appear to favour the doctrine that objects ``endure'' identically through possible worlds, since it seems sensible to identify property extensions with sets of objects, and the \emph{same} objects are added to or subtracted from the extensions in the generation of new worlds.  But trans-world ``perdurance''---the doctrine that trans-world ``identity" (in fact a misnomer, according to the doctrine) is a relation holding between parts of the same trans-world continuant---can be accommodated without difficulty. (It is the commitment to there being a unique, objective trans-world continuation relation, and therefore something for a rigid designator to get a grip on, which distinguishes the perdurantist from those, like Lewis, who favour modal talk of \emph{world-bound} objects: cf.~Lewis ([1986], pp.~218-20).)

\item \emph{Haecceitistic properties.} Perhaps the most obvious form of haecceitism---and perhaps \emph{prima facie} the least attractive---is the acceptance of haecceitistic properties.  According to this view, for every object there is a property uniquely associated with it, which that object and no other necessarily possesses, and which is not necessarily co-extensive with any (perhaps complex) qualitative property.  An imprecise (and even quasi-religious) reading of these properties is as ``inner essences" or ``souls" (hence the view's unpopularity). Perhaps a more precise, and less controversial, reading is that each thing possesses some property which ``makes" it that thing and no other: a property ``in virtue of which" it is that thing (cf.~Adams [1981], p.~13).  Whether or not that in fact more precise, this reading implies a notion of ontological primacy---that the property comes in some sense ``before" the thing---about which we remain silent.  But we disavow this implication.  But we intend our third version of haecceitism to be no more controversial than combinatorial independence; in fact we take this version of haecceitism to be \emph{equivalent} to combinatorial independence.

When we say that according to this view, there is a haecceitistic property corresponding to each object, by `object' is not meant `world-bound object', which would entail a profligate multiplication of properties. Rather, these haecceitistic properties are envisaged as an alternative means to securing trans-world continuation (understood according to either endurantism or perdurantism).  The motivation is as follows.  

Throughout this paper, we are quidditists, meaning that we take for granted the trans-world identity of properties and relations. (We remain agnostic as to whether this trans-world identity is genuine identity, i.e.~endurance of qualities; or whether there is instead a similarity relation applying to qualities between worlds (either second-order, applying directly to the qualities themselves; or else first-order, applying to the qualities indirectly, via the objects that instantiate them), i.e.~perdurance.)  So for the sake of mere uniformity, we may wish to accommodate haecceitism into our quidditistic framework by letting a trans-world monadic property do duty for each trans-world thing.  Haecceitism and anti-haecceitism alike can then be discussed neutrally in terms of world-bound objects and trans-world properties and relations, the difference between them reconstrued in terms of whether or not there are primitive monadic properties which allow one to simulate rigid designation.\footnote{Anti-quidditists (like Black [2001]) can still use our framework, by populating the domain of objects with properties and relations, now treated as the value of first-order variables, and using the new predicate or predicates `has', so that `$Fa$' becomes `$a$ has $F$' (cf.~Lewis [1970]).  Anti-quidditism may then amount to what we later (Section 5) call `anti-haecceitism', but applied to these hypostatized qualities.  However, this purported anti-quidditism must be a ``quidditist" about the `has' relation(s), a situation that clearly cannot be remedied by hypostatizing again, on pain of initiating a Bradley-like regress.   There is no space to pursue the issue here. But we endorse the view of Lewis [2002] that `has' is a `non-relational tie', and speculate that, if it is treated \emph{like} a relation in the logic, then it is better off treated as one whose identity across worlds is not in question---that is: treated ``quidditistically".}

We can characterise this position syntactically as one that demands that, in a language adequate for expressing all facts, the primitive vocabulary contains a 1-place predicate `$N_ax$' for each envisaged trans-world object $a$.\footnote{To simplify we assume that a haecceitist is a haecceitist about every object.  This leaves some (uninteresting) logical space between haecceitism and what we later characterise as anti-haecceitism (see the end of Section \ref{the4}, below).} That a monadic property \emph{can} do duty for a trans-world object---or, to rephrase in terms of the object-language, that a monadic predicate can do duty for a rigid designator---without any loss (or gain!)~in expressive adequacy, is well known (cf.~Quine [1960], \S38).  Given a haecceitistic predicate `$N_ax$', one can introduce the corresponding rigid designator by definition: $a := \riota x. N_ax$; and conversely, given a rigid designator `$a$' and the identity predicate `=', one can likewise introduce the corresponding haecceitistic predicate: $\forall x (N_ax \equiv x=a)$.  We therefore urge the view that the difference between the two stronger versions of haecceitism is \emph{merely notational}, meaning that the addition of the haecceitistic predicate `$N_ax$' to one's primitive vocabulary ontologically commits one to no more, and no less, than the addition of the name `$a$', together with all the machinery of rigid designation.\footnote{By saying that we intend the acceptance of haecceitistic properties to be equivalent to combinatorial independence, we do not intend to rule out as counting as a (strong) version of the acceptance of haecceitistic properties the view described above, viz.~that the haecceities are in some sense ontologically ``prior" to the objects, and serve to ``ground their identity" (whatever that may mean).  That view would \emph{entail} combinatorial independence between objects and other objects, and between objects and the qualitative properties and relations (though not, of course, between objects and their haecceities), and would equally well be served by rigid designators as by the addition of haecceitistic predicates to the primitive vocabulary.  The point is that our second and third versions of haecceitism match completely in logical strength, so they are equally consistent with more metaphysically ambitious views which seek to ``ground" trans-world identity in non-qualitative properties.} 
\end{enumerate}

From now on, we will in this paper take haecceitism in a sense stronger than the first, Lewisian version.  The second and third, stronger versions are equivalent, but we favour the notational trappings of the third version, i.e.~the acceptance of haecceitistic properties. Thus when we later ban names from the object-language (in Section \ref{HBsec}), this ought to be seen \emph{not} as a substantial restriction against the haecceitist, but merely a convenient narrowing of notational options for the sake of a more unified presentation.  We simply require the haecceitist to express her position though the adoption of haecceitistic predicates, though all are at liberty to read each instance of the haecceitist's `$N_ax$' as `$x=a$'.

A concern may remain: how can ontological commitment to a \emph{property} be equivalent to ontological commitment to a (trans-world) \emph{object}?  There is no mystery, once we lay down some principles for what ontological commitment to properties involves.  We take it that ontological commitment to a collection of properties, relations and objects entails a commitment to all the logical constructions thereof, because instantiation of the logical constructions can be defined away without residue in terms of instantiation and the existence of the original collection.  (For example.:~for any object, the complex predicate `$Fx \wedge Gx$' is satisfied by that object just in case both `$Fx$' and `$Gx$' are satisfied by it.) So ontological commitment to logical constructions is no \emph{further} commitment at all.  Now, ontological commitment to certain objects is revealed clearly enough: one need only peer into the domain of quantification to see if they are there.  Ontological commitment to complex properties and relations is equally straightforward, being a matter of commitment to their components.  But what about ontological commitment to the properties and relations taken as \emph{primitive}, i.e.~not as logical constructions---what does that involve?  Well, since ontological commitment to objects is clear enough, let us use that: let us say that ontological commitment to the primitive properties and relations is a commitment to their being \emph{instantiated} by some object or objects.

With these principles laid down, we can now prove that commitment to the trans-world continuant $a$ and commitment to the haecceitistic property, \emph{being $a$}, entail each other.  \emph{Left to right:} We can take two routes.  First route: Commitment to any object at all (i.e.~a non-empty domain) entails commitment to the identity relation, which is instantiated by everything. So commitment to the trans-world continuant $a$ entails commitment to the identity relation, since $a = a$.  But the property  \emph{being $a$} is just a logical construction out of the identity relation and the trans-world continuant $a$, so commitment to the trans-world continuant $a$ entails commitment to the property \emph{being $a$}.  Second route: Commitment to the trans-world continuant $a$ entails commitment to the property  \emph{being $a$} being instantiated.  But that is just to say that commitment to the trans-world continuant $a$ entails commitment to the property  \emph{being $a$}. \emph{Right to left:}  Commitment to the property \emph{being $a$} entails either: (i) a commitment to its being instantiated (if taken as primitive); or (ii) a commitment to the entities of which it is a logical construction (if taken as a logical construction).  If (i), then we are committed to something's being $a$, that is, the existence of $a$.  If (ii), and `being $a$' is understood properly as containing a rigid designator, not as an abbreviated definite description \`a la Russell, then we are committed to its logical components, i.e.~the identity relation, and $a$ itself. \emph{QED.}

To sum up: the acceptance of haecceitistic differences (our first version of haecceitism, above) need not commit one to either combinatorially independent trans-world objects (our second version), or to non-qualitative properties that could do duty for them (our third version), though the two latter doctrines are perhaps the most natural way of securing haecceitistic differences, and may themselves be considered as notational variants of each other.   We stipulate that we mean by `haecceitism' one of the stronger versions, and for reasons purely to do with uniformity of presentation, we stipulate that the advocacy of this stronger version of haecceitism be expressed through the acceptance of haecceitistic properties.  This version of haecceitism will be further developed, along with our three other metaphysical theses, in Section \ref{4metaph}.  

\section{A logical perspective on identity}\label{logicidy}
\subsection{The Hilbert-Bernays account}\label{HBsec}
What we will call the Hilbert-Bernays account of the identity of objects, treated as (the values of
variables) in a first-order language, goes as follows; (cf. Hilbert and Bernays ([1934], \S5: who in fact do not endorse it!),~Quine ([1970], pp. 61-64) and Saunders ([2003a], p. 5)). The idea is that there being only finitely many primitive predicates enables us to capture the idea of identity in a single axiom. In fact,
the axiom is a biconditional in which identity is equivalent to a long conjunction of
statements that predicates are co-instantiated. The conjunction exhausts, in a
natural sense, the predicates constructible in the language; and it caters for
quantification in predicates' argument-places other than the two occupied by $x$ and $y$.

 In detail, we proceed as follows.
Suppose that $F^1_i$ is the $i$th 1-place predicate, $G^2_j$ is the $j$th 2-place predicate,
and $H^3_k$ is the $k$th 3-place predicate; (we will not need to specify the ranges of
$i,j,k$).  Suppose that the language has no names, or function symbols, so that predicates
are the only kind of non-logical vocabulary. Then the biconditional will take the following
form:
$$
\begin{array}{cc}
	\begin{array}{rcll}
		\forall x\forall y \bigg\{x = y &\equiv& \Big[\ \ldots \ \wedge \
		(F^1_ix \equiv F^1_iy)
		\ \wedge\ \ldots\\\\
		&&\quad \ldots \ \wedge \ \forall z \left(
		(G^2_jxz \equiv G^2_jyz) \wedge
		(G^2_jzx \equiv G^2_jzy)\right)
		\ \wedge \ \ldots\\\\
		&&\quad \ldots \ \wedge \ \forall z \forall w \Big(
		(H^3_kxzw \equiv H^3_kyzw) \wedge
		(H^3_kzxw \equiv H^3_kzyw)
		\\ &&\qquad\qquad\qquad\qquad\qquad\qquad \wedge \
		(H^3_kzwx \equiv H^3_kzwy)\Big)
		\ \wedge \ \ldots \Big]\bigg\}
	\end{array}
	&
(HB)
\end{array}
$$
(For primitive predicates, we usually omit the brackets and commas often used to indicate
argument-places.)  Note that for each 2-place predicate, there are two biconditionals to
include on the
right-hand side; and similarly for a 3-place predicate.  The general rule is: $n$
biconditionals for an $n$-place predicate.\footnote{We
do not need to include explicitly on the right-hand side clauses with repeated instances of
$x$ or $y$, such as $(G^2_jxx \equiv G^2_jyy)$, since these clauses are implied by the
conjunction of other relevant biconditionals.  For example: From $ \forall z (G^2_jxz
\equiv G^2_jyz)$ we have, in particular, that   $(G^2_jxx \equiv G^2_jyx)$.  And from $
\forall z (G^2_jzx \equiv G^2_jzy)$ we have, in particular, that $ (G^2_jyx \equiv G^2_jyy)$.
It follows that $(G^2_jxx \equiv G^2_jyy)$.  A similar chain of arguments applies for an
arbitrary $n$-place predicate.}

This definition of the Hilbert-Bernays account prompts three comments.
\begin{enumerate}
\item {\em Envisaging a rich enough language}:--- The main comment is the obvious one: since
the right hand side (in square brackets) of (HB) defines an equivalence relation---which from now
on we will call  `indiscernibility' (or for emphasis: `indiscernibility by the primitive
vocabulary')---discussion is bound to turn on the issue whether this relation is truly
identity of the objects in the domain. Someone who advocates (HB) is envisaging a vocabulary rich enough (or: a domain of objects that is varied enough, by the lights of the vocabulary chosen) to discern any two distinct objects, and thereby force the equivalence relation given by the right hand side of (HB) to be identity. This comment can be developed in six ways.

\begin{itemize}
\item [(i)] Indiscernibility has the formal properties of identity expressible in first-order logic, i.e. being an equivalence relation and substitutivity. (Cf.~eq.~(\ref{schema}) in (i) of comment 3 below; and for details, Ketland ([2009], Lemma 12)). 
\item [(ii)] Let us call an interpretation of the language, comprising a domain $D$ and various subsets of $D, D^2$ etc. a `structure'; (since in the literature on identity, `interpretation' also often means `philosophical interpretation'). Then: if in a given structure, the identity relation is first-order definable, then it is defined by indiscernibility, i.e. the right hand side of (HB) (Ketland ([2009], Theorem 16)).
\item [(iii)] On the other hand: assuming `=' is to be interpreted as the identity relation  (cf.~(i) in 3 below), there will in general be structures in which the leftward implication of (HB) fails; i.e.~structures  with at least
two objects indiscernible from each other. We say `in general' since from the view-point of pure logic, `=' might itself be one of the 2-place predicates $G^2_j$: in which case, the leftward implication trivially holds.
\item [(iv)] The last sentence leads to the wider question whether the interpretation given to some of the predicates $F, G, H$ etc.  somehow presupposes identity, so that the Hilbert-Bernays account's reduction of identity is, philosophically speaking, a charade. This question will sometimes crop up below (e.g. for the first time, in footnote \ref{fn:spirithaec}); but we will not need to address it systematically. Here we can just note as an example of the question, the theory of pure sets. It has only one primitive predicate, `$\in$', and the axiom (HB) for it logically entails the axiom of extensionality.  But one might well say that this only gives a genuine reduction of identity if one can understand the intended interpretation of `$\in$' without a prior understanding of `$=$'.
\item [(v)] There is the obvious wider question, why we restrict our discussion to first-order languages. Here our main reply is twofold: (a) first-order languages are favoured by the incompleteness of higher-order logics, and we are anyway sympathetic to the view that first-order logic suffices for the formalisation of physical theories (cf.~Boolos \& Jeffrey [1974], p.~197; Lewis [1970], p.~429); and (b) the Hilbert-Bernays account is thus restricted---and though, as we emphasized, we do not endorse it, it forms a good spring-board for discussing kinds of indiscernibility. There are also some basic points about identity in second-order logic, which we should register at the outset. Many discussions (especially textbooks: e.g.~van Dalen [1994], p.~151; Boolos \& Jeffrey [1974], p.~280) take the principle of the identity of indiscernibles to be expressed by the second-order formula $\forall P\big((Px \equiv Py) \supset x = y\big)$: which is a theorem of (any deductive system for) second-order logic with a sufficiently liberal comprehension scheme. But even if one is content with second-order logic, this result does not diminish the interest of the Hilbert-Bernays account (or of classifications of kinds of discernibility based on it). For the second-order formula is a theorem simply because the values  of the predicate variable include singleton sets of elements of the domain (cf.~Ketland [2006], p.~313). And allowing such singleton sets as properties of course leads back both to haecceitism, discussed in Section \ref{Convent}, and to (iv)'s question of whether understanding the primitive predicates requires a prior understanding of identity. In any case, we will discuss the principle of the identity of indiscernibles from a philosophical perspective in Section \ref{4metaph}.
\item [(vi)] Finally, there is the question what the proponent of the Hilbert-Bernays account is to do when faced with a language with infinitely many primitive predicates.  The right hand side of (HB) is constrained to be finitely long, so in the case of infinitely many primitives it is prevented from capturing all the ways two objects can differ.  Here we see two roads open to the Hilbert-Bernays advocate: (a) through parametrization, infinitely many $n$-adic primitives, say $R_1xy, R_2xy, \ldots$ may be subsumed under a single, new $(n+1)$-adic primitive, say $Rxyz$, where the extra argument place is intended to vary over an index set for the previous primitives; and (b) one may resort to infinitary logic, which allows for infinitely long formulas---suggesting, as regards (HB), in philosophical
terms, a supervenience of identity rather than a reduction of it.  (We will consider infinitary logic just once below, towards the end of Section \ref{parag2}.)

\end{itemize}

To sum up these six remarks: we see the Hilbert-Bernays account as
intending a reduction of identity facts to qualitative facts---as proposing that there are no
indiscernible pairs of objects.
This theme will recur in what follows. Indeed, the next two comments relate to the
choice of language.

\item {\em Banning names}:--- From now on, it will be clearest to require the language
to have no individual constants, nor function symbols, so that the non-logical vocabulary contains only predicates. But this will not affect our arguments: they would
carry over intact if constants and function symbols were allowed.

As we see matters, it is only the haecceitist who is likely to object to this apparent limitation in expressive power. But here, Section \ref{Convent}'s discussion of haecceitism comes into its own. For even with no names, the haecceitist has to hand her thisness predicates $N_a x, N_b x$ etc., with which to refer to objects by definite descriptions. Thus we  propose, following Quine ([1960], \S\S 37-39), that we, and in particular the haecceitist, replace
proper names by 1-place predicates; (each with an accompanying uniqueness axiom;
and with the predicates then shoe-horned into the syntactic form of singular terms, by invoking
Russell's theory of descriptions).\footnote{Then any sentence $S =: \Phi(a)$ containing the name $a$, where the 1-place formula $\Phi(x)$ contains no occurrence of $a$, may be replaced by the materially equivalent sentence $\forall x\left(N_ax \supset \Phi(x)\right)$, which contains no occurrence of the name $a$.  Note also that Saunders ([2006], p.~53) limits his inquiry to languages without names; but no Quinean trick is invoked.  Robinson ([2000], p.~163) calls such languages `suitable'.} And as emphasised in Section \ref{Convent}, $N_a x$ etc.~are to be thinly construed: the predicate $N_a x$ commits one to nothing beyond what the predicate $a = x$ commits one to. Thus the presence of these predicates in the non-logical vocabulary means that 
  the haecceitist should have no qualms about endorsing the Hilbert-Bernays
account---albeit in letter, rather than in spirit.\footnote{For  the spirit of the Hilbert-Bernays account is a
\emph{reduction} of identity facts to putatively qualitative facts, about the
co-instantiation of properties.  Indeed: according to the approach in which `=' is not  a
logical constant (cf. comment 3(ii)), acceptance of (HB)
entails that the language with the equality symbol `=' is a definitional extension of the
language without it.  However, representing each haecceity by a one-place primitive predicate,
accompanied by a uniqueness axiom, assumes the concept of identity through the use of `=' in the axiom: making this
reduction of identity, philosophically speaking, a charade. By contrast, if only genuinely qualitative properties are expressed by the non-logical vocabulary, the philosophical
reduction of identity facts to qualitative facts can
succeed---provided, of course, that the language is rich enough or the domain varied enough. \label{fn:spirithaec}}

A point of terminology: Though we ban constants from the formal language, we will still use $a$ and $b$ as names in the {\em meta}-language (i.e. the language in which we write!) for the one, or
two!, objects, with whose identity or diversity we are concerned. We will also always use `=' in the meta-language to
mean identity!

\item {\em `=' as a logical constant?}:--- It is common in the philosophy  of logic to
distinguish two approaches by which formal languages and logics treat identity:
\begin{itemize}
\item [(i)] `=' is a logical constant in the sense that it is  required, by the 
definition of the semantics, to be interpreted, in
any domain of quantification, as  the identity relation. Then for
any formula (open sentence) $\Phi(x)$ with one free variable $x$, the formula
\be
	\forall x\forall y \left(x = y \supset (\Phi(x) \equiv \Phi(y))\right)
\label{schema}
\ee
is a logical truth (i.e. is true in every structure). Thus on this approach, the
rightward implication in (HB) is a logical truth.

 But the leftward implication in (HB) is not. For as discussed in comment 1, the
language may not be discerning enough. More precisely: there are
structures in which (no matter how rich the language!)~the leftward implication
fails.\footnote{Such is Wiggins' ([2001], pp.~184-185) criticism of the Hilbert-Bernays account as formulated by Quine ([1960], [1970]).} 

\item [(ii)] `=' is treated like any other 2-place predicate,
so that its properties flow entirely from the theory with which we are concerned, in
particular its axioms if it is an axiomatized theory. Of course, we expect our theory to
impose on `=' such properties as being an equivalence relation (cf.~Ketland [2009], Lemmas 8-10). We can of course go 
further towards capturing the intuitive idea of
identity by imposing every instance of the schema, eq. \ref{schema}.
So on this approach, the Hilbert-Bernays account can be regarded as a proposed finitary
alternative to imposing  eq. \ref{schema}: a proposal whose plausibility depends on the
language being rich enough. Of course, independently of the language being rich:
imposing (HB) in a language with finitely many primitives implies as a
theorem the truth of eq. \ref{schema}.\footnote{Hilbert and Bernays ([1934], p.~186)  show that `=' as defined by (HB) is (up to co-extension!)~the \emph{only} (non-logical) two-place predicate to imply reflexivity and every instance of the schema, eq. \ref{schema}.  The argument is reproduced in Quine ([1970], pp.~62-63).}
\end{itemize}

\end{enumerate}

\indent To sum up this Subsection: the conjunction on the right hand side of (HB)
makes vivid how, on the
Hilbert-Bernays account, objects can be distinct  for different reasons, according to which
conjunct fails to hold. In Section \ref{4kinds}, we will introduce a taxonomy of these kinds.
In fact, this taxonomy will distinguish different ways in which a single conjunct can fail to
hold. The tenor of that discussion will be mostly syntactic. So we will  complement it  by
first discussing identity, and the Hilbert-Bernays account, in terms of permutations on the
domain of quantification.

\subsection{Permutations on domains}\label{permnslogic}
We now discuss how permutations on a domain of objects can be used to express
qualitative similarities and differences between the objects. More precisely: we
will define what it is for a permutation to be a {\em symmetry} of an interpretation of the
language, and relate this notion to the Hilbert-Bernays account.

\subsubsection{Definition of a symmetry}\label{defsymmy}
Let $D$ be a domain of quantification, in which the predicates $F^1_i, G^2_j, H^3_k$ etc. get
interpreted.   So writing `ext' for `extension', ext($F^1_i$) is, for each $i$, a subset of
$D$; and ext($G^2_j$) is, for each $j$, a subset of $D \times D = D^2$; and ext($H^3_k$) is,
for each $k$, a subset of $D^3$ etc. For the resulting interpretation of the language, i.e.~$D$ together with these assigned extensions, we write $\cal I$. We shall also call such an
interpretation, a `structure'.

Now let $\pi$ be a permutation of $D$.\footnote{We note {\em en passant} that since a permutation is
a bijection, the definition of permutation involves the use of the `=' symbol; so that which
functions are considered to be permutations is subject to one's treatment of identity.  But no worries: as we noted in comment 2 of Section \ref{HBsec}, this definition is cast in the meta-language!}
We now define
a symmetry as a permutation that ``preserves all properties  and relations''. So we say that
$\pi$ is a {\em symmetry} (aka \emph{automorphism}) of $\cal I$ iff all the extensions of all the predicates are
invariant under $\pi$. That is, using $o_1, o_2, ...$ as meta-linguistic variables: $\pi$ is
a symmetry iff:
\be \label{symdef}
	\begin{array}{rcl}
		\forall o_1, o_2, o_3 \in D, \forall i, j, k: \qquad o_1 \in \mbox{ext}(F^1_i) &
\mbox{iff} & \pi(o_1) \in \mbox{ext}(F^1_i);\ \mbox{and}\\
		 \langle o_1, o_2 \rangle  \in \mbox{ext}(G^2_j) & \mbox{iff} & \langle \pi(o_1),
\pi(o_2) \rangle \in
\mbox{ext}(G^2_j);\ \mbox{and}\\
		 \langle o_1, o_2, o_3 \rangle  \in \mbox{ext}(H^3_k) & \mbox{iff} & \langle\pi(o_1),
\pi(o_2), \pi(o_3)\rangle \in \mbox{ext}(H^3_k);
	\end{array}
\ee
and similarly for predicates with four or more argument-places.\footnote{So a haecceitist (cf. comment 2 in Section \ref{HBsec}) will take only the identity map as a symmetry---unless they stipulate that haecceitistic properties are exempt from the definition of symmetry. \label{fn:exempthaec}}

\subsubsection{Relation to the Hilbert-Bernays account}\label{HBandsymmy}
Let us now compare this definition with the Hilbert-Bernays account. For the moment, we make
just two comments, (1) and (2) below. They dispose of natural conjectures, about symmetries leaving invariant
the indiscernibility equivalence classes. That is, the conjectures are  false: the first
conjecture fails because of the somewhat subtle notions of weak and relative
discernibility, which will be central later; and the second is technically a special case of
the first.\footnote{Our two comments agree with Ketland's results and examples ([2006], Theorem (iii) and example in footnote 17; or [2009], Theorem 35(a) and example). But we will not spell out the differences in jargon or examples, except to report that Ketland calls a structure `Quinian' iff the leftward implication of (HB) holds in it, i.e.~if the identity relation is first-order definable, and so (cf.~comment 1(ii) of Section \ref{HBsec}) defined by indiscernibility, i.e.~by the right hand side of (HB).}
In Section \ref{symmrevisit}, we will discuss how the
conjectures can be mended: roughly speaking, we need to replace indiscernibility by
a weaker and   ``less subtle'' notion, called `absolute indiscernibility'.

(1): {\em All equivalence classes invariant?}:--- Suppose that we adopt the relativization to
an arbitrary language: so we do not require the language to
be  rich enough to force indiscernibility to be  identity. Then one might conjecture that,
for each choice of language, a permutation is a symmetry iff it leaves invariant (also known as: {\em fixes}) each
indiscernibility equivalence class. That is: $\pi$ is a symmetry of the structure $\cal I$ iff each member
of each equivalence class $\subset \mbox{dom}({\cal I})$ is sent by $\pi$ to a member of that
same equivalence class.

In fact, this conjecture is false. The condition, leaving invariant the indiscernibility
equivalence classes, is stronger than being a symmetry. We first prove the true implication,
and then give a counterexample to the converse.

So suppose $\pi$ leaves invariant each indiscernibility class; and let $\langle o_1, o_2, ..., o_n
\rangle$ be in the extension ext($J^n$) of some $n$-place predicate $J^n$. Since $\pi$ leaves invariant
the indiscernibility class of $o_1$, it follows that $\langle \pi(o_1), o_2, ..., o_n
\rangle$ is also in ext($J^n$). (For if not, (HB)'s corresponding conjunct, i.e. the
conjunct for the first argument-place of the predicate $J^n$, would discern $o_1$ and
$\pi(o_1)$). From this, it follows similarly that since $\pi$ fixes the indiscernibility
class of $o_2$, $\langle \pi(o_1), \pi(o_2), ..., o_n \rangle$ is also in ext($J^n$). And so
on: after $n$ steps, we conclude that $\langle \pi(o_1), \pi(o_2), ..., \pi(o_n) \rangle$ is
in ext($J^n$). Therefore $\pi$ is a symmetry.

Philosophical remark:  one way of thinking of the Hilbert-Bernays account trivializes this theorem. That is: according to comment 1 of Section \ref{HBsec}, the proponent of this account envisages that the indiscernibility classes are singletons. So only the identity map leaves them all invariant, and  trivially, it is a symmetry. (Compare the discussion of haecceitism in footnotes \ref{fn:spirithaec} and \ref{fn:exempthaec}.)

Counterexample to the converse: Consider the following structure, whose domain comprises four objects, which we label $a$ to $d$.  The primitive non-logical vocabulary consists of just
the 2-place relation symbol $R$.  $R$ is interpreted as having the extension
$$\mbox{ext}(R) = \left\{
\langle a, b \rangle,
\langle a, c \rangle,
\langle a, d \rangle,
\langle b, a \rangle,
\langle b, c \rangle,
\langle b, d \rangle
\right\}\ .$$ Now let `=' be defined by the Hilbert-Bernays axiom (HB).   From this
the reader can check that `=' has the extension
$$\mbox{ext}(=) = \left\{
\langle a, a \rangle,
\langle b, b \rangle,
\langle c, c \rangle,
\langle d, d \rangle,
\langle d, c \rangle,
\langle c, d \rangle
\right\} \ ;$$i.e.~$c$ and $d$ are indiscernible. So `='
is \emph{not} interpreted as identity (cf. the relativization of `=' to the language, discussed in
3(ii) of \S\ref{HBsec}).
The relation `=' carves the domain into three equivalence classes: $[a] = \{a\}, [b] = \{b\}$
and $[c] = [d] = \{c, d\}$.

\begin{figure}[h!]
   \centering
   \includegraphics{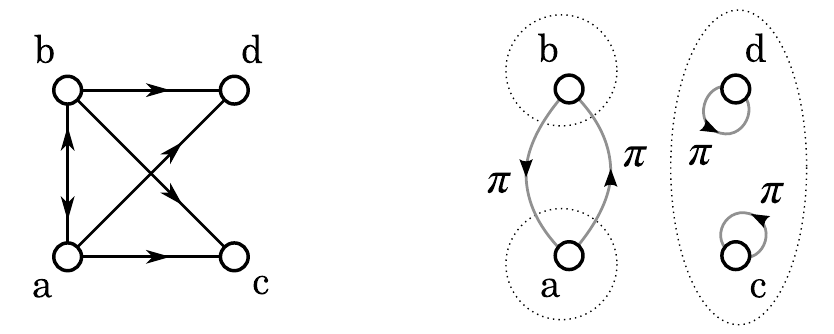}
   \caption{A counterexample to the claim that symmetries leave invariant the indiscernibility
classes.  In this structure, drawn on the left, the permutation $\pi$ which swaps $a$ and $b$
is a symmetry; but $a$ and $b$ form their own separate equivalence classes under `=', as
shown on the right.}
   \label{fig: sym_ind}
\end{figure}

Now consider the permutation $\pi$, whose only effect is to interchange the objects $a$ and
$b$; i.e. $\pi := \left(\begin{array}{c} abcd\\bacd \end{array} \right) \equiv (ab)$.  This
permutation is a symmetry, since it preserves the extension of $R$ according to the
requirement (\ref{symdef}); yet it does \emph{not} leave invariant the equivalence classes
$[a]$ and $[b]$ (Fig.~\ref{fig: sym_ind}).\footnote{
Two remarks. (1): Agreed, this counterexample could be simplified.  A structure with just $a$ and $b$, with ext$(R) = \left\{\langle a,b\rangle,\langle b,a\rangle\right\}$ has $a, b$ discernible, but the swap $a \mapsto b, b \mapsto a$ is a symmetry.  But our example will also be used later.  (2) Accordingly, in our counterexample, $(ab)(cd)$ would work equally well: i.e. it also is a symmetry that does not leave invariant $[a]$ and $[b]$. Looking ahead: Theorem 1 in Section \ref{parag1} will imply that $\{a, b\}$ and $\{c, d\}$ are each subsets of absolute indiscernibility classes; in fact, each is an absolute indiscernibility class.}

(2): {\em Only the trivial symmetry?}:--- Suppose now that in $\cal I$, the predicate `=' {\em is} interpreted as
identity; and that (HB) holds in $\cal I$. So we are supposing that the objects are
various enough, the language rich enough, that indiscernibility  in $\cal I$ is identity. Or
in other words: the indiscernibility classes are singletons. On these suppositions, one might
conjecture that that the only symmetry is the trivial one, i.e. the identity map $id: D \raw
D$.  (Such structures are often called `rigid' (Hodges [1997], p.~94).)

\begin{figure}[h!]
   \centering
   \includegraphics{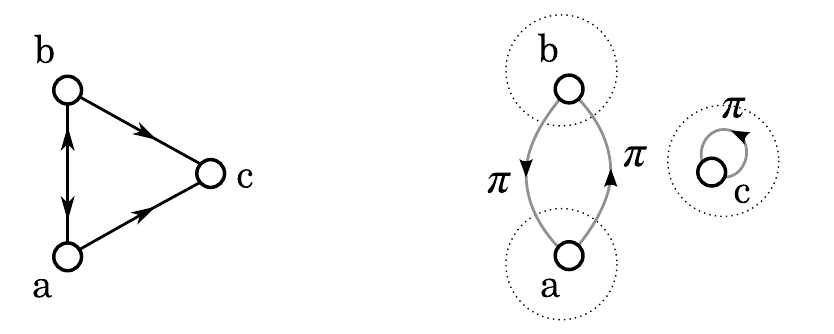}
   \caption{A counterexample to the claim that if the indiscernibility classes are
singletons, the only symmetry is the identity map.  In this structure, the indiscernibles of Figure \ref{fig: sym_ind} have been identified.  $a$ and $b$ are still distinct (they are discernible) but are swapped by
the symmetry $\pi$. }
   \label{fig: sym_ind3}
\end{figure}

In fact, this is false. Comment (1) has just shown that there are symmetries that do not
leave invariant the indiscernibility classes. Our present suppositions have now collapsed
these indiscernibility classes into singletons. So  there will be symmetries which do not
leave invariant the singletons, i.e. are not the identity map on $D$. To be explicit, consider the structure, and its indiscernibility classes, drawn in
Fig. \ref{fig: sym_ind3}.

\section{Four kinds of discernment}\label{4kinds}

We turn to defining the different ways in which two distinct objects can be discerned in a structure.  These kinds of discernment will be developed (indirectly) in terms of which conjuncts on the RHS of the Hilbert-Bernays axiom (HB) are false in the structure.\footnote{Other authors, notably Muller and Saunders [2008], consider the discernment of two objects \emph{by a theory} (say $T$), so that the RHS of (HB), applied to the two objects in question, is a theorem of $T$; whereas our concern is the discernment of two objects \emph{in a structure}.  Our focus on structures is necessary: given our ban on names, the sentence expressing the satisfaction of the RHS of (HB) by the two objects cannot even be written!}

\subsection{Three preliminary comments}\label{4comms}
\indent  (1) {\em Precursors}: Broadly speaking, our four kinds of discernment follow the
discussions by Quine ([1960], [1970], [1976]) and Saunders ([2003a], [2003b], [2006]). Quine ([1960], p. 230)
endorses the Hilbert-Bernays account of identity and then distinguishes what he calls
absolute and relative discernibility. His absolute discernibility will correspond to (the
disjunction of) our first two kinds---and we will follow him by calling this disjunction
`absolute'. Besides, his relative discernibility will correspond to (the disjunction of) our
third and fourth kinds. But we will follow Saunders ([2003a], p. 5; [2003b], pp. 19-20; [2006], p. 5) by reserving
`relative' for the third kind, and using `weak' for the fourth. (So for us `non-absolute'
will mean `relative or weak'.)\footnote{Quine ([1976], p. 113) defines what
he calls grades of discriminability, which is a spectrum of strength.  Saunders ([2006], pp. 19-20) agrees that there is such a spectrum of strength, although in his [2003a] (p. 5) he makes the three categories `absolute', `relative' and `weak' mutually exclusive.  We say `kinds' not `categories' or `grades' to avoid the connotation of mutual exclusion or a spectrum of strength.}

\indent (2) {\em Suggestive labels}: We will label these kinds with words like `intrinsic'
which are vivid, but also connote metaphysical doctrines and controversies (e.g. Lewis 1986,
pp. 59-63). We disavow the connotations: the official meaning is as defined, and so
is relative to the interpretation of
the non-logical vocabulary.

\indent (3): {\em Two pairs yield four kinds}: Our four kinds of discernment arise from two
pairs.  We begin by distinguishing between a formula with one free variable (labelled {\bf
1}) and a formula with two free variables (labelled {\bf 2}).\footnote{We thank Leon Horsten for the observation that the notion of discernibility may be parameterized to other objects (so that, e.g., we might say that $a$ is discernible from $b$ relative to $c, d, \ldots$), which would involve formulas with more than two variables.  The idea seems to us workable, but we will not pursue it here.}  Each of these cases is then
broken down into two subcases (labelled {\bf a} and {\bf b})   yielding four cases in all:
labelled {\bf 1(a)} to {\bf
2(b)}. We will also give the four cases mnemonic labels: e.g. {\bf 1(a)} will also be called
{\bf (Int)} for `intrinsic'.

The intuitive idea that distinguishes sub-cases will be different for  {\bf 1} and {\bf 2}. For {\bf 1}, the idea is to distinguish whether discernibility depends on a relation to another
object; while for {\bf 2}, the idea is to distinguish whether
discernibility depends on an asymmetric relation. Both these ideas are semantic, and even a
bit
vague. But the definitions of the sub-cases will be syntactic, and precise---and we will
therefore remark that they do not completely match the intuitive idea.
But we will argue in Section \ref{parag3} in favour of adopting our syntactic definitions.

As announced in comment 2 of Section \ref{HBsec}, $a$ and $b$ will be names in the meta-language (in which we are writing!) for the one or two objects with which we are concerned.

\subsection{The four kinds defined}\label{4defined}
\begin{description}
	\item [1(a) (Int)] $\!\!$ {\em 1-place formulas with no bound variables,  which apply to only one of
the two  objects $a$ and $b$.} This of course covers the case of primitive 1-place
predicates,
e.g.  $Fx$: so that for example, $a \in$ ext($F$) but $b \notin$ ext($F$), or vice versa.
But we also intend this
case to cover 1-place formulas arising by slotting into a polyadic formula more than one
occurrence of a single free variable, while the polyadic formula nevertheless does {\em not}
contain \emph{any} bound variables.

The intent here is to exclude formulas which quantify over objects other than the two we are concerned with.  So this case will include formulas such as: $Rxx$ and $Fx
\wedge Hxxx$; ($R, H$ primitive 2-place and
3-place predicates, respectively; not abbreviations of more complex open sentences).  But it
will exclude formulas such as  $\forall z(Fz \supset Rzx)$, which contain  bound
variables.\footnote{Recall our ban on individual constants ((2) in
Section \ref{HBsec}). If we had instead allowed them, this sub-case {\bf 1(a)} would be
defined so as to also exclude all formulas containing any constant, including $a$ and $b$.
The
exclusion of formulas such as $Rcx$, which  refer to a third object, is obviously desirable, given the intuitive idea
of discernment by intrinsic properties.   However, the exclusion of formulas involving only
the constants $a$ and-or $b$ may be more
puzzling.  Our rationale is that, for any
formula of the type $Rax$, $Rbx$, etc. which is responsible for discerning two objects, there
will be an alternative formula (either $Rxx$ or $Rxy$) which we would instead credit for the
discernment, and which falls under one of the three other kinds. \label{fn:Gax}}

  We will say that two objects that do not share some monadic formula 
 in this sense are discerned \emph{intrinsically}, since their
distinctness does not rely on any relation either object holds to any other.  An everyday
example, taking `is spherical' as a primitive 1-place
predicate,  is given by a ball and a die.
Another example, with $Rxy$ the primitive 2-place predicate `loves' (so that $Rxx$ is the 1-place predicate `loves his- or herself') is
\textit{Narcissus} $\in$ ext$(R**)$, but (alas) \textit{Echo} $\notin$ ext$(R**)$.

We shall say that any pair of objects  discerned other than by {\bf 1(a)}---i.e. discerned by
{\bf 1(b)},
or by {\bf 2(a)}, or by {\bf 2(b)} below---are {\em extrinsically} discerned.

	\item [1(b) (Ext)]$\!\!$ {\em 1-place formulas with bound variables,  which apply to
only one of the two objects $a$ and $b$.} That is: this kind
contains polyadic formulas that {\em do} contain bound variables. So it contains formulas
such as  $\forall z(Fz \supset Rzx)$. And  $a$ and $b$ are discerned by such formulas if: for
example, $a \in$ ext($\forall z(Fz \supset Rz*)$) but $b \notin$ ext($\forall z(Fz
\supset Rz*)$). We will say that they have been discerned \emph{externally}.  An
example,
taking $F$ = `is a man', $R$ = `admires' is: \textit{Cleopatra} $\in$ ext($\forall z(Fz \supset
Rz*)$)  but \textit{Caesar} $\notin$ ext($\forall z(Fz \supset Rz*)$).  (Recall that we
reserve the term \emph{extrinsic} to cover all three kinds {\bf 1(b), 2(a), 2(b)}: so
external discernment is more specific than extrinsic.)

The intent is that in this kind of discernment, diversity follows from the
relations the two objects $a$ and $b$ have to \emph{other} objects. However, as we said in (3) of Section \ref{4comms}, the precise syntactical definition cannot be expected to match exactly the intuitive idea. And indeed, there are examples of external discernment where the relevant value of the bound variable in the
discerning formula is in fact $a$ or $b$, even though this is invisible from the syntactic
perspective.  (In the example just given, the universal quantifier in $\forall z(Fz \supset Rzx)$ quantifies over a domain that includes Caesar himself.)\footnote{As to our ban on individual constants: if we had instead allowed them, this sub-case {\bf 1(b)} would be
defined so as to also exclude formulas containing $a$ and $b$, but to allow other constants $c, d$ etc., so as to capture the idea of discernment by relations to other objects. But as in the case of bound variables, it could turn out that the ``third'' object picked out is in fact $a$ or $b$. Cf.~footnote \ref{fn:Gax}.}

 \end{description}

 We will say that two objects discerned by a formula either of kind {\bf (Int)} or of kind {\bf (Ext)} are
\emph{absolutely discerned}.
Note that a pair of objects could be both intrinsically and externally
discerned. But since \textbf{(Ext)} is intuitively a ``weaker'' form of discernment, we shall sometimes
say that a pair of objects that are externally, but not intrinsically, discerned, are {\em
merely} externally discerned.

\begin{figure}[h!]
   \centering
   \includegraphics{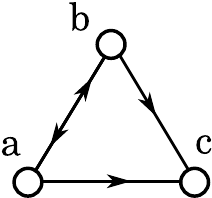}
   \caption{An object's being an individual requires its being absolutely discerned from
\emph{all} others, but not their being absolutely discerned from
anything else.  Here, $c$ is absolutely discerned from both
$a$ and $b$, e.g. by the formula $\forall z(Rzx \supset Rxz)$, while $a$ and $b$ are
themselves non-individuals.}
   \label{fig: non-ind}
\end{figure}

\noindent\emph{Interlude: Individuality and absolute discernment.} \quad
 We will also say that an object that is absolutely discerned from {\em all} other objects is
an {\em individual} or {\em has individuality}.  Note that if an object is an individual,
some or all of the other objects might themselves fail to be individuals (cf. figure
\ref{fig: non-ind}).

Being an individual is tantamount to being the bearer of a uniquely instantiated definite
description: where `tantamount to' indicates a qualification. The idea is: given an individual, we take {\em seriatim} the formulas that absolutely discern it from the other
objects in the structure, and conjoin them and so construct a definite description that is instantiated only by the given individual. The qualification is that in an
infinite domain, there could be infinitely many ways that a given individual was absolutely
discerned from all the various others: think of how a finite vocabulary supports
arbitrarily long formulas, and so denumerably many of them. Thus in an infinite domain the
above ``{\em seriatim}'' procedure might yield an infinite conjunction---preventing a
finitely long uniquely instantiated definite description.\footnote{
 A terminological note: Saunders ([2003b], p.~10) says that an object that is the bearer of a
uniquely instantiated definite description is `referentially determinate', and Quine ([1976], p.~113) calls such an object `specifiable'. So, modulo our
qualification about infinite domains, these terms correspond to our (and Muller \& Saunders' [2008]) use of `individual'. Cf.~also comment 1(vi) in Section \ref{HBsec}. \label{fn:specifiable}}

We will examine absolute discernibility  in Section \ref{symmrevisit}. For the moment, we return to our four kinds of discernibility: i.e.
to presenting the last two kinds. {\em End of Interlude.}

 \begin{description}
 	\item [2(a) (Rel)]$\!\!$ {\em Formulas with two free variables, which are satisfied by
the two
objects $a$ and $b$ in one order, but not the other.} For example, for the formulas $Rxy$ and
$\exists z Hxzy$, we have: $\langle a, b \rangle \in$ ext($R$), but $\langle b, a \rangle \notin$ ext($R$); and $\langle a, b \rangle \in$ ext($\exists z H*z\bullet$), but $\langle b, a \rangle \notin$ ext($\exists z H*z\bullet$).  Here the  diversity of $a$ and $b$ is an extrinsic matter (both intuitively, and
according to our definition of `extrinsic', which is discernment by any means other
than {\bf (Int)}), since it follows from their relation to each
other. But it is not a matter of a relation to any third object.  Following Quine ([1960], p. 230), we will
say that objects so
discerned are \emph{relatively} discerned.

And as above, we will say that objects that are relatively discerned but neither
intrinsically nor externally discerned, are {\em merely} relatively discerned. Merely
relatively discerned objects are never individuals in our sense (viz. absolutely discerned
from all other objects).\footnote{Note that this kind of discernment does not require the
discerning formula, for example $Gxy$, to be asymmetric  for all its instances; i.e., we do not
require $\forall x\forall y(Gxy \supset \neg Gyx)$. In this we agree with Saunders (e.g. [2003a], p.~5)
and Quine ([1960], p.~230). Our rationale is that
intuitively, this kind of discernment does not require anything about the global pattern of
instantiation of the relation concerned. The same remark
applies to {\bf (Weak)} below, where now we differ from Saunders (e.g. [2003a], p.~5), who demands that the discerning relation be irreflexive. Nevertheless, we adopt Saunders'/Quine's word `weak'.}

	\item [2(b) (Weak)]$\!\!$ {\em Formulas with two free variables,  which are satisfied by
the two
objects $a$ and $b$ taken in either order, but not by either object taken twice.}  For example, for the formulas $Rxy$ and $\exists zHxzy$, we have: $\langle a,a \rangle, \langle b,b \rangle \notin$ ext($R$), but $\langle a,b \rangle, \langle b,a \rangle \in$ ext($R$); and   $\langle a,a \rangle, \langle b,b \rangle \notin$ ext($\exists zH*z\bullet$), but $\langle a,b \rangle, \langle b,a \rangle \in$ ext($\exists zH*z\bullet$). (We say `but not by {\em either} object taken twice' to prevent $a$ and $b$ being intrinsically discerned.) Again, the diversity
of $a$ and $b$ is
extrinsic, but does not depend on a third object; rather diversity follows from their
pattern of instantiation of the relation $R$.  We call objects so discerned
\emph{weakly} discerned. And we will say that objects that are weakly discerned but neither
intrinsically nor externally nor relatively discerned (i.e. fall outside {\bf (Int), (Ext),
(Rel)} above), are {\em merely} weakly discerned.

Objects which are discerned {\em merely} weakly are not individuals, in our
sense (since they are not absolutely discerned).  Max Black's famous example of two spheres a
mile apart ([1952], p. 156) is an example of two such objects. For the two spheres bear the relation `is
a mile away from',  one to another; but not each to itself. The irony is that Black,
apparently unaware of weak discernibility, proposes his duplicate spheres as a putative example of
two objects that are qualitatively indiscernible (and therefore as a counterexample to the
principle of the identity of indiscernibles).\footnote{That is, assuming that space is not closed.  In a closed universe, an object may be a non-zero distance from itself, so the relation `is one mile away from' is not irreflexive, and cannot be used to discern.  French's ([2006], \S 4) and Hawley's ([2009], p.~109) charge of circularity against Saunders [2003a] enters here: it seems that the irreflexivity of
`is one mile away from' relies on the prior guarantee that the two spheres are indeed
distinct; but their distinctness is supposed, in turn, to be grounded by that very relation
being irreflexive.  The openness or closedness of space would decide the matter, of course,
but that too seems to stand or fall with the irreflexivity or otherwise of distance
relations---between spatial or spacetime points, if not material objects.  Cf.~also comment 1(iv) in Section \ref{HBsec}.}
 \end{description}

We will also say that two objects that are not discerned by any of our four kinds (i.e. by no 1-place or 2-place formula whatsoever) are {\em indiscernible}. For emphasis, we will sometimes call such a pair {\em utterly indiscernible}. In particular, we will say `utter indiscernibility' when contrasting this case with the failure of only one (or two or three) of our four kinds of discernibility. Of course, utter indiscernibles are only accepted by someone who denies the Hilbert-Bernays account.

This quartet of definitions prompts the question: What are the logical implications between these kinds of discernibility?  We pursue this in the second Appendix (Section 7).

\section{Absolute indiscernibility: some results}\label{symmrevisit}
In Section \ref{HBandsymmy}, we saw that a permutation leaving invariant the indiscernibility classes must be a symmetry; then we gave a counterexample to the converse statement, and to a related conjecture that if indiscernibility is identity, there is only the trivial symmetry. But now that we have defined absolute discernibility (viz. as the disjunction, {\bf (Int)} or {\bf (Ext)}), we can ask about the corresponding claims that use instead the absolute concept. That is the task of this Section. (But its results are hardly needed for the discussions and results in later Sections.)

We will prove that with the absolute concept, Section \ref{HBandsymmy}'s converse statement is ``resurrected'', i.e.  a symmetry leaves invariant the absolute indiscernibility classes  (Section \ref{parag1}). Then we will give some illustrations, including a  counterexample to the converse of {\em this} statement (Section \ref{paragButton}). Then we will show that for a {\em finite} domain of quantification, absolute indiscernibility of two objects is equivalent to the existence of a symmetry mapping one object to the other (Section \ref{parag2}).\footnote{Absolute discernibility and individuality are closely related to \emph{definability} in a formal language; (for example, an object that is definable is an individual in the sense of Section \ref{4defined}'s Interlude).  The interplay between definability (and related notions) and invariance under symmetries is given a sophisticated treatment by da Costa \& Rodrigues [2007], who consider higher-than-first-order structures.  Some of their results have close affinities with our two theorems; in particular their theorems 7.3-7.7.  As in footnote \ref{fn:KetlandCosta}, we thank N.~da Costa.} 

But first, beware of an ambiguity of English.
For relations of indiscernibility, we have a choice of two usages. Should we use `absolute indiscernibility' for just `not
absolutely discernible' (which will therefore include pairs of objects that are discernible,
albeit by other means than absolutely)? Or should we use `absolute indiscernibility' for some
kind (species) of indiscernibility---as, indeed, the English adjective `absolute' connotes?
(And if so, which kind should we mean?)\footnote{This sort of ambiguity is of course not specific to discernment: it is common enough: should we read `recalcitrant immobility' as `not-(recalcitrant mobility)' or as `recalcitrant not-mobility'?}

\begin{figure}[h!]
   \centering
   \includegraphics{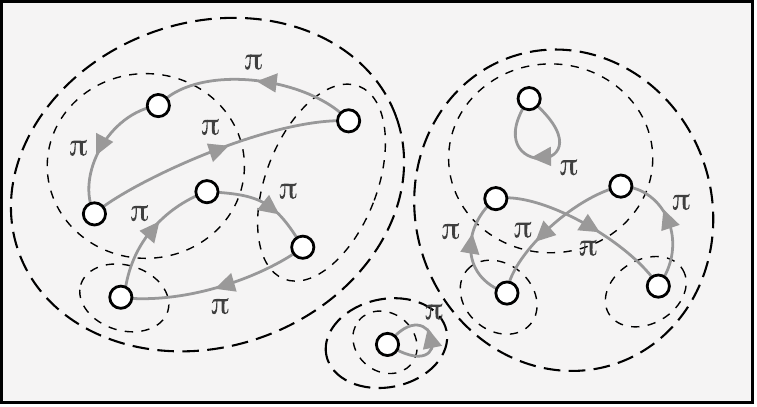}
   \caption{Preview to Section \ref{parag1}'s theorem.  A generic symmetry $\pi$ acting on a domain must preserve the absolute indiscernibility classes (thick broken lines), but may break the indiscernibility classes (thin broken lines). Note also that the object at centre-bottom,  alone in its absolute indiscernibility class and therefore an individual, must be sent to itself under $\pi$.}
   \label{fig:classes}
\end{figure}

 We stipulate that we mean the {\em former}. Then: since absolute discernibility  is a kind of (implies) discernibility, we have, by contraposition: indiscernibility implies absolute indiscernibility (in our usage!). Since both indiscernibility and absolute indiscernibility are equivalence relations, this implies that the absolute indiscernibility classes are unions of the indiscernibility classes; cf.~Figure \ref{fig:classes}. With this definition, Section \ref{parag1}'s theorem will be: a symmetry leaves invariant the absolute indiscernibility classes.
 
Besides, for later use, we make the corresponding stipulation about the phrases `intrinsic indiscernibility', `external indiscernibility' etc. That is: by `intrinsic indiscernibility' and `intrinsically indiscernible', we will mean `not-(intrinsic discernibility)' and `not-(intrinsically discernible)', respectively; and so on for other phrases.

We also admit that instead of Section 3's syntactic approach to defining kinds of discernibility in terms of formulas, one could frame definitions in terms of the semantic ideas of permutations and symmetries.  In the second Appendix (Section 8), we explore the consequences for alternatives, especially as regards this Section's results.

\subsection{Invariance of absolute indiscernibility classes}\label{parag1}
Recall that in Section \ref{HBandsymmy}, we saw that leaving invariant (fixing) the indiscernibility classes was sufficient, but not necessary, for being a symmetry. That is: Section \ref{HBandsymmy}'s counterexample showed that being a symmetry is not sufficient for leaving invariant the indiscernibility classes. This situation prompts the question, what being a symmetry {\em is} sufficient for. More precisely: is there a natural way to weaken Section \ref{HBandsymmy}'s sufficient condition for being a symmetry---viz. indiscernibility invariance---into being instead a necessary condition? In other words: one might conjecture that
leaving invariant some supersets of the indiscernibility classes yields a necessary  condition of being a symmetry.

In fact, our concept of absolute indiscernibility is the natural weakening.  (N.B.~Our ban on names is essential to its being a weakening: allowing names in a discerning formula makes (the natural redefinition of) absolute discernment \emph{equivalent} to weak discernment.)  That is: being a symmetry implies leaving invariant the absolute indiscernibility classes.  Cf.~Figure \ref{fig:classes}. We will first prove this, and then give a counterexample to the converse statement: it will be similar to the counterexample used in Section \ref{HBandsymmy} against that Section's converse statement.

{\bf Theorem 1}: For any structure (i.e. interpretation of a first-order language): if a permutation is a symmetry, then it leaves
invariant the absolute indiscernibility classes.

{\em Proof}: We prove the contrapositive: we assume that  there is some element $a$ of the domain  which is absolutely discernible from its image $b := \pi(a)$ under the permutation $\pi$, and we prove that $\pi$ is not a symmetry.  (Remember that `$a$' and `$b$' are names in the metalanguage only; we stick to our ascetic object-language demands set down in comment 2 of Section \ref{HBsec}.\footnote{We are extremely grateful to Leon Horsten for making us aware of the problems with a previous version of this proof, in which we reintroduced names into the object language.}) So our assumption is that for some object $a$ in the domain, with $b :=\pi(a)$, there is some formula $\Phi(x)$ with one free variable for which $a \in$ ext($\Phi$) while $b \notin$ ext($\Phi$), or vice versa ($a \notin$ ext($\Phi$) while $b \in$ ext($\Phi$)):

The proof proceeds by induction on the logical complexity of the absolutely discerning formula $\Phi$.  From the assumption that $a \in$ ext($\Phi$) iff $b \notin$ ext($\Phi$), we will show that, whatever the main connective or quantifier used in the last stage of the stage-by-stage construction of $\Phi$, there is some logically simpler open formula, perhaps with more than one ($n$, say) free variable, $\Psi(x_1, \ldots x_n)$, and some objects $o_1, \ldots o_{n} \in D$ (not necessarily including $a$, and not necessarily $n$ in number, since maybe $o_i = o_j$ for some $i \neq j$) such that we have:    $\langle o_1, \ldots o_{n}\rangle \in$ ext($\Psi$) iff $\langle \pi(o_1), \ldots \pi(o_{n})\rangle \notin$ ext($\Psi$), where $\pi(o_i)$ is the image of $o_i$ under the permutation $\pi$.  That is,
we continue to break $\Psi$ down to logically simpler formulas until we obtain some \emph{atomic} formula whose differential satisfaction by some sequence of objects and the sequence of their images under the permutation $\pi$ directly contradicts $\pi$'s being a symmetry.

Our proof begins by setting $\Psi := \Phi(x)$ (so to start with, the adicity $n$ of our formula equals 1 and our objects $o_i$ comprise only $a$). We then reiterate the procedure until we reach an atomic formula. Thus:---

\emph{Step one.} We have that $\langle o_1, \ldots o_n\rangle \in$ ext($\Psi$) iff $\langle \pi(o_1), \ldots \pi(o_n)\rangle \notin$ ext($\Psi$).  (Remember that to start with, we set $n=1$ and $o_1=a$.  And the `iff' means only material equivalence.)

\emph{Step two.}  Proceed by cases:
\begin{itemize}
	\item If $\Psi(x_1, \ldots x_n) = \neg\xi(x_1, \ldots x_n)$, then we have \\
	$\langle o_1, \ldots o_n\rangle \in$ ext($\neg\xi$) iff $\langle \pi(o_1), \ldots \pi(o_n)\rangle \notin$ ext($\neg\xi$); that is,\\
	 $\langle o_1, \ldots o_n\rangle \in$ ext($\xi$) iff $\langle \pi(o_1), \ldots \pi(o_n)\rangle \notin$ ext($\xi$);\\
	  so $\xi$ is our new, simpler $\Psi$.
	
	\item If $\Psi$ is a conjunction, then we can write\\
	$\Psi(x_1, \ldots x_n) = \big(\xi(x_{i(1)},\ldots x_{i(l)}) \ \wedge\ \eta(x_{j(1)}, \ldots x_{j(m)}\big)$,\\
	where $l,m \leqslant n$ and $l+m\geqslant n$, and $i: \{1,2,\ldots l\} \to \{1,2,\ldots n\}$ and $j: \{1,2,\ldots m\} \to \{1,2,\ldots n\}$ are injective maps.   First of all, we recognise that \\
	$\langle o_1, \ldots o_n\rangle \in$ ext($\Psi$)\quad iff\quad  $\langle o_{i(1)}, \ldots o_{i(l)}\rangle \in$ ext($\xi$) and $\langle o_{j(1)}, \ldots o_{j(m)}\rangle \in$ ext($\eta$). \\
	 Then, given step one, namely\\
	 $\langle o_1, \ldots o_n\rangle \in$ ext($\Psi$) iff $\langle \pi(o_1), \ldots \pi(o_n)\rangle \notin$ ext($\Psi$),\\
	 this is equivalent to\\
	 $\langle o_{i(1)}, \ldots o_{i(l)}\rangle \in$ ext($\xi$) and $\langle o_{j(1)}, \ldots o_{j(m)}\rangle \in$ ext($\eta$) \ iff \\
	  $\langle \pi\left(o_{i(1)}\right), \ldots \pi\left(o_{i(l)}\right)\rangle \notin$ ext($\xi$) or $\langle \pi\left(o_{j(1)}\right), \ldots \pi\left(o_{j(m)}\right)\rangle \notin$ ext($\eta)$.\\
That is:\\
$\langle o_{i(1)}, \ldots o_{i(l)}\rangle \in$ ext($\xi$) iff $\langle \pi\left(o_{i(1)}\right), \ldots \pi\left(o_{i(l)}\right)\rangle \notin$ ext($\xi)$, \ or \\
	   $\langle o_{j(1)}, \ldots o_{j(m)}\rangle \in$ ext($\eta$) iff $\langle \pi\left(o_{j(1)}\right), \ldots \pi\left(o_{j(m)}\right)\rangle \notin$ ext($\eta)$.\\
	So either the formula $\xi(x_1, \ldots x_l)$ or $\eta(x_1, \ldots x_m)$ is our new formula $\Psi$; with adicity $l$, respectively $m$, replacing the adicity $n$; and the objects $o_{i(1)}, \ldots o_{i(l)}$, respectively $o_{j(1)}, \ldots o_{j(m)}$ replacing the objects $o_1, \ldots o_n$.  (This is an inclusive `or': if either formula suffices, imagine that only one is chosen to continue the inductive procedure.  Heuristic remarks. (1): It is only in this clause that the process can reduce the number of variables occurring in $\Psi$, and hence the number $n$ of objects under consideration. (2): The next three cases can be dropped in the usual way, if we suppose the language to use just $\neg, \wedge$ as primitive connectives.)
	
	\item If $\Psi = \big(\xi \vee \eta\big)$, then continue with $\Psi = \neg\big(\neg\xi \wedge \neg\eta\big)$.
	
	\item If $\Psi = \big(\xi \supset \eta\big)$, then continue with $\Psi = \neg\big(\xi \wedge \neg\eta\big)$.
	
	\item If $\Psi = \big(\xi \equiv \eta\big)$, then continue with $\Psi = \Big(\neg\big(\xi \wedge \neg\eta\big) \ \wedge \ \neg\big(\eta \wedge \neg\xi\big)\Big).$
	
	\item If $\Psi(x_1, \ldots x_n) = \exists z\xi(z,x_1, \ldots x_n)$, then we have, using `$*$' to mark the $n$-component argument-place,\\
	$\langle o_1, \ldots o_n\rangle \in$ ext($\exists z\xi(z, *)$) iff $\langle \pi(o_1), \ldots \pi(o_n)\rangle \notin$ ext($\exists z\xi(z, *)$).\\
	So we have\\
	$\langle o_1, \ldots o_n\rangle \in$ ext($\exists z\xi(z, *)$) and $\langle \pi(o_1), \ldots \pi(o_n)\rangle \in$ ext($\neg\exists z\xi(z, *)$), or\\
	$\langle o_1, \ldots o_{n}\rangle \in$ ext($\neg\exists z\xi(z, *)$) and $\langle \pi(o_1), \ldots \pi(o_{n})\rangle \in$ ext($\exists z\xi(z, *)$).\\
	 That is:\\
	$\langle o_1, \ldots o_{n}\rangle \in$ ext($\exists z\xi(z, *)$) and $\langle \pi(o_1), \ldots \pi(o_{n})\rangle \in$ ext($\forall z\neg\xi(z, *)$), or\\
	$\langle o_1, \ldots o_{n}\rangle \in$ ext($\forall z\neg\xi(z, *)$) and $\langle \pi(o_1), \ldots \pi(o_{n})\rangle \in$ ext($\exists z\xi(z, *)$).
	\begin{itemize}
	\item The first disjunct entails that there is some object in $D$---call it $c$---for which\\
	$\langle c, o_1, \ldots o_{n}\rangle \in$ ext($\xi$) and $\langle \pi(c), \pi(o_1), \ldots \pi(o_{n})\rangle \in$ ext($\neg\xi$), i.e.\\
	$\langle c, o_1, \ldots o_{n}\rangle \in$ ext($\xi$) and $\langle \pi(c), \pi(o_1), \ldots \pi(o_{n})\rangle \notin$ ext($\xi$).\\
	(The second conjunct holds for $\pi(c)$, since it holds for \emph{all} objects in $D$.)
	\item The second disjunct entails that there is some object in $D$---call it $d$---for which\\
		$\langle \pi^{-1}(d), o_1, \ldots o_{n}\rangle \in$ ext($\neg\xi$) and $\langle d, \pi(o_1), \ldots \pi(o_{n})\rangle \in$ ext($\xi$), i.e.\\
		$\langle \pi^{-1}(d), o_1, \ldots o_{n}\rangle \notin$ ext($\xi$) and $\langle d, \pi(o_1), \ldots \pi(o_{n})\rangle \in$ ext($\xi$).\\
	(The first conjunct holds for $\pi^{-1}(d)$, since it holds for \emph{all} objects in $D$.)
	\end{itemize}
	But we can give $\pi^{-1}(d)$ the name $c$; so that we can recombine the disjuncts and conclude that, for some object $c$ in $D$,
	$\langle c, o_1, \ldots o_{n}\rangle \in$ ext($\xi$) iff $\langle \pi(c), \pi(o_1), \ldots \pi(o_{n})\rangle \notin$ ext($\xi$).\\
	So the formula $\xi(x_1, \ldots x_{n+1})$  is our new $\Psi$, $n+1$ is our new adicity, and $c, o_1, \ldots o_{n}$ are our new objects. (Heuristic remark: It is only in this clause that the process can increase, by one, the adicity of $\Psi$, and hence the number of objects under consideration.)
	
	\item If $\Psi(x_1, \ldots x_n) = \forall z\xi(z,x_1, \ldots x_n)$, then continue with $\Psi(x_1, \ldots x_n) = \neg\exists z\neg\xi(z,x_1, \ldots x_n)$. (Heuristic remark: This case can be dropped in the usual way, if we suppose the language to use just $\exists$ as the primitive quantifier.)
	
	\item If $\Psi$ is an atomic formula, then:
	\begin{itemize}
	\item either $\Psi = F^1_i$ for some primitive 1-place predicate $F^1_i$, in which case:
	\begin{center} $o_1$ ext($F^1_i$) \quad iff\quad $\pi(o_1) \notin$ ext($F^1_i$); \end{center}
	\item or $\Psi = G^2_j$ for some primitive 2-place predicate $G^2_j$, in which case:
	\begin{center} $\langle o_1, o_2\rangle \in$ ext($G^2_j$)\quad iff\quad $\langle \pi(o_1), \pi(o_2)\rangle \notin$ ext($G^2_j$)  \end{center}
	(we emphasize that this case includes the 2-place predicate $G^2_j$ being `=', i.e.~equality);
	\item and so on for any 3- or higher-place predicates.
	\end{itemize}
	Each case directly contradicts the original assumption that $\pi$ is a symmetry (cf.~(\ref{symdef})).  

\end{itemize}

\noindent\emph{End of proof}

{\bf Corollary 1}: An individual is sent to itself by any symmetry.\\
Proof: Section \ref{4defined} defined an individual as an object that is absolutely discerned from every other object. So its absolute indiscernibility class is its singleton set. QED.\\
 \indent This implies, as a special case,  the ``resurrection'' of Section \ref{HBandsymmy}'s second conjecture. That is, we have

{\bf Corollary 2}: If all objects are individuals, the only symmetry is the identity map.

\subsection{Illustrations and a counterexample}\label{paragButton}
We will illustrate Theorem 1 and Corollary 2, with examples based on those in Section \ref{HBandsymmy}. Roughly speaking, these examples will show how Section \ref{HBandsymmy}'s counterexamples to its two conjectures are ``defeated'' once we consider absolute indiscernibility instead of utter indiscernibility. Then we will give a counterexample to the  converse of Theorem 1.

\begin{figure}[h!]
   \centering
   \includegraphics{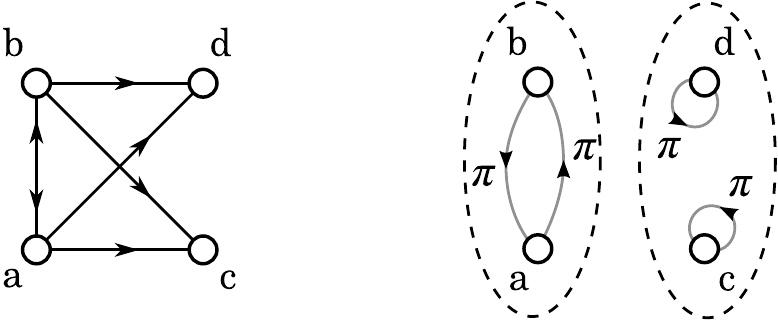}
   \caption{Illustration of Theorem 1, viz. that a symmetry leaves invariant the equivalence classes for the relation `is absolutely indiscernible
from'. Here, the symmetry $\pi$ from Fig.~\ref{fig:
sym_ind} leaves invariant the absolute indiscernibility classes, shown on the
right.}
   \label{fig: sym_ind2}
\end{figure}

{\em Theorem 1 illustrated}:--- In Section \ref{HBandsymmy}'s counterexample (1), $a$ and $b$ are absolutely indiscernible. Thus Figure \ref{fig: sym_ind2} illustrates the theorem.

\begin{figure}[h!]
   \centering
   \includegraphics{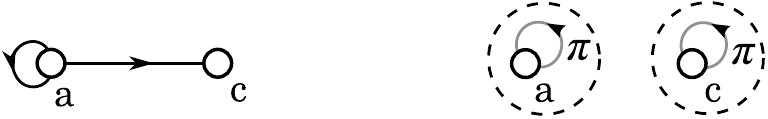}
   \caption{Illustration of Corollary 2. When `is not absolutely discernible from' is taken as identity, the only symmetry for each structure is the identity map.}
   \label{fig: sym_ind4}
\end{figure}

{\em Corollary 2 illustrated}:---  
 We similarly illustrate Corollary 2 by modifying Section \ref{HBandsymmy}'s second counterexample, i.e. counterexample (2) (Figure \ref{fig: sym_ind3}) to Section \ref{HBandsymmy}'s second conjecture. The rough idea is to identify absolute indiscernibles; rather than just {\em utter} indiscernibles (as is required by (HB)). But beware: identifying absolute indiscernibles that are not utter indiscernibles will lead to a contradiction. Figure \ref{fig: sym_ind3} (and also Figure \ref{fig: sym_ind2}) is a case in point: it makes true $Rab$ and $\neg Raa$, so that if you identify $a$ and $b$, you are committed to the contradiction between $Raa$ and $\neg Raa$.  But by increasing slightly the extension of $R$, turning absolute indiscernibles into utter indiscernibles, we can give an illustration of Corollary 2, based on Figure \ref{fig: sym_ind3}, which avoids contradiction. Namely, we  require  that $Raa$ and $Rbb$; this makes $a$ and $b$  utterly indiscernible, not merely absolutely indiscernible. Then we identify $a$ and $b$, yielding Figure \ref{fig: sym_ind4}.

\begin{figure}[h!]
   \centering
   \includegraphics{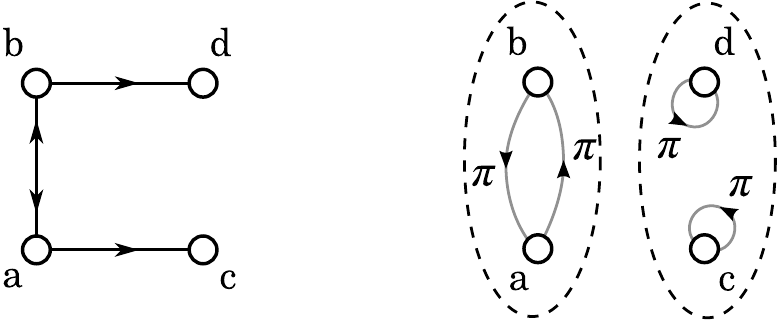}
   \caption{A counterexample to the claim that if the absolute indiscernibility classes are left invariant by the permutation $\pi$, then $\pi$ is a symmetry.}
   \label{fig: sym_ind5}
\end{figure}

{\em Against the Theorem's converse}:---
We turn to showing that Theorem 1's converse does not hold: there are structures for which there are permutations which preserve the absolute indiscernibility classes, yet which are not symmetries.  Consider the structure in Figure \ref{fig: sym_ind5}.\footnote{We thank Tim Button for convincing us of this, and for giving this counterexample.}  In this structure the relation $R$ has the extension ext$(R)=\left\{
\langle a,b \rangle,
\langle b,a \rangle,
\langle a,c \rangle,
\langle b,d \rangle
\right\}$. So as in Fig \ref{fig: sym_ind} (i.e. the counterexample in (1) of Section \ref{HBandsymmy}), $a, b$ are weakly discernible, and $c,d$ are \ind. But $a$ and $b$ are \abs ly \ind.
 (Proof using Theorem 1: the permutation $(ab)(cd)$ is a symmetry, so $\{a,b\}$ and $\{c,d\}$ must each be (subsets of) absolute indiscernibility classes.)
 Then the familiar permutation $\pi$, which just swaps $a$ and $b$, clearly preserves the absolute indiscernibility classes.
 Yet $\pi$ is not a symmetry, since e.g. $\langle a,c \rangle \in$ ext$(R)$, but $\langle \pi(a),\pi(c) \rangle = \langle b,c \rangle \notin$ ext$(R)$.

Figure \ref{fig: sym_ind5} illustrates the general reason why the class of symmetries is a subset of the class of permutations that leave invariant the absolute indiscernibility classes. Namely: for a permutation to be a symmetry, it is not enough that it map each object to one absolutely indiscernible from it; it must, so to speak, drag all the related objects along with it.  For example, in Figure \ref{fig: sym_ind5}, it is not enough to swap $a$ and $b$; the objects ``connected" to them, namely $c$ and $d$ respectively, must be swapped too.  (In more complex structures, we would then have to investigate the objects ``connected" to these secondary objects, and so on). 

To sum up: this counterexample, together with Theorem 1 and the results of Section \ref{HBandsymmy},  place symmetries on a spectrum of logical strength, between two varieties of permutations defined using our notions of utter indiscernibility and absolute indiscernibility.  That is: for a given structure, we have:

\begin{center}
\begin{minipage}{3.5cm} \begin{center} $\pi$ leaves invariant the indiscernibility classes \end{center}\end{minipage} $\xLongrightarrow{(\textrm{cf.}~\S2.2.2)}$
\begin{minipage}{3cm} \begin{center} $\pi$ is a symmetry \end{center}\end{minipage} $\xLongrightarrow{(\textrm{cf.~Th.~1})}$ 
\begin{minipage}{3.5cm} \begin{center} $\pi$ leaves invariant the absolute indiscernibility classes  \end{center}\end{minipage}
\end{center}

\subsection{Finite domains: absolute indiscernibility and the existence of 
symmetries}\label{parag2}
For structures with a finite domain of objects, there is a partial converse to Section \ref{parag1}'s Theorem 1: viz. that if $a$ and $b$ are absolutely indiscernible, then there is a symmetry that sends $a$ to $b$. To prove this, we will temporarily expand the language to contain a name for each object. We will also use the Carnapian idea of a {\em state-description} of a structure (Carnap [1950], p.~71). This is the conjunction of all the true atomic sentences, together with the negations of all the false ones. But for us, the state-description should also include the conjunction of all the true statements of non-identity between the objects in the domain. This will ensure that a map that we will need to define in terms of the state-description is a bijection (and thereby a symmetry).
 
We see no philosophical or dialectical weakness in the proof's adverting to these non-identity statements. But we agree that the reason it is legitimate to include them is different, according to whether you adopt the Hilbert-Bernays account of identity or not. Thus the opponent to the Hilbert-Bernays account will include in the state-description 
all the non-identity sentences holding between any two objects in the 
domain; for the state-description is to be a \emph{complete} description of the structure, so these
non-identity facts should be included. For the proponent of the Hilbert-Bernays account, on the other hand, facts about identity and non-identity are entailed by the 
qualitative facts, in accordance with (HB). So a description of a structure (in particular, a Carnapian state-description) can be complete, i.e. express all the facts, without explicitly including all the true non-identity sentences.  But it is also harmless to include them as conjuncts in the state-description.\footnote{Harmless, that is, provided the HB-advocate is clear-headed.  Recall from comment 1 of Section \ref{HBsec}, that the proponent of the Hilbert-Bernays account assumes that the language is rich enough, or that the domain is varied
enough, for each object  to be  discerned in some way (maybe: relatively or weakly) from every other.  Thus one could also argue that this assumption involves no loss of generality: for if it does not hold for a structure, then the indiscernible objects are to be identified; or else---on pain of contradiction for the HB-advocate---the primitive vocabulary is to be expanded so as to discern them, and the proof is then run again with a structure of discerned objects.}

We will state the Theorem as a logical equivalence, although one implication (the leftward one) is just a restatement of Section \ref{parag1}'s Theorem 1, and so does not need the assumption of a finite domain.

{\bf Theorem 2}: In any finite structure, for any two objects $x$ and $y$:\\
$x$ and $y$ are absolutely indiscernible $\Longleftrightarrow$ there is some symmetry $\pi$ such that $\pi(x) = y$.

{\em Proof}: \emph{Leftward}:
This direction is an instance of Section \ref{parag1}'s Theorem 1, that symmetries leave invariant the absolute indiscernibility classes.\\ 
\indent\emph{Rightward:} Consider an arbitrary finite structure with $n$ distinct objects, $o_1, o_2, \ldots o_i, \ldots o_n$ in its domain $D$, and any two absolutely indiscernible objects in that domain, $o_1, o_2$  (so we set $x = o_1, y = o_2$).  We temporarily expand the language to include a name $\hat{o}_i$ for each object $o_i$.\footnote{Some readers---especially those worrying about our sudden use of names in the object-language---may like to convince themselves that one achieves the same results if, within each absolute indiscernibility class, names are permuted among their denotations.  Therefore the proof is invariant under permutations of non-individuals.}  Then:
\begin{enumerate}
	
	\item Construct the state-description $\mathbf{S}$ of the structure. Thus for example, if the language has just one primitive 1-place predicate $F$ and one primitive 2-place predicate $R$, we define:
	$$
	\mathbf{S} := \bigwedge_{i,j} S_{ij} \ \wedge \ \bigwedge_i S_{i} \ \wedge \
	\bigwedge_{i < j}\hat{o}_{i} \neq \hat{o}_{j}\ \wedge \ \forall z_0\left(\bigvee_{i} z_0=\hat{o}_i\right)
	$$

	$$
	\mbox{where} \qquad S_{ij}:=\left\{
	\begin{array}{rl}
		R\hat{o}_i\hat{o}_j & \mbox{if} \ \langle o_i, o_j \rangle \in \mbox{ext}(R)\\
		\neg R\hat{o}_i\hat{o}_j & \mbox{if} \  \langle o_i, o_j \rangle \notin \mbox{ext}(R)\\
	\end{array}
	\right.
	$$

	$$
	\mbox{and} \qquad S_{i}:=\left\{
	\begin{array}{rl}
		F\hat{o}_i & \mbox{if} \ o_i \in \mbox{ext}(F)\\
		\neg F\hat{o}_i & \mbox{if} \ o_i \notin \mbox{ext}(F)\\
	\end{array}
	\right.  .
	$$

	\item Define the $n$-place formula $\varsigma$ from $\mathbf{S}$ by replacing all 
instances of each name with an instance of a free variable $z_1,..., z_n$. Writing $\frac{\hat{o}_1}{z_1}$ for the substitution of $\hat{o}_1$ by $z_1$ etc.,  we define:
	$$
	\varsigma := \varsigma(z_1,z_2,\ldots z_n) := 
\mathbf{S}\left(\frac{\hat{o}_1}{z_1},\frac{\hat{o}_2}{z_2},\frac{\hat{o}_3}{z_3},\ldots \frac{\hat{o}_n}{z_n} \right).
	$$
	Note that $\mathbf{S}$ is $\varsigma(\hat{o}_1,\hat{o}_2,\hat{o}_3 \ldots \hat{o}_n)$.

	\item From $\varsigma(z_1,\ldots z_n)$, we define $n$ one-place formulas, the $i$th being the ``finest description" of the $i$th object in $D$, i.e.~$o_i$, by existentially quantifying over all but the $i$th variable, which is itself replaced by another variable, say $x$:
	$$
	\Sigma_{i}(x) := \exists z_1\ldots\exists z_{i-1}\exists z_{i+1}\ldots\exists z_n\ \varsigma(z_1, \ldots z_{i-1}, x, z_{i+1}, \ldots z_n).
	$$

	\item The structure clearly makes  true  $\Sigma_1(\hat{o}_1)$, since it is entailed by $\mathbf{S}$.  But by our assumption, ${o}_1$ is absolutely indiscernible from ${o}_2$.  This means that $o_1$ and $o_2$ satisfy the same one-place formulas.  So the structure also makes true  $\Sigma_1(\hat{o}_2)$.  So $\mathbf{S}$ also entails $\Sigma_1(\hat{o}_2)$.

	\item We apply existential instantiation to every existential quantifier in $\Sigma_1(\hat{o}_2)$. Formally, this introduces $n-1$ {\em new} names, say $\alpha_1,...,\alpha_{n-1}$. But  we know (from the last conjunct in $\mathbf{S}$) that there are at most $n$ objects in $D$; so each of the $\alpha_k$ must name one of the $o_1,..., o_n$.  And the non-identity conjuncts in $\mathbf{S}$ also entail that any pair of names among the $\alpha_1, \ldots \alpha_{n-1}$ denote distinct objects, all of which are distinct from $o_2$. Thus we infer that $\mathbf{S}$ entails the sentence
	$$
	\varsigma(\hat{o}_2, \hat{o}_{\alpha(1)}, \hat{o}_{\alpha(2)} \ldots \hat{o}_{\alpha(n-1)})
	$$
where the bijective map $\alpha: \{1, 2,3, \ldots n-1\} \to \{1, 3,4, \ldots n\}$ is defined by the fact that for each $k = 1, \ldots n-1$ the new name $\alpha_k$ refers to $o_{\alpha(k)}$.
	
	\item We now construct a map $\pi$ on $D$ using the sentences $\mathbf{S}$ and $\varsigma(\hat{o}_2, \hat{o}_{\alpha(1)}, \hat{o}_{\alpha(2)} \ldots \hat{o}_{\alpha(n-1)})$ as follows:
	$$
	\begin{array}{rcrllllll}
		\mathbf{S}&\equiv&  \varsigma(& \hat{o}_1, &  \hat{o}_2, & \hat{o}_3, & \ldots  & \hat{o}_n & )\\\\
		&& &\bigg\downarrow\pi&\bigg\downarrow\pi&\bigg\downarrow\pi&\cdots&\bigg\downarrow\pi \\\\
		&&  \varsigma(& \hat{o}_2, & \hat{o}_{\alpha(1)}, & \hat{o}_{\alpha(2)}, & \ldots & \hat{o}_{\alpha(n-1)} & )

	\end{array}
	$$
That is: $\pi$ maps $o_1$ to $o_2$, and $o_2$ to $o_{\alpha(1)}$, and $o_3$ to  $o_{\alpha(2)}$, and so on for 
all the objects in $D$.
	
	\item Then $\pi$ is a bijection, since all the $o_2, o_{\alpha(1)}, \ldots 
o_{\alpha(n-1)}$ are distinct.  And $D$ is the range of $\pi$.  So $\pi$ is  a permutation.  But by construction $\pi$ is also a symmetry.  This is because: (i) it induces a map from the state-description $\mathbf{S}$ to another state-description $\varsigma(\hat{o}_2, \hat{o}_{\alpha(1)}, \hat{o}_{\alpha(2)}, \ldots \hat{o}_{\alpha(n-1)})$; (ii) the latter sentence is entailed by the former and, since both are state-descriptions, both sentences have maximal logical strength; so (iii) the sentences are materially equivalent; and (iv) this material equivalence, together with their maximal logical strength, entails that the extensions of all primitive predicates are preserved under $\pi$.  Thus $\pi$ is the symmetry sought.
	
\end{enumerate}
{\em End of proof}

We now sketch two examples showing the need for finiteness in the statement of Theorem 2. The first uses a countable domain and is sufficient on its own to prove the need for finiteness in Theorem 2; the second, which involves an uncountably infinite domain, is unnecessary, but illustrative.

The countable example is familiar (e.g.~Boolos \& Jeffrey [1974], p.~191).  We use the fact that first-order arithmetic, in the language $\mathcal{L_A}$ with primitive symbols $s, +, \times$,\footnote{Remember (comment 2 in Section \ref{HBsec}) that we ban names from primitive vocabularies, so we take $\mathcal{L_A}$ not to contain $\mathbf{0}$, the name assigned to 0 (the number zero) in the standard model of arithmetic, as a primitive.  The standard results which we use here are not affected, for we can \emph{introduce} $\mathbf{0}$ by description, in terms of the other primitive vocabulary, i.e. $\mathbf{0} := \riota x\ \neg\exists y\ x = s(y)$. \label{fn: zero}} is not $\aleph_0$-categorical, by finding two elementarily equivalent but non-isomorphic structures, which we then use to create a single structure with an absolutely indiscernible pair not related by any symmetry.  Take the standard model of arithmetic $\mathfrak{N}:= \langle\mathbb{N}, 0, s^\mathfrak{N}, +^\mathfrak{N}, \times^\mathfrak{N}\rangle$ and a non-standard model $\mathfrak{N^*}:= \langle\mathbb{N}^*, \emptyset, s^\mathfrak{N^*}, +^\mathfrak{N^*}, \times^\mathfrak{N^*}\rangle$, where $\mathbb{N^*}$ consists of an initial segment, whose first element is an object $\emptyset$ and which is isomorphic to $\mathbb{N}$, followed by countably infinitely many $\mathbb{Z}$-chains, densely ordered without a greatest or least $\mathbb{Z}$-chain.  (So $\mathbb{N^*}$ has the same order type as $\mathbb{N} + \mathbb{Z} \cdot \mathbb{Q}$.)

We then create a new structure $\mathfrak{M} = \langle\mathcal{N}, 0, \emptyset, s^\mathfrak{M}, +^\mathfrak{M}, \times^\mathfrak{M}\rangle := \langle\mathbb{N} \cup \mathbb{N^*}, 0, \emptyset, s^\mathfrak{N} \cup s^\mathfrak{N^*}, +^\mathfrak{N} \cup +^\mathfrak{N^*}, \times^\mathfrak{N} \cup \times^\mathfrak{N^*}\rangle$; (let us assume that the domains $\mathbb{N}$ and $\mathbb{N^*}$ are disjoint, so that $\emptyset \neq 0$).  Now, the structures $\mathfrak{N}$ and $\mathfrak{N}^*$ are elementarily equivalent, so any one-place formula in $\mathcal{L_A}$ which is true of $0$ in $\mathfrak{N}$ is also true of $\emptyset$ in $\mathfrak{N^*}$, and \emph{vice versa}.  So then any one-place formula in $\mathcal{L_A}$ which is true of $0$ in $\mathfrak{M}$ will also be true of $\emptyset$ in $\mathfrak{M}$, and \emph{vice versa}.\footnote{That is of course not to say that the \emph{same propositions} will be true of $0$ in $\mathfrak{M}$ as in $\mathfrak{N}$ (and similarly for $\emptyset$ and $\mathfrak{N^*}$). In particular, the description proposed in fn.~\ref{fn: zero} as a definition of $\mathbf{0}$, namely $\riota x\ \neg\exists y\ x = s(y)$, is \emph{false} of $0$ in $\mathfrak{M}$, because the uniqueness claim fails; but of course the description is also false of $\emptyset$ in $\mathfrak{M}$.}  In that case, $0$ and $\emptyset$ are absolutely indiscernible in $\mathfrak{M}$.  But $\mathfrak{N}$ and $\mathfrak{N^*}$ are not isomorphic; therefore $\mathfrak{M}$ has no symmetries which map $0$ to $\emptyset$ or \emph{vice versa}.  So we have two objects which are absolutely indiscernible but which are not related by any symmetry.\footnote{If we expand $\mathcal{L_A}$ to the \emph{infinitary} language $\mathcal{L_A}(\omega_1, \omega)$ (which allows countably infinite conjunction and disjunction, but only finitary quantification), $0$ and $\emptyset$ become absolutely discernible, since we now have the linguistic resources to distinguish between the order types of $\mathbb{N}$ and $\mathbb{N^*}$.  Specifically, if we define the relation $x<y := \exists t (x+s(t)=y)$, and recursively define the function $s^0(x) := x;\ s^{n+1}(x) := s\left(s^n(x)\right)$, then $0$ and $\emptyset$ are absolutely discerned by the formula
$
\Phi(x) := \forall y \left(x < y \supset \bigvee_{n\in\omega} y = s^n(x)\right)
$, since $0 \in \mbox{ext}_\mathfrak{M}(\Phi)$ but $\emptyset \notin \mbox{ext}_\mathfrak{M}(\Phi)$. Intuitively, the formula $\Phi(x)$ says that every element greater than $x$ lies only finitely far away from $x$.}

Now we provide an example of an uncountably infinite structure in which two objects $a_0$ and $b_0$ are absolutely indiscernible, but no symmetry maps one to the other. In this example, there is just one 2-place relation $R$. We require that $a_0$ bears $R$ to each of denumerably many distinct $a_1, a_2, \ldots a_i, \ldots$; and each of these $a_i$ bears $R$ only to itself.\footnote{So the $a_i$s are weakly discerned from each other by  $\neg Rxy$; and so will be distinct even for a proponent of the Hilbert-Bernays account.} We also require that $b_0$ bears $R$ to each of {\em continuum many} distinct $b_1, b_2, \ldots b_j, \ldots$; and each of these $b_j$ bears $R$ only to itself. Thus we have: 
$$
\mbox{ext}(R) = \left\{
\langle a_0, a_1 \rangle,
\langle a_1, a_1 \rangle,
\langle a_0, a_2 \rangle,
\langle a_2, a_2 \rangle,
\ldots
\langle b_0, b_1 \rangle,
\langle b_1, b_1 \rangle,
\langle b_0, b_2 \rangle,
\langle b_2, b_2 \rangle,
\ldots
\right\}.
$$
with
$$
\mbox{card}(\{x : \langle a_0, x \rangle \in \mbox{ext}(R)\}) = \aleph_0 \ ; \qquad
\mbox{card}(\{x : \langle b_0, x \rangle \in \mbox{ext}(R)\}) = 2^{\aleph_0}.
$$
So $a_0$ and $b_0$ are each like the centre of a wheel, with relations to the $a_i, b_j$ respectively like their wheel's spokes.

\begin{figure}[h!]
   \centering
   \includegraphics{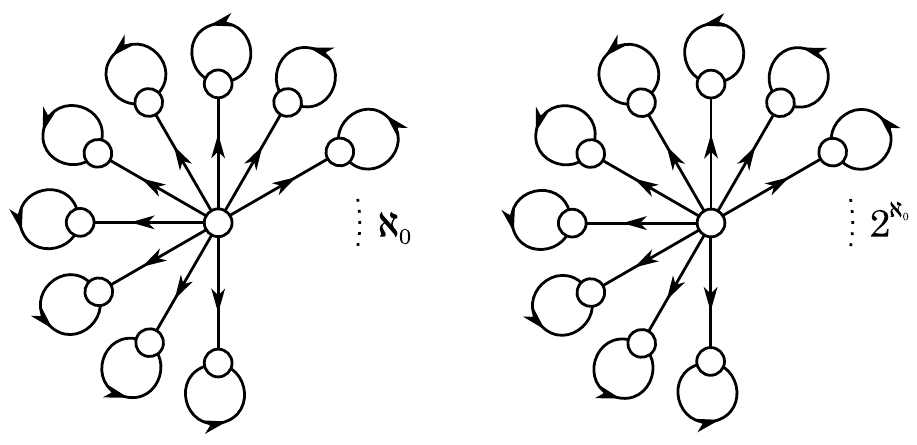}
   \caption{A structure with an uncountable domain in which two objects are absolutely indiscernible, yet no symmetry relates them.}
   \label{fig: uncountable}
\end{figure}

These different cardinalities imply that there is no symmetry  $\pi$ such that $\pi(a_0) = b_0$ or $\pi(b_0) = a_0$. (Recall the discussion after Figure \ref{fig: sym_ind5}, at the end of Section \ref{paragButton}.) But on the other hand, the fact that the ``only difference'' between $a_0$ and $b_0$ is the order of the infinity of the objects to which they are related means that they are absolutely indiscernible; for the L\"owenheim-Skolem theorem implies that no first-order $\Phi(x)$ can express this difference. More precisely: the L\"owenheim-Skolem theorem states that no set of formulas of a first-order language has only models each with denumerably many objects. From this it follows that no first-order $\Phi(x)$ can express the fact that $a_0$ (and not $b_0$) has only denumerably many relata. And this implies that no first-order $\Phi(x)$ can discern $a_0$ and $b_0$; so they are absolutely indiscernible.

So much by way of developing various results about absolute discernibility. 
As we mentioned in this Section's preamble, we discuss alternative \emph{semantic} definitions of absolute discernibility in the second Appendix, Section 8.
Here, we now turn to metaphysical matters.

\section{Four metaphysical theses}\label{4metaph}
We will now state and discuss four rival metaphysical theses, framed in terms of Section \ref{4defined}'s four kinds. We will first define the theses and make some comments about them (Section \ref{the4}). Then we describe how they disagree in  distinguishing possibilities, with one thesis distinguishing two possibilities where another thesis sees only one---thereby condemning the first as proposing a distinction without a difference, or as trading in chimeras (Section \ref{hypercube}). We will not be committed to any of our four theses---our aim is merely to describe!  In Section \ref{structuralism}, we describe a salient recarving of the four theses into two umbrella positions. One of these positions, which we call \emph{structuralism}, is of particular interest, since it prompts formal procedures that are to be found in a variety of modern physical theories (and which we discuss in other papers).

 It will be obvious that several other metaphysical theses could be formulated as readily as our four. But for our purposes, these are the salient ones. 

From now on, this paper shifts its focus from syntax and logic to
semantics and ontology. Accordingly, we imagine that all parties to the debate about identity and indiscernibility envisage a language that is adequate for expressing all facts, in particular facts of identity and diversity (cf.~Adams [1979], p.~7; Lewis [1986], \S4.4); and if they adopt or even just consider the Hilbert-Bernays account, they will imagine deploying it for such a language. (Agreed, the parties will differ about what this language, in order to be adequate, must contain. In particular, we noted in comment 2 of Section \ref{HBsec} that the haecceitist will require the language to express haecceities with predicates $N_a x, N_b x$ etc.)

Three further comments about this shift of focus. They concern, respectively: how uncontentious we will be; the jargon we will use; and our ignoring some modal issues.

 (1): In (2) of Section \ref{4comms}, we officially disavowed any
connotations that words like `intrinsic' or `external' might have, while allowing ourselves
to use them as mnemonic labels. Now that we are entering on metaphysics, this disavowal is
muted, in so far as we now envisage an adequate language.

(2):  Given the discussion of jargon at the end of Section \ref{Convent}, we now adopt the term `individual' in its fully proper sense, as defined in Section \ref{4defined}, as an object that is absolutely discerned from all others, relative to the envisaged ``ideal" language which is adequate for expressing all facts.  Thus `individual' can be used by all parties: though they may well differ about the
vocabulary of this language, and also about which objects (if any!) are individuals.

(3): We emphasise that we will not pursue all the issues about modality that are raised by our theses and our comments on them.  We will mention only briefly and in passing some implications for the identity of objects across structures (our formal representatives of ``possible worlds''); we will not consider modal languages and quantification over structures, and we will steer clear of the issue of essential properties. Of course much of the literature {\em is} thus concerned. But we have plenty to do, while setting aside these issues: this self-imposed limitation will be prominent in some comments in Section \ref{the4}, and in Section \ref{hypercube}.

\subsection{The four theses}\label{the4}
The four theses
fall into two pairs.  Broadly speaking, each thesis of the first pair is part of a reductive account of identity; while the third and fourth theses are explicitly non-reductive about identity. (We will shortly see another, equally good, way to sort the theses into pairs.)

For each thesis, we will mention an author or two who endorses it, or who is sympathetic to it. But beware: some authors formulate their positions in terms of whether facts of identity and diversity are reduced to or grounded in other facts (qualitative or not). But `reduced' and `grounded' are vague, and we will for simplicity instead say `implied by', etc.
 
 The first pair are both versions of the Principle of the Identity of Indiscernibles
(PII), and both conform to the Hilbert-Bernays account of identity. The first is strong; so we write SPII.  It vetoes, not only objects that are indiscernible one from another, but also objects that are merely relatively or merely weakly discerned. So it amounts to a substantial demand on the primitive vocabulary, that it be rich enough to discern any pair of objects absolutely. (One might say: `discern any pair of objects in any structure'. But as discussed in (3) just above, we will not emphasise this sort of universal quantification over structures.) So, according to this thesis, all objects are individuals. This thesis seems to be what Hacking [1975] had in mind, given his treatment of his examples.  Saunders ([2003b], p.~17) ascribes a strengthening of the view to Leibniz, in view of Leibniz's thesis that relational properties are always reducible to the intrinsic properties of the relata; (cf.~comment (2) at the end of this Section).\footnote{The fact that absolute discernibility does not imply intrinsic discernibility, and therefore that a ``Leibnizian" SPII is a strengthening of SPII, was first pointed out by Ladyman ([2007]).}  Adams ([1979], p.~11) says of what appears to be SPII, that it is ``a most interesting thesis, but much more than needs to be claimed in holding that reality must be purely qualitative". 

 The second is a weak version of PII: written WPII. It understands `indiscernibility' as covering all of our four kinds, i.e. as what we called at the end of Section \ref{4defined}, `utter indiscernibility'. So this weak  PII corresponds to the leftward implication in (HB), that indiscernibility implies identity. Since this implication is  the contentious half of (HB), this weak  PII is in effect a metaphysical statement of the
Hilbert-Bernays account. Like the first thesis, this amounts to
a demand on the primitive vocabulary, albeit a weaker one: viz. that it be rich enough to discern any pair of objects, by one or other of our four kinds.  In the words of Adams ([1979], p.~10), the demand on the primitive vocabulary is that each ``thisness" be ``analyzable into, equivalent with, or even identical with, purely qualitative properties or suchnesses".
So, according to this thesis, there can be objects that are not individuals: objects that are only relatively or weakly discerned from at least one other object. To put it as a slogan: there can be 
diversity without individuality. Our central examples of authors who advocate this thesis are Quine ([1960], [1970], [1976]) and Saunders ([2003a], [2003b], [2006]). Other authors who are, or who have been, sympathetic are Robinson ([2000]), Ladyman ([2005]) and Button ([2006]).\footnote{Muller [2011] offers an alternative definition of PII, in which a theory satisfies PII iff $T \vdash \forall x\forall y\left(\mbox{AutInd}(x,y) \supset x = y\right)$, where `AutInd$(x,y)$' satisfied just in case $x$ and $y$ fail
to be even weakly discernible by properties and relations whose extensions are invariant under all automorphisms (what we call symmetries).  The idea is to impose a restriction on the properties
that are permitted to discern.  (Note that it excludes haecceitistic properties, but fails to exclude the use of the identity relation itself.)  It is an alternative way to focus on the purely qualitative properties and relations, other than our strategy of banning names. Therefore it is akin to our WPII, except that Muller's PII applies to theories rather than classes of structures.}

 On the other hand, the third and fourth theses conflict with the Hilbert-Bernays account, in spirit if not in letter. The third thesis is our version of Haecceitism, introduced in Section \ref{subhaec}.  As we noted in footnote \ref{fn:spirithaec} of Section \ref{HBsec}, the existence of haecceities contradicts the spirit of the Hilbert-Bernays account.\footnote{Agreed, our tactic of banning names, and instead using  predicates $N_a x$ etc., enabled the haecceitist to agree with the letter of the Hilbert-Bernays account.} For the latter seeks a reduction of identity to qualitative facts; whereas  the haecceitist denies that identity is in all cases reducible to qualitative facts---some pairs of objects differ just by their non-qualitative thisnesses. We shall take our version of haecceitism to hold that any pair of objects differ by their thisnesses, and thus are absolutely discerned---so that every object is an individual.  (This does not imply that the discerning work is being done by properties rather than objects: recall our earlier remarks in Section \ref{subhaec}.) Authors who advocate this thesis include Kaplan ([1975], pp. 722-3; cf.~footnote \ref{fn:haecKap}) and Adams ([1979], p. 13).  Cf.~also Wittgenstein's \emph{Tractatus}, remarks 2.013 and 4.27.\footnote{Remark 2.013: ``Each thing is, as it were, in a space of possible states of affairs.  This space I can imagine empty, but I cannot imagine the thing without the space."  We interpret this as a commitment to the type of combinatorial independence described in paragraph 2 of Section \ref{subhaec}.
 Also witness remark 4.27: ``For $n$ states of affairs, there are $K_n = \sum_{\nu=0}^n \left(\begin{array}{c}n\\\nu\end{array}\right)$ [$= 2^n$] possibilities of existence and non-existence.  Of these states of affairs any combination can exist and the remainder not exist."  The commitment to haecceitism may not be obvious from this remark, but compare our discussion of a specific example in Section \ref{ss:haec}.}

We note {\em en passant} that the first and third theses---the strong version of PII, and Haecceitism---are perhaps the two main traditional positions in the philosophy of identity. In terms of our usage of the term `individual', they agree that all objects are individuals. But they disagree about whether qualitative indiscernibility is sufficient for being one and the same object (or equivalently, for them: one and the same individual).

Finally, the fourth thesis accepts that a pair of objects can be merely relatively, or merely weakly, discerned---but goes on to deny (HB)'s leftward implication. That is, it denies that indiscernibles must be identical: there are pairs of objects that are utterly indiscernible from one another. In other words: there are utter indiscernibles. But unlike Haecceitism, this thesis does not permit non-qualitative properties which could absolutely discern qualitatively indiscernible objects; so not all objects are individuals. So unlike Haecceitism, this thesis is a denial of the Hilbert-Bernays account in letter as well as in spirit.  Here our ban on names comes in particularly useful: according to the thesis under discussion, `$x=y$' is a legitimate primitive formula of the language, representing a particular relation (namely identity); but `$x=a$', which we demand be written as `$\forall y (N_ay \equiv y=x)$', is illegitimate: it represents \emph{no} property at all, since there are no haecceitistic properties.\footnote{In Robinson's ([2000], pp.~163, 173) terminology, the appropriate language for a proponent of QII is \emph{suitable}, since it vetoes thisness predicates, but not \emph{extra-suitable}, since it takes the equality symbol as a primitive.}

So far as we can tell, this fourth thesis has only recently been formulated, in part in response to Saunders' advocacy of WPII. Ladyman ([2007], p. 37) calls this thesis `contextual
ungrounded identity'. But we will label it QII, standing for `Qualitative Individuality with
Indiscernibles'; (so here, the `II'  does {\em not} mean `identity of indiscernibles'!). Other formulations, close or identical to our QII, are in: Esfeld [2004], Pooley ([2006], [2009]), Esfeld and Lam [2006], Ketland [2006], Ladyman [2007] and Leitgeb \& Ladyman [2008].\footnote{We also think the quasi-set theory developed by Krause and co-authors (e.g.~Krause [1992], Dalla Chiara, Giuntini \& Krause [1998], French \& Krause [2006] (ch.~7), Krause and French [2007] and da Costa \& Krause [2007]) is a kindred position: `kindred' since it allows objects to be permuted without engendering any change (it is therefore apt to describe it as a theory of non-individuals); but only kindred, i.e.~not the same position, since 
it vetoes talking of identity as a relation among such objects, i.e.~`$x = y$' is not a wff if $x$ and $y$ are taken to range over so-called ``$m$-atoms".\label{fn:qset}} 
So this fourth thesis, QII, is:\\
\indent (i) like haecceitism, and unlike the two versions of PII, in that it disagrees with the Hilbert-Bernays account (and also it denies any other reduction of identity to qualitative facts); but also \\
\indent (ii) like the weak version of PII, and unlike both the strong version and Haecceitism, in that it allows there to be objects that are not individuals.

Note that while (i) corresponds to pairing the first two theses, in contrast to the last two (as we announced at the start of this Subsection), (ii) corresponds to pairing the first and third theses (both demand individuals, in our sense), in contrast to the second and fourth theses (which both allow non-individuals, in our sense). 

In our opinion, these two ways of pairing the theses are equally natural. Thus we can consider the four theses as illustrating the four possible combinations of answers to two equally natural Yes-No questions. Namely, the questions:\\
\indent \indent (a): Is indiscernibility sufficient for identity? (Or equivalently, the contrapositive: is diversity (non-identity) sufficient for discernibility? Here `discernibility' is to mean `qualitative discernibility', i.e. it excludes appeal to haecceities.)\\
\indent \indent  (b): Is every object an individual (in our sense of being absolutely discerned from every other object)?\\
Thus the four metaphysical
theses can be placed in the Table below.

\begin{center}
\begin{tabular}{|c|c|cc|} \hline\multicolumn{2}{|c|}{}& \multicolumn{2}{c|}{ Is qualitative indiscernibility sufficient for identity?}
\\ \cline{3-4}
\multicolumn{2}{|c|}{}& \phantom{MMMMM}Yes\phantom{MMMMM} & No \\\hline Is every object & Yes & SPII & Haecceitism \\ an individual? & No & WPII & QII \\ \hline\end{tabular}
\end{center}

To sum up, here is the official statement of our four metaphysical theses.
\begin{description}
	\item [SPII]: {\em The Strong Principle of the Identity of Indiscernibles} \\
Any two objects are absolutely discernible, in Section \ref{4defined}'s sense, by qualitative properties or relations.  Therefore every object is an individual.  The possibility of objects discernible merely relatively or merely weakly is denied.

	\item [WPII]: {\em The Weak Principle of the Identity of Indiscernibles} \\
Any two objects are discernible, through {some} 1- {or} 2-place formula involving only qualitative properties and relations, according to the Hilbert-Bernays axiom.  The gap between discernibility and absolute discernibility allows for objects which are non-individuals; such objects are either merely relatively or merely weakly discernible.

	\item [Haecceitism]: \\
Any object $a$ has a haecceity $N_a x$, a property with no other instances. (But we construe each of these properties ``thinly'': specifically, as identical with the property expressed by `$x=a$'.) This means that the diversity of  objects is primitive in the sense that two objects need not be at all qualitatively discernible.  And every object is an individual. Individuality is also primitive in the sense that an object's absolute discernibility from all others may rely solely on the haecceities, and need not involve qualitative properties or relations.

	\item [QII]: {\em Qualitative Individuality with Indiscernibles} \\
	The diversity of objects is primitive in the sense that two objects can be qualitatively utterly indiscernible.  Individuality, by contrast, is \emph{not} primitive: each object's individuality requires absolute discernment from all others by purely qualitative properties or relations.  The gap between diversity and individuality allows non-individual objects.  So a non-individual is either merely relatively discernible, or merely weakly discernible, or else utterly indiscernible, from at least one other object.
	
\end{description}

Finally, we emphasise that our four theses are not the only occupants of their respective quadrants of logical space, as carved out by the two questions in the Table. Examples are as follows.\\
\indent (1): We saw already in Section \ref{Convent} that there are different versions of haecceitism: how thickly should haecceities be construed?\\
\indent (2): SPII affords another example. One can say Yes to both our questions in different ways than our SPII. One obvious way is by requiring every primitive predicate of the envisaged adequate language to be 1-place.\footnote{Such a language is envisaged by Saunders ([2003b], p.~17) for a modern treatment of Leibniz's metaphysics.  See also footnote \ref{fn:Leibniz}.} And our previous discussion has shown two other less obvious ways:\\
\indent \indent (a): One could require that every object be specifiable in Quine's ([1976]) sense (cf.~the Interlude, and footnote \ref{fn:specifiable}, in Section \ref{4defined});\\
\indent \indent (b): One could allow only those structures that have only the trivial symmetry, i.e.~for which the identity map is the only symmetry. Corollary 2 of Theorem 1 (Section \ref{parag1}) means that in infinite domains this requirement is weaker than all objects being individuals (i.e. our SPII). For it is only in finite domains that they are equivalent (cf. Theorem 2 in Section \ref{parag2}).\\
\indent (3): WPII affords another example. That is: one can say Yes to `Is qualitative indiscernibility sufficient for identity?, and No to `Is every object an individual?', other than by endorsing WPII. For one could allow relative discernibles, but forbid structures containing mere weak discernibles.\\
\indent (4): Finally, QII affords another example. One can say No to both our questions in different ways than our QII: for example, by saying that only some objects have haecceities---this again secures the No answers, that not all objects are individuals, and that there are utter indiscernibles.

\subsection{The theses' verdicts about what is possible}\label{hypercube}
We turn to illustrating exactly what structures each of our four theses permit, for the simple case of exactly two objects $a$ and $b$ and one (qualitative) relation $R$. (This case will be complicated enough!) This exercise will be governed by three main rules.\\
\indent (1): All the state-descriptions will have common elements which we will \emph{not} show explicitly.  Specifically, every state-description contains conjuncts which constrain the cardinality of the domain to be two: $\left(x \neq y\ \wedge\ \forall z (z=x \vee z=y)\right)$, with $x$ and $y$ bound by existential quantification.  For WPII and SPII, the `=' symbol is of course governed by the Hilbert-Bernays axiom (HB); for those metaphysical theses, (HB) should be considered as an additional implicit conjunct in the state-descriptions written below.\\
\indent (2): As usual, the haecceitist is asked to add two haecceitistic predicates, `$N_a$' and `$N_b$', to the primitive vocabulary, one for each object, and associated uniqueness conditions (again implicit in the state-descriptions below).\\
\indent (3): We characterise the structures syntactically; that is, by the logically strongest sentences for which the structures are models.  For finite structures, this is equivalent to characterising structures up to isomorphism.  So when we ``count the structures" allowed by each metaphysical thesis, we are in fact counting the \emph{equivalence classes} of mutually isomorphic structures.  Sometimes we will say `structure' when we mean `equivalence classes of structures'---when counting it is useful only to mean the latter!

\subsubsection{Haecceitism}\label{ss:haec}
We start with Haecceitism, which will turn out to be the most ``generous" metaphysical position, in the sense of allowing more possibilities.  The inclusion of the haecceitistic predicates in the primitive vocabulary allows us always to discern the two objects in any structure.  Consider the state-description sentences
$$
\exists x\exists y (N_ax \wedge N_by \wedge \pm Rxx \wedge \pm Rxy \wedge \pm Ryx \wedge \pm Ryy)
$$
where $\pm Rxy$ designates either $Rxy$ or its negation.\footnote{Here we see how our ban on names (comment 2 in Section \ref{HBsec}) has not lost the haecceitist any expressive power.  This expression is equivalent, in a language with names, to the template sentence $(\pm Raa \wedge \pm Rab \wedge \pm Rba \wedge \pm Rbb)$.}

A decision, for each of the $R$-formulas, whether it is asserted or denied, together with the haecceitistic uniqueness conditions, suffices to completely describe a structure (up to isomorphism).  There are 4 $R$-formulas, so there are $2^4 = 16$ distinct (equivalence classes of) structures.

\begin{figure}[h!]
   \centering
   \includegraphics[scale=0.75]{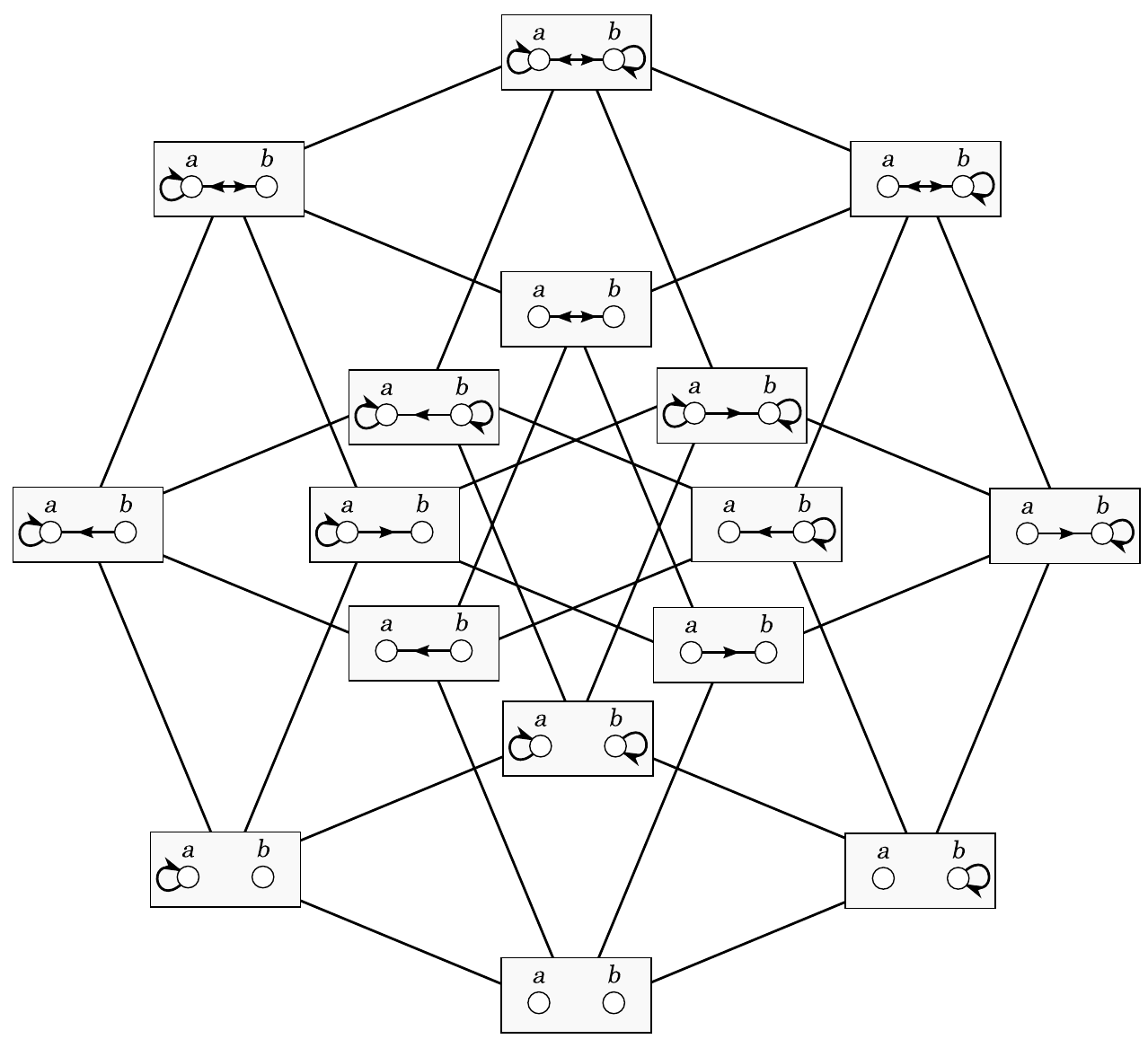}
   \caption{The lattice of distinct, permitted two-object structures, according to the haecceitist.}
   \label{fig: haec}
\end{figure}

The structures may be partially ordered, according to which $R$-formulas are asserted in their descriptions. That is: we can partially order, by `is a subset of', the power-set of the set of $R$-formulas; this induces a partial order on the structures, by associating each structure with the subset of $R$-formulas that are true in it. For the Haecceitist case, this partial order produces a lattice of structures. (A {\em lattice} is a partially ordered set, in which any two elements have a greatest lower bound, and a least upper bound.) It is shown in Figure \ref{fig: haec}. 

In Figure \ref{fig: haec}, lines connecting structures indicate a single alteration in the assertion/denial of an $R$-formula. The Figure also merits two further comments:\\
\indent (1): It adopts an unusual convention about how to display the partial order. It is   usual to draw a lattice by putting all the elements that are immediately greater than the least element together on a horizontal row (often called `rank'), above the least element; and all the elements that are immediately greater than any of them, on a next higher rank; and so on. Figure \ref{fig: haec} does \emph{not} do that. Instead, we have chosen to display, using the four different \emph{directions} of the various upward lines (e.g.~the four different lines emanating from the least element), the four different truth-value flips that one can make, so as to make more $R$-formulas true.
(An example in tribute to Hitchcock and Cary Grant: the direction North by North-West corresponds to affirming the formula $Rba$.) \\
\indent (2): An assignment of a truth-value, 0 or 1, to each of the four $R$-formulas (a ``fourfold decision''), can be thought of as a vertex of the 4-dimensional unit hypercube $[0,1]^4 \subset \mathR^4$. There are $2^4 = 16$ such vertices, and Figure \ref{fig: haec} is a parallel projection of the hypercube onto the plane of the paper.

Note how the use of haecceitistic predicates distinguishes structures that would otherwise be equivalent. For example, the (equivalence class of) two-object structures picked out by the sentence
$$
\exists x\exists y  (\neg Rxx \wedge Rxy \wedge \neg Ryx \wedge \neg Ryy)
$$
splits, for the haecceitist, into two equivalence classes: those specified by
$$
\exists x\exists y (N_ax \wedge N_by \wedge \neg Rxx \wedge Rxy \wedge \neg Ryx \wedge \neg Ryy)
$$
and those specified by
$$
\exists x\exists y (N_ax \wedge N_by \wedge \neg Rxx \wedge \neg Rxy \wedge Ryx \wedge \neg Ryy) \ .
$$

\subsubsection{QII}
\begin{figure}[h!]
   \centering
   \includegraphics[scale=0.75]{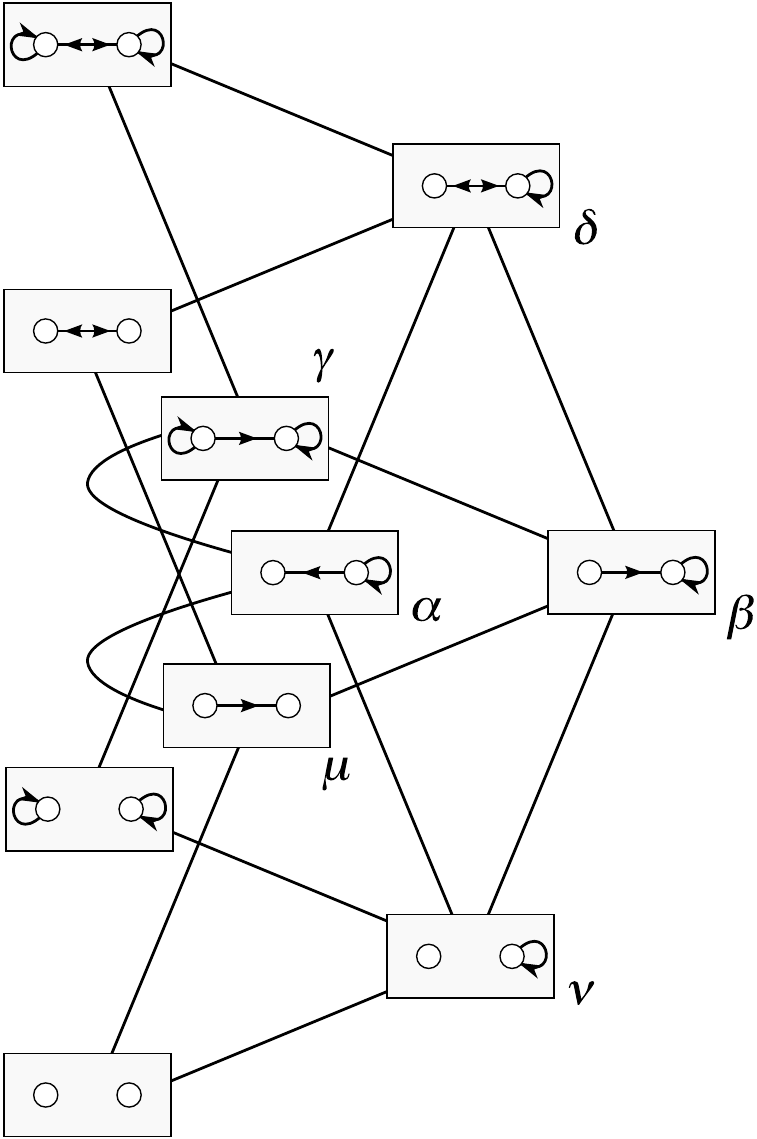}
   \caption{The poset of distinct, permitted two-object structures, according to the proponent of QII.}
   \label{fig: qii}
\end{figure}

The proponent of QII has no recourse to haecceitistic predicates, but still allows the distinctness of the two objects to be primitive in the sense of allowing utter indiscernibles.  Her state-description sentences for describing the two-object structures are
$$
\exists x\exists y (x \neq y \wedge \pm Rxx \wedge \pm Rxy \wedge \pm Ryx \wedge \pm Ryy) \ .
$$
The lack of any way to \emph{absolutely} discern one object from the other, except by the pattern of instantiation of $R$, means that the number of distinguishable structures is less than the \emph{prima facie} number, 16.  Any choice of the combination of $R$-formulas to assert that renders the two objects absolutely discernible is double-counted, since we obtain an equivalent sentence by the interchange of the bound variables $x$ and $y$.  Since there are two structures in which the objects are utterly indiscernible, and two in which they are merely weakly discernible, and none in which they are merely relatively discernible, we conclude that $16 - 2 - 2 - 0 = 12$ of our 16 \emph{prima facie} allowable structures are double-counted.  So there are in fact only $2+2+\frac{12}{2} = 10$ distinct structures.  The partially ordered set---unlike for Haecceitism, it is not a lattice---of these structures is shown in Figure \ref{fig: qii}.  (Note: in accordance with our discussion, this Figure lacks names for the objects.)  Thus the effect of eliminating haecceitistic differences is to ``fold over'' the haecceitist's lattice, given in Figure \ref{fig: haec}, along its central vertical axis.

The result is no longer a lattice, since it is no longer the case that any two structures have a unique join and meet.  This seemingly unremarkable technical result underlies a significant metaphysical one: namely that there is no unique transworld identity relation.  We can see this as follows.  Take, for example, the two (equivalence classes of) structures $\alpha$ and $\beta$ that lie in the middle row of Figure \ref{fig: qii}.  What is the supremum of these two structures, i.e.~the structure in which all the relations that hold in either of these two structure hold, and no more?

Without haecceitistic predicates, the answer depends on how else we decide to cross-identify the objects. We cannot, as with Haecceitism, simply take the union of ext$(R)$ in each of the two structures (i.e.~the union of two sets of ordered pairs) to obtain a unique, new ext$(R)$, since for QII (and WPII and SPII, below) it is not a set of ordered pairs but an (isomorphism) \emph{equivalence class} of such sets that represents a single world---and the unions of two isomorphism equivalence classes does not yield an isomorphism equivalence class.  So we must employ an alternative strategy: to form the union \emph{first}, for two structures \emph{haecceitistically conceived}, and \emph{then} to take equivalence classes under isomorphism.  But which pair of haecceitistically-conceived structures do we choose?  Each of $\alpha$ and $\beta$ is an equivalence class containing two structures, so there are $2\times 2 = 4$ pairs to choose from, which, after taking equivalence classes, yields two candidates for the supremum.  They are $\gamma$ and $\delta$ in Figure \ref{fig: qii}.

In other words: we must decide on some QII-acceptable way to cross-identify the objects between structures.  In our example, we could either take our transworld individuals to be identified by the formulas $\pm Rxx$, (yielding as supremum the equivalence class $\delta$), or by the formulas $\pm\exists y (y\neq x\ \&\ Rxy)$ (yielding as supremum the equivalence class $\gamma$).

So under QII the partial order is not a lattice.  But using formulas to cross-identify objects between structures brings out four further differences from the situation under Haecceitism: differences which will also be shared by WPII and SPII, below. 
\begin{enumerate}
\item[(i)] There is no \emph{uniquely salient} option as to which qualitative properties and relations ground the transworld identity relation.

\item [(ii)] One may need to use individuating formulas that are \emph{disjunctive} in the primitive vocabulary to cross-identify objects between structures whose equivalence classes of properties- and relations-in-extension are disjoint. (Cf.~e.g.~structures $\mu$ and $\nu$ in Figure \ref{fig: qii}.  There are two rival candidate pairs for trans-structural objects, and it is hard to say which cross-identification criteria are most natural.)

\item[(iii)] Even allowing disjunctive individuating formulas, there will not always be cross-identifying formulas, since at least one of the structures may contain \emph{non-individuals} (cf.~e.g.~any of the four unlabelled (i.e.~leftmost) structures in Figure \ref{fig: qii} with any other in the poset).  (In such a case, cross-identification is anyway unnecessary to yield a unique supremum and infimum.)

\item[(iv)] Cross-identification of an object between structures by the same formula fails to be transitive, since the formula may fail to be uniquely instantiated in every structure.  Therefore we shouldn't, strictly speaking, speak of transworld \emph{identity} at all; and talk of counterparts \`a la Lewis ([1986], Ch.~4) seems far more appropriate.\footnote{On a related note, we thank Leon Horsten for drawing our attention to the fact, shown by Quine [1960] and F\o llesdal [1968], that the use of definite descriptions to individuate objects in quantified modal logics leads to a collapse of modal distinctions, if we require that any two definite descriptions which are actually co-instantiated are necessarily co-instantiated.  This latter requirement (which Quine ([1960], p.~198) calls a ``disastrous assumption") is a last-gasp solution to the problem of referential opacity for modal operators.  We suggest an alternative, seemingly more radical, escape: to veto as ill-formed any sentence in which modal operators are put in the scope of a quantifier.  This apparently onerous restriction means that first-order variables are no longer rigid designators that come for free.  We welcome this result: for Haecceitists this is no restriction at all, since for transworld identification there is always the recourse to haecceitistic predicates; and for everyone else  transworld identification is provided for, if at all, by qualitative matters.}
\end{enumerate}

\subsubsection{WPII} \label{ss:WPII}
We saw that the QII proponent complains that the Haecceitist double-counts some structures. The proponent of PII agrees with this accusation against Haecceitism, but in turn complains that both the Haecceitist and the QII proponent ``over-count'' some structures by counting them at all, i.e.~by accepting them as possible---or at least as possible \emph{two}-object structures.

This equivocation between the elimination of two-object structures, and the re-description of them as one-object structures corresponds to the two ways of understanding the Hilbert-Bernays axiom (HB), discussed in comment 3 of Section \ref{HBsec}.  On one hand, if we take `=' as a logical constant, then we take the specification of the cardinality of the domain as logically independent of the specification of the pattern of instantiation of the qualitative properties and relations, and therefore as possibly in contradiction with it, taken in conjunction with (HB).  In such a case we take (HB) as eliminating these problematic structures.  On the other hand, if we take `=' as \emph{defined} by (HB), then we have no business fixing the cardinality of the domain independently of the pattern of instantiation: the problematic structures just mentioned are then construed as clumsily described one-object structures.  For definiteness, let us talk from now on of elimination rather than re-description.

\begin{figure}[h!]
   \centering
   \includegraphics[scale=0.75]{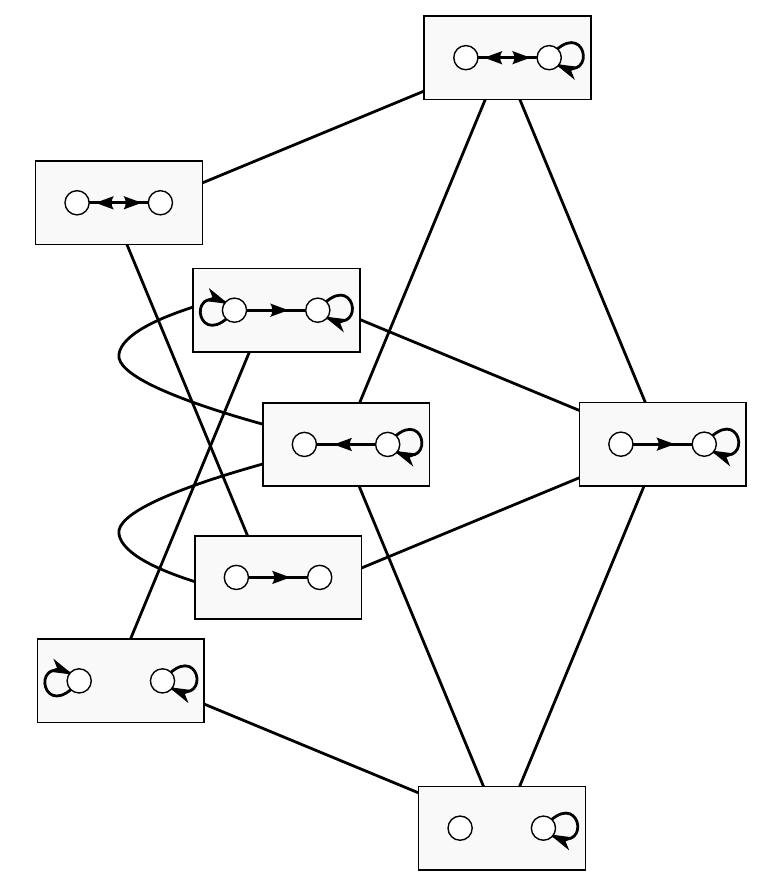}
   \caption{The poset of distinct, permitted two-object structures, according to the proponent of WPII.}
   \label{fig: wpii}
\end{figure}

The proponent of PII has at her disposal for the specification of structures only the state-description sentences
$$
\exists x\exists y (\pm Rxx \wedge \pm Rxy \wedge \pm Ryx \wedge \pm Ryy) \ .
$$
But, as just discussed, the specifications obtained from these sentences are hostage to the Hilbert-Bernays axiom (HB); consequently not all of them will describe structures of as many as two objects.  Those structures for which the PII proponent is committed to identify the objects are excluded as \emph{genuine} two-object structures.  Which structures these are depends on which version (strength) of the principle of identity of indiscernibles (PII) is endorsed.

The most liberal PII proponent, the proponent of WPII (Weak PII), rules out only those structures in which there are any \emph{utterly} indiscernible objects.  There are only two such structures in our example: viz. corresponding to all four of the $R$-formulas being asserted, or all four being denied.  This reduces the number of distinct, permitted structures from QII's ten to eight; see Figure \ref{fig: wpii}.

\subsubsection{SPII}
\begin{figure}[h!]
   \centering
   \includegraphics[scale=0.75]{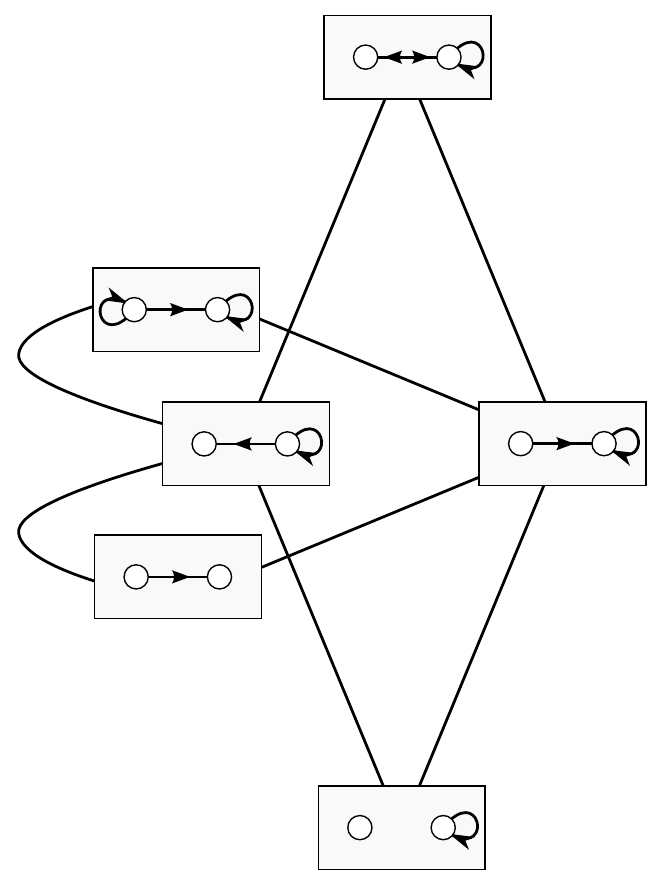}
   \caption{The poset of distinct, permitted two-object structures, according to the proponent of SPII.}
   \label{fig: spii}
\end{figure}

The partially ordered set of structures permitted by WPII, Figure \ref{fig: wpii}, is further diminished if we endorse SPII: i.e.~the strengthening of PII to eliminate structures that contain absolute indiscernibles.  Figure \ref{fig: wpii} has two such structures: those in which the objects are merely weakly discernible.  By eliminating these two structures, we are left with only six distinct structures; see Figure \ref{fig: spii}.\footnote{Returning to the ``Leibnizian" proponent of SPII:  if we follow Saunders' ([2003b], p.~17) idea that the ``Leibnizian" requires all objects to be be \emph{intrinsically} discernible, this would reduce the number of permitted structures still further to {\em four}. These are the structures in which only one of the two objects satisfy $Rxx$.  We will discuss this view, which we call \emph{qualitative intrinsicalism}, in Section \ref{structuralism}.\label{fn:Leibniz}}  Note that in the move from WPII to SPII, the elimination of structures \emph{cannot}, as in the move from QII to WPII, be instead understood as re-description.   For, even if we refrain from independently specifying the cardinality of the domain, a state-description on its own may contradict the requirement that any two objects be absolutely discernible.  This fact underscores the logical strength of SPII.

\subsubsection{A glance at the classification for structures with three objects}\label{512struct}
Finally, we glance at structures with three objects. The existence of a third object means that we here get the ``first'', i.e. most elementary, illustrations of external discernibility in the \emph{intuitive} sense that requires a third object.  We also see here the first cases of mere relative discernibility. We have no space for details. We just report that, as in the preceding Subsections, one can count and classify the structures according to each of their symmetries.  Out of the 512 structures accepted by the Haecceitist, 420 have no non-trivial symmetries (and therefore lie in isomorphism equivalence classes with cardinality 6), 84 have only pair-wise swaps as non-trivial symmetries (with 3 structures per isomorphism equivalence class), 4 have only cyclic permutations as non-trivial symmetries (with 2 structures per isomorphism equivalence class), and there are 4 totally symmetric structures (for which isomorphism equivalence classes are singletons).  Half of all of the structures with non-trivial symmetries contain indiscernible objects. 

From this information, we may compute the following totals: 
\begin{center}
\begin{tabular}{rc}
\textit{Metaphysic} & \textit{Number of structures}\\\hline\hline
Haecceitism &  $512$\\
QII & $104$\\
WPII &  $88$\\
SPII & $70$ 
\end{tabular}
\end{center}
And overall, ninety-two structures have non-trivial symmetries. But sufficient unto the day is the counting thereof! 

\subsection{Structuralism and intrinsicalism}\label{structuralism}

The four metaphysical theses just discussed have various similarities and differences, but one distinction in the logical space is of particular interest to us: we believe it meshes with much of the recent logico-philosophical literature falling under the buzz-words `structuralism' and `structural realism',\footnote{See e.g.~MacBride [2005], Button [2006], Ladyman ([2005], [2007]), Leitgeb \& Ladyman [2008] and Ketland ([2006], [2009]).} and also with modern philosophical discussions of physical theories---in particular, general relativity and non-field-theoretic quantum mechanics.\footnote{See e.g.~Saunders ([2003a], [2003b], [2006]) Esfeld \& Lam [2006], Pooley [2006], Ladyman [2007], Muller \& Saunders [2008] and Muller \& Seevinck [2009].  We add to this discussion in Caulton \& Butterfield [2011].}

This distinction is about \emph{individuality} (a main topic of Section \ref{symmrevisit}): specifically, whether the individuality of an object relies on, or is associated with, \emph{qualitative} and (typically) \emph{relational}, differences to other objects; or whether instead an object's individuality relies on, or is always associated with, purely \emph{intrinsic} differences (defined in Section \ref{4defined}; which may be qualitative \emph{or} solely haecceitistic).  The position we wish to call \emph{structuralism} holds that: either there may be objects which are not individuals; or at least, if every object in every possible world is an individual, then it is not in all cases due to differences which are purely intrinsic.  So the denial of structuralism holds that every object, in every world, is an individual due to purely intrinsic differences. We call this position (for want of a better word) \emph{intrinsicalism}.

Why are these positions philosophically salient? That is: what unites the various, more specific views that fall under the umbrellas `structuralist' or `intrinsicalist'?  To this we have two answers: one historical, one philosophical.  Firstly, intrinsicalism encompasses what we think are two central views which have dominated, up until recently, philosophical debate about identity and individuality; namely, haecceitism and a strengthening of SPII, in which all objects are taken to possess a unique, though maybe very complex, individuating essence; (the `bundle theory' of objects being one such view: cf.~Adams [1979], p.~7).  On the other hand, structuralism encompasses a variety of alternatives to intrinsicalism which have recently been articulated.\footnote{Hintikka \& Hintikka [1983] present a similar dichotomy to our structuralism vs.~intrinsicalism (which they call the ``structural'' and ``referential system", respectively), and bemoan the bias towards intrinsicalism in the historical development of formal logic.  We agree that the historical dominance of intrinsicalism is unfortunate and unjustified; and note with interest their psychological evidence (from studies with children) that the evidence may be sex-linked.}

Secondly, a common feature of structuralist, as against intrinsicalist, views is that they prompt a certain interpretation of, or even a revision of, traditional formalisms in both semantics and physical theories.  We end the paper with a brief discussion of this (Section \ref{subs:semstr}).  But first (Section \ref{subs:relto4}), we relate these two new positions to the four metaphysical theses of Section \ref{the4}.

\subsubsection{Relation to the four metaphysical theses} \label{subs:relto4}

Structuralism, as we define it, includes some of our four metaphysical theses (roughly two ``and a half" of them), which may be linearly ordered on a spectrum, according to their strictness, i.e.~what they rule out.  (This is already suggested by the possible world diagrams of Section \ref{hypercube}, in which we pass from haecceitism to SPII by a successive vetoing of worlds.)

At the weak end of this spectrum we find QII, for which the numerical diversity of objects is accepted as primitive, just as it is by the haecceitist.  QII is structuralist both because it allows non-individuals and because it allows external discernibility as a basis for individuality. 

Along the way to the strong end of the spectrum, we find versions of structuralism that ground diversity, as well as individuality, in the differential instantiation of qualitative properties and relations.  In doing so, we find a commitment to the Hilbert-Bernays account of identity; so here we find WPII.  WPII is structuralist both in allowing for non-individuals (though not indiscernible non-individuals) and in allowing for external discernibility.

At the strong end of the spectrum, we find that our previous metaphysical thesis, SPII, has been split.  For SPII demands that every object be an individual.  This does not yet commit one to intrinsicalism: for intrinsicalism requires also that the individuating properties be intrinsic.  We may therefore distinguish between an SPII which \emph{allows} for objects individuated by merely external formulas---that is, a structuralist SPII---and an SPII which does \emph{not} allow for such objects---an intrinsicalist SPII.  To repeat the main idea: to deserve the label `structuralism' (in our usage),  the individuality of objects must in some cases involve their absolute discernment by \emph{extrinsic}, qualitative properties, not by \emph{intrinsic}, qualitative properties.\footnote{We should point out that even intrinsicalist SPII has a whiff of structuralism about it.  Consider, for example, the three-object structure $\mathfrak{A} := \langle \{a,b,c\}, F^\mathfrak{A}, G^\mathfrak{A}, H^\mathfrak{A} \rangle$, with three primitive monadic predicates with extensions $F^\mathfrak{A} = \{a, b\}, G^\mathfrak{A} = \{b, c\}, H^\mathfrak{A} = \{a, c\}$.  This structure is permitted under intrinsicalist SPII, the objects being individuated by the monadic predicates $F\wedge G, G\wedge H, F\wedge H$.  The whiff of structuralism is obvious when we consider the combinatorial theory of logical possibility, which takes the structure $\mathfrak{A}$ to indicate the (logical) possibility of other structures, some of which have $a,b$ and $c$ allocated to the extensions of $F, G$ and $H$ in ways that do not secure their qualitative individuality---one such case is $a,b,c \in$ ext$(F \wedge \neg G \wedge \neg H)$.  It should not come as a surprise that SPII (whether intrinsicalist or structuralist) restricts na\"ive, combinatorially-conceived possibility, since the admissibility under SPII of affirming or denying a given elementary proposition depends on global features of the structure description (namely, whether certain other elementary propositions are affirmed or denied).  Note that, in contrast, QII and, perhaps, WPII do \emph{not} restrict possibility in this way.  For, as explained in Section \ref{ss:WPII}, QII and (given the reductionist attitude to identity) WPII may be taken as \emph{redescribing} rather than \emph{vetoing} the structures permitted under unrestricted Haecceitism.}

\begin{figure}[h!]
   \centering
   \includegraphics[scale=1]{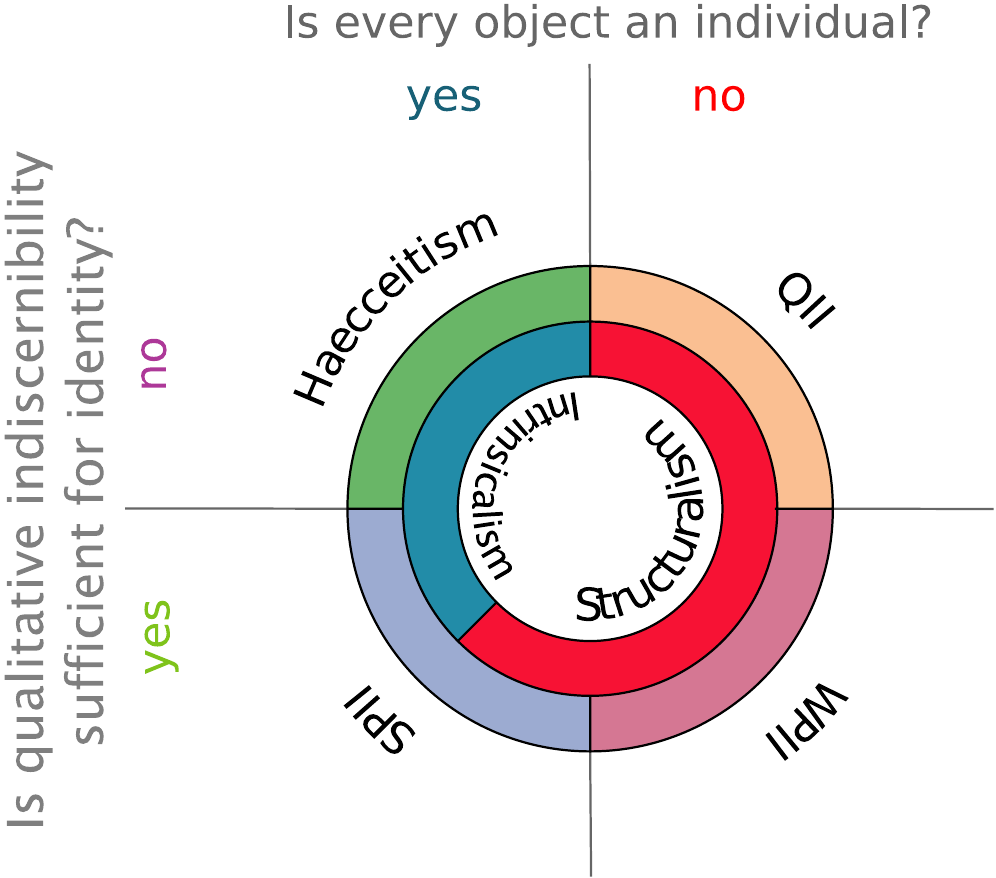}
   \caption{Four metaphysical theses and two salient positions.}
   \label{fig:circleofviews}
\end{figure}

Finally, we note that haecceitism is a form of intrinsicalism, since haecceities are intrinsic individuating properties \emph{par excellence}.   Thus our taxonomy is summarised in Figure \ref{fig:circleofviews}, below.

\subsubsection{The semantics of the structuralist}\label{subs:semstr}
As we have seen, grounding individuality qualitatively has consequences for the specification of possible states of affairs.  Thus a structuralist cannot make sense of what Lewis ([1986], p.~221) called `haecceitistic differences': differences to do with which object occupies which role in the mosaic of qualitative matters of fact.  Since an object's individuality is provided, if it all, by the role that it plays in the structure of which it is a part, there simply is nothing else to grasp it with, that could be used to give sense to, for example, the object swapping roles with another.  A structuralist simply does not agree with the haecceitist that a permutation of objects ``underneath" the mosaic makes any sense.

But here the structuralist runs up against an aspect of formal semantics (model theory) which threatens to make her position hard to state.  Model theory allows without demur that an object may be individuated independently of its qualitative role in a structure.\footnote{This point is often emphasised in the philosophy of modality; a good early example is Kaplan [1966].} Or think of Lewis's ([1986], p.~145) `Lagadonian' languages, in which objects are their own names.  In philosophical terms, this trick ensures that haecceities are always at hand to give sense to a permutation of ``bare" objects underneath the qualitative properties and relations. The same point applies in physical theories. Permutations of the objects that the theory treats (e.g.~particles in classical mechanics or quantum mechanics) can be implemented on the theory's state space (e.g.~a classical phase space, or quantum Hilbert space, for several particles); and typically, a state is changed by such a permutation, i.e.~its image is a different state, even if the mosaic of qualitative properties and relations is unchanged.

In response, the structuralist can adopt either of two approaches, which we now briefly describe; (more details are in Caulton \& Butterfield [2011]). The first involves a certain interpretation of the usual formalism (of the formal semantics or physical theory in question). The second involves revising the formalism.

First, the structuralist can accommodate the orthodox practice in model theory and physical theory---that permutations of ``bare'' objects make sense---by an interpretative move.  That is, she can admit that such permutations typically induce a different formal representation of a state of affairs:  a different model in the sense of model theory, or a different state in the physical theory's state-space. But she then expresses her position by saying that such a permutation does not produce a representation of a \emph{distinct} state of affairs.

On this view, it is misleading to say that model-theoretic {structures}, or states in the state-space,  represent possible states of affairs; for the representation relation is not one-to-one, but many-to-one.  Rather, a single state of affairs is represented by the \emph{equivalence class} of mutually isomorphic structures (states), the isomorphism being given by a permutation of the `bare' objects in the structure's domain (a permutation of the physical theory's objects, e.g.~particles).  This ascent from structures to equivalence classes suits a theory which seems ``not to care" how its objects are arranged under the mosaic of properties and relations, so long as the same qualitative pattern is instantiated;  such theories support a structuralist intepretation. 

The second approach proposes to keep the original structuralist prohibition against permutations making sense, by revising the formalism.   The idea is to replace each equivalence class of isomorphic representations by a less structured item, so that permutations do not make even a \emph{formal} difference, or else cannot be made sense of.  It is well established, for both model theory and physical theories. (See Belot [2001], p.~59 for a philosophical introduction.) Thus in elementary model theory, it is straightforward to show that if we quotient a structure by the indiscernibility relation, then (i) the resulting structure is elementarily equivalent to the given one, and (ii) in it, identity is first-order definable, namely by indiscernibility---i.e.~the leftward implication of (HB) holds (Ketland [2009], Theorems 18, 20). In physical theories (both classical and quantum), one considers the quotient of the state space (often called a \emph{reduced state-space}) by various equivalence relations (often going by the name `gauge-equivalence')---not just by isomorphisms induced by permutations of the underlying objects (particles). For more details, from a philosophical perspective, cf.~e.g.~Belot ([2003], especially \S 5) and Butterfield ([2006], \S\S2.3, 7). To close, we shall only note the situation for quantum mechanics, with isomorphisms induced by permutations of the underlying quantum particles. Here also one can quotient: in effect, particles of the various possible symmetry types (including paraparticles) can be represented in a ``reduced Hilbert space'' (in general, of lower dimension than the original, ``na\"ive'' one) in which each particle permutation is represented either trivially or not at all. (We 
 develop this topic in Caulton \& Butterfield [2011].) Here we note only that this second approach (unlike quasi-set theory, cf.~footnote \ref{fn:qset}) retains the orthodox ideas and methods of model theory and classical logic, merely limiting their application to the favoured, i.e.~quotient, structures or states.

{\em Acknowledgements}:--- We heartily thank: Tim Button and Fraser MacBride for comments on previous versions and many conversations; Newton da Costa, Leon Horsten, Jeff Ketland, James Ladyman, \O ystein Linnebo, Fred Muller and Michael Seevinck for correspondence and comments; and audiences in Amsterdam, Bristol, Cambridge, Pittsburgh and Sydney for questions and comments.  One of us (AC) would like to express gratitude to the Arts and Humanities Research Council for financial support.

\section{References} 

\quad\ \ Adams, R. M. [1979]: `Primitive thisness and primitive identity', \emph{Journal of Philosophy} \textbf{76}, pp.~5-26.

Adams, R. M. [1981]: `Actualism and thisness', \emph{Synthese} \textbf{49}, pp.~3-41.

Belot, G. [2001]: `The Principle of Sufficient Reason', \emph{The Journal of Philosophy} \textbf{98}, pp.~55-74.

Belot, G. [2003]: `Notes on symmetries', in Brading and Castellani (eds.), \emph{Symmetries in Physics: Philosophical Reflections}, Cambridge: Cambridge University Press, pp.~393-412.
 
Black, M. [1952]: `The identity of indiscernibles', \emph{Mind} \textbf{61}, pp.153-64. 

Black, R. [2000]: `Against quidditism', \emph{Australasian Journal of Philosophy} \textbf{78}, pp.~87-104.
 
Boolos and Jeffrey [1974]: \emph{Computability and Logic} (4th ed.). Cambridge University Press: London. 
 
Butterfield, J. N. [1993]: `Interpretation and identity in quantum theory', \emph{Studies in the History and Philosophy of Science} \textbf{24}, pp. 443-76. 

Butterfield, J. N. [2006]: `On Symplectic Reduction in Classical Mechanics', in Earman, J. and Butterfield J. N. (eds.), \emph{The Handbook of Philosophy of Physics}, North Holland, pp.~1-131.

Button, T. [2006]: `Realistic Structuralism's Identity Crisis: A Hybrid Solution', {\em Analysis} \textbf{66}, pp.~216-22.

Carnap, R. [1950]: {\em The Logical Foundations of Probability}. Chicago:  Chicago University Press.

Castellani, E. (ed) [1998]: \emph{Interpreting Bodies: Classical and Quantum Objects in Modern Physics}.  Princeton University Press.

Caulton, A.~and Butterfield, J. [2011]: `Symmetries and Paraparticles as a Motivation for Structuralism', forthcoming in \emph{The British Journal for the Philosophy of Science}.  Available online at: http://philsci-archive.pitt.edu/5168/.

da Costa, N. and Krause, D. [2007]: `Logical and Philosophical Remarks on Quasi-set Theory', 
{\em Logic Journal of the IGPL}, {\bf 15}, pp.~1-20.

da Costa, N. and Rodrigues, A. [2007]: `Definability and Invariance', \emph{Studia Logica} \textbf{86}, pp.~1-30.

Dalla Chiara, M.~L.,~Giuntini, R.~and Krause, D. [1998]: `Quasiset Theories for Microobjects: 
A Comparison', in Castellani (1998), pp.~142-152.

van Dalen, D. [1994]: \emph{Logic and Structure} (3rd ed.). Springer Verlag: Berlin.
 
Esfeld, M. [2004]: `Quantum entanglement and a metaphysics of relations' \emph{Studies in the History and Philosophy of Modern Physics}, {\bf 35}, pp. 601-617. 

Esfeld, M.~and Lam, V. [2006]: `Modest structural realism', \emph{Synthese} {\bf 160}, pp.~27-46. 

F\o llesdal, D. [1968]: `Quine on modality', \emph{Synthese} \textbf{19}, pp.~147-157.

French, S. [2006]: `Identity and individuality in quantum theory', \emph{The Stanford Encyclopedia of Philosophy (Spring 2007 Edition),} Edward N. Zalta (ed.). \\ URL = http://plato.stanford.edu/entries/qt-idind.

French, S. and Krause, D. [2006]: \emph{Identity in Physics}.  Oxford: Oxford University Press.

French, S. and Redhead, M. [1988]: `Quantum physics and the identity of indiscernibles', \emph{British Journal for the Philosophy of Science} \textbf{39}, pp. 233-46. 

Hacking, I. [1975]: `The identity of indiscernibles', \emph{Journal of Philosophy,} \textbf{72}, pp.~249-56. 

Hawley, K. [2009]: `Identity and Indiscernibility', \emph{Mind} \textbf{118}, pp.~101-119.

Hilbert, D. and Bernays, P. [1934]:  \emph{Grundlagen der Mathematik}, vol. 1.  Berlin: Springer.

Hintikka, M. J. and Hintikka, J. [1983]: `How can language be sexist?', in S.~Harding and M.~Hintikka (eds.), \emph{Discovering Reality: Feminist Perspectives on Epistemology, Metaphysics, Methodology, and Philosophy of Science}, Dordrecht: Kluwer, pp.~139-148.

Kaplan, D. [1966]: `Transworld Heir Lines', in Loux (ed.), {\em The Possible and the Actual: readings in the metaphysics of modality}, New York: Cornell University Press, pp.~88-109.

Kaplan, D. [1975]: `How to Russell a Frege-Church', \emph{Journal of Philosophy} \textbf{72}, pp. 716-729.

Ketland, J. [2006]: `Structuralism and the identity of indiscernibles', {\em Analysis} {\bf 66}, pp.~303-315.

Ketland, J. [2009]: `Identity and Indiscernibility', {\em Review of Symbolic Logic}, forthcoming.

Krause, D. [1992]: `On a quasi-set theory', \emph{Notre Dame Journal of Formal Logic} \textbf{33} (3), pp. 402-11. 

Krause, D. and French, S. [2007]: `Quantum sortal predicates', \emph{Synthese} \textbf{143} (14) pp. 417-430.

Ladyman, J. [2005]: `Mathematical Structuralism and the Identity of Indiscernibles' {\em Analysis} {\bf 65}, pp.~218-21.

Ladyman, J. [2007]: `On the identity and diversity of objects in a structure',  \emph{Aristotelian Society Joint Sessions Supplementary Volume}, June 2007.

Leitgeb, H. and Ladyman, J. [2008]: `Criteria of identity and structuralist ontology', \emph{Philosophia Mathematica} \textbf{16}, pp.~388-396.

Lewis, D. [1970]: `On the Definition of Theoretical Terms', \emph{Journal of Philosophy} 
\textbf{67}, pp. 427-446.

Lewis, D. [1986]: {\em On the Plurality of Worlds}. Wiley-Blackwell.

Lewis, D. [2002]: `Tensing the copula', \emph{Mind} \textbf{111}, pp.~1-14.

MacBride, F. [2005]: 'Structuralism Reconsidered' in S. Shapiro (ed.), \emph{The Oxford Handbook of Philosophy of Logic \& Mathematics}, Oxford University Press, pp.~563-589.

Margenau, H. [1944]: `The exclusion principle and its philosophical importance', \emph{Philosophy of Science} \textbf{11}, pp. 187-208.

Massimi, M. [2001]: `Exclusion principle and the identity of indiscernibles: a response to Margenau's argument', \emph{British Journal for the Philosophy of Science} \textbf{52}, pp. 303-30.

Muller, F. A. [2011]: `How to discern space-time points structurally', \emph{Philosophy of Science Association 22nd Biennial Meeting (Montr\'eal, Canada) Contributed Papers}.

Muller, F. A. and Saunders, S. [2008]: `Discerning fermions', \emph{British Journal for the Philosophy of Science} \textbf{59}, pp.~499-548.

Muller, F. A. and Seevinck, M. [2009]: `Discerning elementary particles', \emph{Philosophy of Science} \textbf{76}, pp.~179-200. 

Pooley, O. [2006]: `Points, particles, and structural realism', in D. Rickles, S. French \& J. Saatsi (eds.), \emph{The Structural Foundations of Quantum Gravity}, Oxford: Oxford University Press, pp. 83--120.

Pooley, O. [2009, manuscript]: `Substantivalism and Haecceitism'.

Quine, W. V. O. [1960]:  \emph{Word and Object}.  Cambridge: Harvard University Press.

Quine, W. V. O. [1970]:  \emph{Philosophy of Logic}.  Cambridge: Harvard University Press.

Quine, W. V. O. [1976]: `Grades of discriminability', \emph{Journal of Philosophy} \textbf{73}, pp. 113-16.

Robinson, D. [2000]: `Identities, Distinctness, Truthmakers, and Indiscernibility Principles', \emph{Logique \& Analyse} \textbf{43}, pp.~145-183.

Saunders, S. [2003a]: `Physics and Leibniz's Principles', in K. Brading and E. Castellani, (eds.), \emph{Symmetries in Physics: Philosophical Reflections}, Cambridge University Press.

Saunders, S. [2003b]: `Indiscernibles, covariance and other symmetries: the case for non-reductive relationism', in A. Ashtkar, D. Howard, J. Renn, S. Sarkar and A. Shimony (eds.), \emph{Revisiting the Foundations of Relativistic Physics: Festschrift in Honour of John Stachel}, Amsterdam: Kluwer.

Saunders, S. [2006]: `Are quantum particles objects?', \emph{Analysis} \textbf{66}, pp.~52-63.

Wiggins, D. [2001]: \emph{Sameness and Substance Renewed}.  Cambridge: Cambridge University Press.

\section{Appendix: Implications between the kinds}\label{loglrelns}
In this Appendix, we return to  our four kinds of discernibility as defined in Section \ref{4defined}, and give a survey of their mutual logical implications.

\subsection{Clearing the ground}\label{groundclear}
Given four kinds, there are $2^4 = 16$ apparent truth-combinations (i.e.~rows of a
truth-table), each of which needs to be shown possible, or impossible, in a full survey of
the implications between the kinds. One readily sees that all 16 truth-combinations are indeed possible by allowing different predicates to yield different kinds of discernment. For example, for all four discernments holding: two objects $a$ and
$b$ may be discerned intrinsically (by some {\bf (Int)} formula),
externally (by some {\bf (Ext)} formula), relatively (by some {\bf (Rel)} formula) \emph{and}
weakly (by some {\bf (Weak)} formula); just because each discernment may be due to a
\emph{different} property or relation.

So let us instead ask which combinations are possible, and which impossible, when a \emph{single}
property or relation is considered. We now address this. That is, we focus on a language with
only \emph{one} primitive predicate: a 2-place relation $R$ (so that, in particular, `=' is not a primitive predicate).

There are two ways to proceed; which we will follow in order, in Sections
\ref{abundant} and \ref{sparse}. First, one can consider discernments by {\em any} formula
involving $R$, no matter how long it might be (Section \ref{abundant}). Or second, one can restrict oneself to
discernment by one of a smaller set of formulas. In Section \ref{sparse}, we will  restrict ourselves to the formula $Rxy$ and a circumscribed subset of its logical
descendants.

  We shall call these two ways of proceeding, `abundant' and `sparse', respectively; adopting
the now-standard jargon for two broad metaphysical approaches to properties and
relations (Lewis 1986, p.~59f.).  We use
these terms with metaphysical innocence: for example, by `sparse' we mean merely that the
formula in
question falls in our circumscribed subset. But we should add that we think the abundant case is more important philosophically, since it makes no disputable choice of what is an appropriate way to define such a subset. 

For both the abundant and sparse cases, the structures we shall consider will be formed from
three objects: $a$, $b$ and $c$.
As usual, $a$ and $b$ will
be the two objects being discerned: $c$ is included as our third object ``underpinning" external
discernment, and illustrating the occasional importance of global structure.
Considering three-object structures is sufficiently general: discernment is a
relation between a pair, $a$ and $b$, which may rely on other objects; and for the sake of simplicity we can collapse these into a single third object $c$.

Since the structures have three objects, and the non-logical
vocabulary contains only one 2-place predicate $Rxy$, any structure may
be completely specified by asserting the truth or falsity of the $3^2 = 9$ atomic sentences
$Raa, Rab, \ldots$.  There are $2^9 = 512$ distinct such complete specifications, and so 512
distinct structures.\footnote{That is, according to standard logic.  But we see in Section \ref{512struct} that the various metaphysical theses about identity disagree about how many of these
512 structures are genuinely distinct, allowable structures.} And in each structure, $a$ and $b$ are discerned in one or more of our four ways---or are not discerned in any such way.  Accordingly, the taxonomies of
implications below, in Section \ref{abundant} and \ref{sparse}, amount to a partition of this
set of 512 structures into (up to $2^4 = 16$) cells representing the possible combinations of kinds of discernment.  We will represent this partitioning in Venn diagrams (Figure \ref{fig: discabund}
and Figure \ref{fig: discs}).

\subsection{Implications: the abundant case}\label{abundant}
In this case, the implications turn out to be simple: a chain of one-way implications. We have ${\bf (Int)
\Raw (Ext) \Raw (Rel) \Raw (Weak) \Raw (All)}$. Here ${\bf (All)}$ is our label for the set of all
structures: in our three-object scenario, the 512 structures. So the last implication, ${\bf (Weak)
\Raw (All)}$, just means that there are structures containing indiscernible objects.

\begin{figure}[h!]
   \centering
   \includegraphics[width=\textwidth]{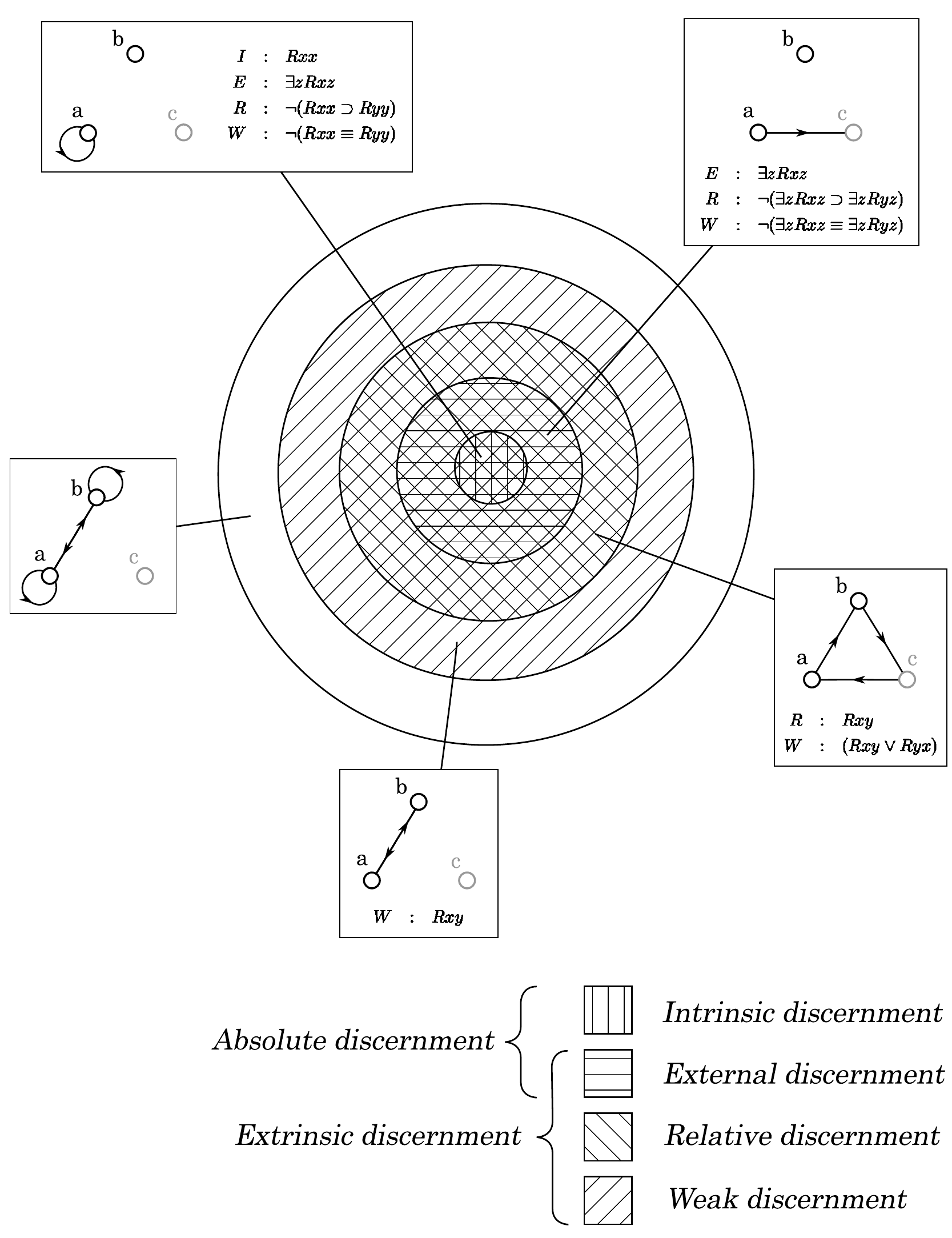}
   \caption{Implications between kinds of discernment: the abundant case. For each annulus
(ring), an exemplifying structure is shown.}
   \label{fig: discabund}
\end{figure}

In each of our arguments for these implications, we will comment that the converse
implication fails, and mention an illustrative counterexample structure. (Besides, Section
\ref{sparse} will discuss these structures in more detail---albeit in a sparser atmosphere!) Figure \ref{fig: discabund} summarizes these results, giving the counterexamples to the converses in boxes.

\begin{description}
	\item[${\bf (Int) \Raw (Ext)}$ :] If $Rxx$ discerns $a$ and $b$ intrinsically, then there will always be a formula (with only $R$ as its non-logical vocabulary) which discerns them.  For example, in a finite structure, the ``finest description" $\Sigma_a(x)$ (defined in Section \ref{parag2}), less the conjuncts containing the `=' symbol (if necessary), will discern $a$ from $b$ externally, if $a$ is discerned from $b$ intrinsically.  Of course,  the externally discerning formula need not always be so complex.  For example, in the two object structure with domain $D = \{a,b\}$, in which ext$(R) = \left\{\langle a,a \rangle\right\}$, $a$ is discerned from $b$ intrinsically by $Rxx$ and externally by $\exists z Rxz$.
	
	But there are structures in which $\exists z Rxz$ still discerns $a$ and $b$ externally, yet $a$ and $b$ (indeed all objects)
are \emph{intrinsically in}discernible. For example, Fig.~\ref{fig: example_3_ab}. Here and in subsequent diagrams, $c$ is grey, rather than black like $a$ and $b$, to emphasise that $a$ and $b$ are the objects being discerned.

\begin{figure}[h!]
   \centering
   \includegraphics{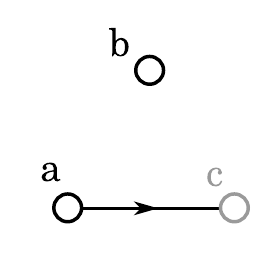}
   \caption{External, but not intrinsic discernment in the abundant case.}
   \label{fig: example_3_ab}
\end{figure}

	\item[${\bf (Ext) \Raw (Rel)}$ :] If a formula $\Phi(x)$ discerns $a$ and $b$ absolutely (= intrinsically or
externally), then $\neg\left(\Phi(x) \supset \Phi(y)\right)$ discerns them relatively. But there
are structures in which no pair of objects are externally discernible, yet all pairs are  relatively
discernible.  For example, the cyclical structure in Fig.~\ref{fig: example_5_ab}. Here the cycle $(abc)$ (i.e. $a \mapsto b \mapsto c \mapsto a$) is a symmetry; and so by Theorem 1, all of $a, b, c$ are absolutely indiscernible.

\begin{figure}[h!]
   \centering
   \includegraphics{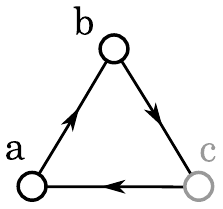}
   \caption{Relative, but not external discernment in the abundant case.}
   \label{fig: example_5_ab}
\end{figure}

	\item[${\bf (Rel) \Raw (Weak)}$ :] If a formula $\Psi(x,y)$ discerns $a$ and $b$ relatively, then:
	\begin{itemize}
		\item if $\langle a, a \rangle \notin$ ext($\Psi$) and $\langle b, b \rangle \notin$ ext($\Psi$), then $\left(\Psi(x,y) \vee
\Psi(y,x)\right)$ discerns them weakly;\footnote{Muller and Saunders (2008, \S4.4, p.~529) combine this implication with \textbf{(Ext) $\Rightarrow$ (Rel)} into a single implication: so that, if $\Phi(x)$ discerns absolutely, then $\left(\neg\left(\Phi(x) \supset \Phi(y)\right) \vee \neg\left(\Phi(y) \supset \Phi(x)\right)\right)$ discerns weakly.  They work with just two kinds of discernibility, called `absolute' and `relative': which are respectively the (quantum) analogues of our `absolute' (i.e.~intrinsic or external) and our `relative or weak'. So their result, in their jargon, is: `absolute implies relative'.}
		\item if $\langle a, a \rangle \in$ ext($\Psi$) and $\langle b, b \rangle \in$ ext($\Psi$), then $\left(\neg\Psi(x,y) \vee
\neg\Psi(y,x)\right)$ discerns them weakly;
		\item if only one of (i) $\langle a, a \rangle \in$ ext($\Psi$) and (ii) $\langle b, b \rangle \in$ ext($\Psi$) holds (so that our two objects are
absolutely discerned), then $\neg\left(\Psi(x,x) \equiv \Psi(y,y)\right)$ discerns them weakly.
	\end{itemize}
	But there are structures in which the objects are merely weakly discerned, but not
relatively discerned.  Cf.~Fig.~\ref{fig: example_6_ab}.\footnote{As mentioned in Section \ref{prosp}, this situation, i.e.~merely weak discernment, is centre-stage in the recent literature about identity in quantum theory: for particles of the same species provide an example.}

\begin{figure}[h!]
   \centering
   \includegraphics{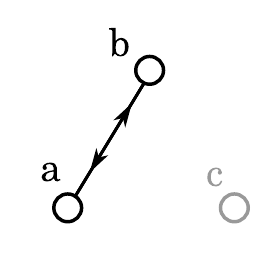}
   \caption{Weak, but not relative discernment in the abundant case.}
   \label{fig: example_6_ab}
\end{figure}

	\item[${\bf (Weak) \Raw (All)}$ :] For there are structures with objects $a$ and $b$ that are not even merely
weakly discernible. Cf.~Fig.~\ref{fig: example_1_ab}.\footnote{This structure is a slight variation on the example discussed by Button (2006, p.~218) and Ketland (2006, p.~309).  Ladyman (2007, p.~34) considers an equivalent case, in which effectively ext$(R) = \O$.}

\begin{figure}[h!]
   \centering
   \includegraphics{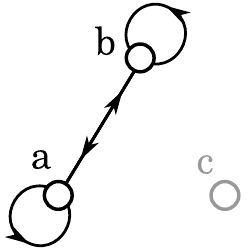}
   \caption{Complete failure of discernment in the abundant case.}
   \label{fig: example_1_ab}
\end{figure}

\end{description}

This chain of implications is summarised by the Venn diagram in Fig.~\ref{fig: discabund}:
which also displays the counter-example  structures (i.e.~representatives of a possible
combination of kinds of discernment) just mentioned.

\subsection{Implications: the sparse case}\label{sparse}
We turn to the sparse case; again working with just one 2-place relation $R$. We first define precisely the set of formulas we will now consider as available to make discernments. The motivating idea is that these definitions should reflect the literature's emphasis on relative and weak discernment. This will mean that for these two kinds of discernment, our definition will be very restrictive. For external discernment, on the other hand, our definition will follow the abundant case in allowing arbitrary formulas. This is natural since external discernment in general involves a third object; and since names are banned, this object will be referred to by a description, which could be arbitrarily complex. As to intrinsic discernment, the abundant case was already restrictive, with only $Rxx$ available (given our tiny vocabulary $\{ R \}$). 

 Thus we stipulate that for the sparse case, we allow discernment to use:\\
\indent (i): For {\bf (Int)}: $Rxx$ and $\neg Rxx$;\\
\indent (ii): For {\bf (Rel)} and {\bf (Weak)}: $Rxy$ and $\neg Rxy$;\\
\indent (iii): For {\bf (Ext)}: any one-place formula.\\

 Recall from the end of Section \ref{groundclear} that naively,
there are $2^9 = 512$ structures to consider.
 But we will not go through these one by one(!). Nor will we need to pause long on
each of the 16 apparently possible combinations of kinds of discernment. For in fact, one can readily check that
three  implications  between our kinds eliminate eight combinations as impossible; as follows (cf.~the definitions of the four kinds of discernment in Section \ref{4defined}):\footnote{Each implication, between two of our
four kinds, eliminates four combinations, viz.~those defined by the four apparently
possible combinations of the other two kinds. But each pair of implications overlap in
the combinations they eliminate, so that overall only eight, rather than twelve,
combinations are eliminated. Our choice of order, in (1) to (3), is thus arbitrary.}\\
\indent (1): Intrinsic discernment implies external discernment, so any \textbf{(Int)} $\wedge\ \neg\!$ \textbf{(Ext)}
truth-combination (``row of the truth-table'') is impossible. The reason is as in the abundant case: if $Rxx$ discerns intrinsically, then some 1-place formula, constructed with $R$ as its only non-logical primitive, discerns externally. \\
\indent (2): Relative and weak discernment exclude each other, since weak discernment demands that $R$ (or $\neg R$) be satisfied by the two objects taken in either order, while relative discernment demands that $R$ be satisfied by the objects taken in one order only; which eliminates a further
three truth-combinations (since the fourth has already been ruled impossible under (1));\\
\indent (3):  Intrinsic and weak discernment exclude each other, since weak discernment demands that $R$ (or $\neg R$) not be satisfied by \emph{either} of the two objects taken twice; which eliminates one further
truth-combination (three others having been already eliminated).

We will now show that all eight remaining truth-combinations are indeed possible, giving  a
representative structure for each. We hope by displaying these to convey the
``flavour'' of the logical relations between our four kinds of discernment.  We will then summarize the implications in a Venn diagram analogous to Figure \ref{fig: discabund}.

{\em (1): Intrinsic and external, but neither relative nor weak, discernment}:

\noindent\begin{minipage}{0.75\textwidth}
In this example (right), the formula  $Rxx$ does the
intrinsic discerning; and the formula  $\exists z Rxz$, or $\exists z Rzx$, does the external
discerning (external even though it is in fact the pair $\langle a, a\rangle$ in ext$(R)$ which is responsible for the discerning). But with our restriction that relative and weak discernment must be by $Rxy$ or $\neg Rxy$, $a$ and $b$ are neither relatively nor weakly discerned.
\end{minipage} \quad\ \ 
\begin{minipage}{0.2\textwidth}
   \includegraphics{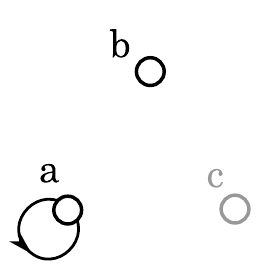}
\end{minipage}\\

{\em (2): External, but  neither intrinsic, nor relative, nor weak, discernment}:

\noindent\begin{minipage}{0.75\textwidth}
In our second case, the externally discerning formula is $\exists z Rxz$.  (Note that this formula also serves to discern externally in the example above; but that here, the ordered pair in ext$(R)$ that is responsible for the discernment is $\langle a,c \rangle$.) Again, our restrictiveness about which formulas can relatively or weakly discern means that in this example, there is none.
\end{minipage} \quad\ \ 
\begin{minipage}{0.2\textwidth}
   \includegraphics{example_3}
\end{minipage}\\

{\em (3): External and relative, but neither intrinsic nor weak, discernment}:

\noindent\begin{minipage}{0.75\textwidth}
In our third case, the relation $Rxy$ applies only in one direction between $a$ and $b$, so that $a$ and $b$ are
\emph{relatively} discerned, while the familiar one-place formula $\exists z Rxz$ applies
uniquely to $a$, making it also externally discernible from $b$.

\
\end{minipage} \quad\ \ 
\begin{minipage}{0.2\textwidth}
   \includegraphics{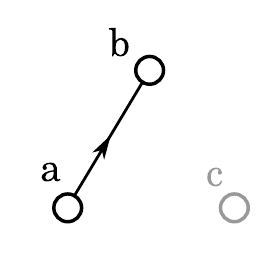}
\end{minipage}\\

{\em (4): Relative, but neither intrinsic nor external nor weak, discernment; also known as:
mere relative discernment}:

\noindent\begin{minipage}{0.75\textwidth}
Our fourth case, which Section \ref{4defined} abbreviated as `mere relative discernment', is
shown right.  By our Theorem 2 (in Section \ref{parag2}), the failure of absolute discernibility between $a$, $b$ and $c$ is associated with the existence of non-trivial symmetries: in this case the permutation $a \mapsto b \mapsto c \mapsto a$ and its inverse preserve the extension of $R$.
\end{minipage} \qquad\ \
\begin{minipage}{0.2\textwidth}
   \includegraphics{example_5}
\end{minipage}\\

{\em (5): Weak, but neither intrinsic nor external nor relative, discernment; also known as:
mere weak discernment}:

\noindent\begin{minipage}{0.75\textwidth}
Our fifth case, which Section \ref{4defined} abbreviated as `mere weak discernment', is
shown right.  Once again, the failure of absolute discernment between $a$ and $b$ is associated with a symmetry with relates them: in this case the swap $a \mapsto b, b \mapsto a$.

\
\end{minipage} \quad\ \ 
\begin{minipage}{0.2\textwidth}
   \includegraphics{example_6}
 \end{minipage}\\

{\em (6): External and weak, but neither intrinsic nor relative, discernment}:

\noindent\begin{minipage}{0.75\textwidth}
In our sixth case, the formula required to externally discern $a$ from $b$ is more complex: we need to refer to $c$, the objects which $a$ $R$s but $b$ does not, under the embargo on names.  $c$ is the only object that does not bear $R$ to anything, so the formula $\neg\exists zRxz$ picks it out.  Thus the formula $\exists z(Rxz \wedge \neg\exists wRzw)$ suffices to discern $a$ from $b$. So in this example, we exploit our allowance that any one-place formula can externally discern.
\end{minipage} \qquad\ \ 
\begin{minipage}{0.2\textwidth}
   \includegraphics{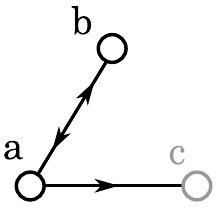}
 \end{minipage}\\

{\em (7): Intrinsic, external and relative, but not weak, discernment}:
 
\noindent\begin{minipage}{0.75\textwidth}
In our seventh case, the formula $\exists zRxz$ suffices to externally discern $a$ from $b$, $a$ being the only object in the structure which bears $R$ to anything.

\
\end{minipage} \quad\ \ 
\begin{minipage}{0.2\textwidth}
   \includegraphics{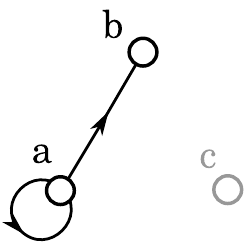}
 \end{minipage}\\

{\em (8): No discernment at all!}:

\noindent\begin{minipage}{0.75\textwidth}
Here,  all four kinds of discernment fail to hold: we have neither intrinsic, nor external, nor relative, nor weak discernment.   $a$ and $b$ are utterly indiscernible.

\
\end{minipage} \quad\ \ 
\begin{minipage}{0.2\textwidth}
   \includegraphics{example_1}
 \end{minipage}\\

\begin{figure}[h!]
   \centering
   \includegraphics[width=\textwidth]{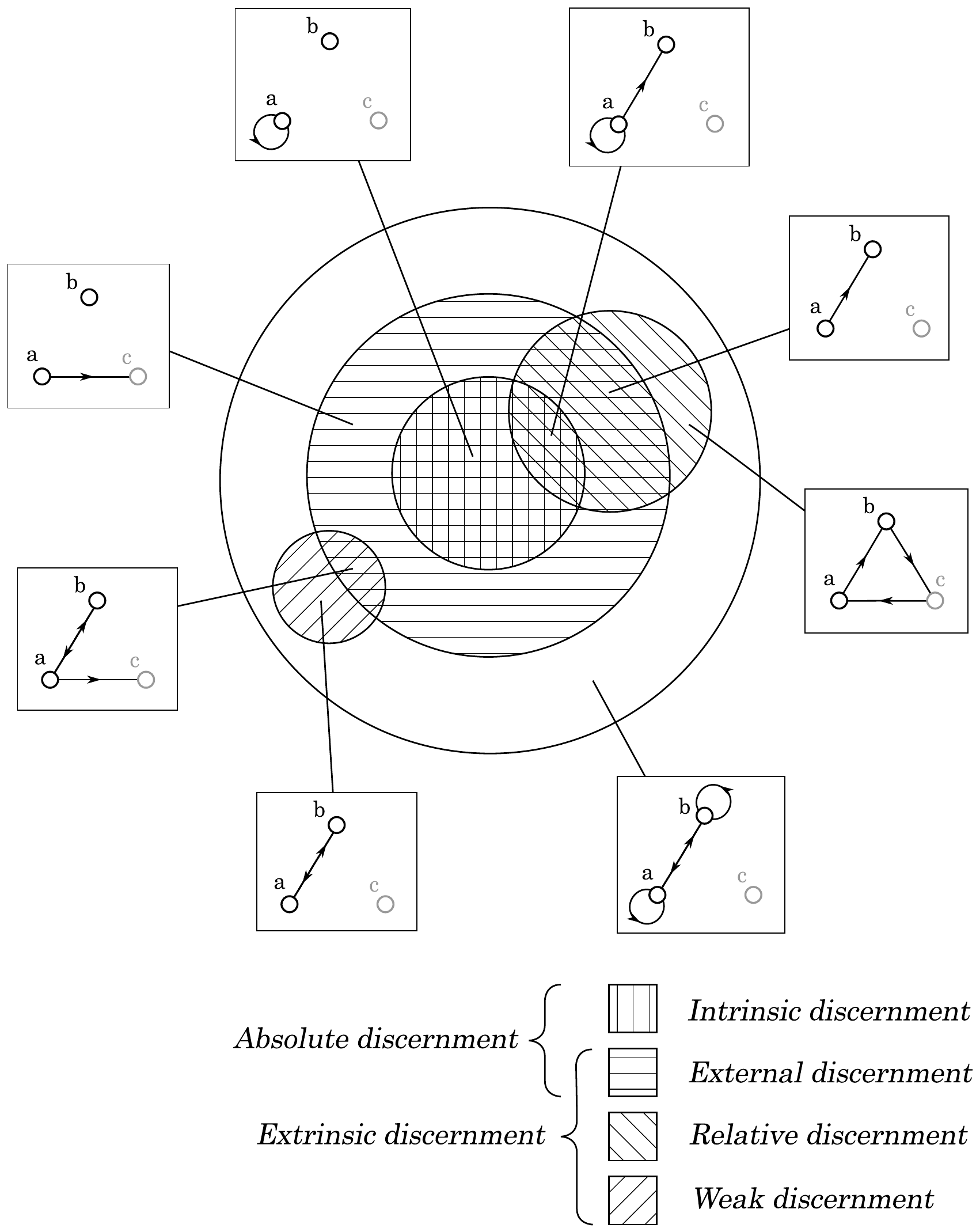}
   \caption{Implications between kinds of discernment: the sparse case. For each combination,
an exemplifying structure is shown.}
   \label{fig: discs}
\end{figure}

We now combine implications (1)-(3) noted at the beginning of this Subsection (\textbf{(Int)} $\Rightarrow$ \textbf{(Ext)}, etc.) with these eight possible combinations of discernment \emph{(1)-(8)} in the Venn diagram of Figure \ref{fig: discs}.

\section{Appendix: Semantic definitions of discernibility}\label{parag3}
In this paper, we have followed the lead of the Hilbert-Bernays account in emphasising a syntactic approach to defining kinds of discernibility. But one can instead emphasise semantic ideas, and seek definitions of indiscernibility in terms of permutations and symmetries. In this Appendix, we will explore this, for absolute discernibility.
  In fact there are two reasonable semantic definitions of absolute discernibility. The first will not preserve \emph{the wording of} our preceding results: as we will show by suitable examples (Section \ref{undermine}). But the second will preserve our preceding results (Section \ref{vindicate}).

\subsection{A definition that ``loses'' the results}\label{undermine}
We noted in \textbf{(Ext)} of Section \ref{4defined} 
that on our syntactic definition of external discernibility, an externally discerning
formula may not in fact refer to a third object.  If one laments this, one might envisage an alternative, \emph{semantic} definition, whereby external discernibility means the satisfaction of some two-place
relation by the pair $\langle a,c \rangle$, but not by the pair $\langle b, c \rangle$ (where we require that $a, b, c$ are all distinct!).  This
definition is more in line with the intuitive understanding of `external', since  it
will of course apply only to structures whose domains contain three or more objects.

This line of thought prompts a corresponding semantic definition of absolute discernibility. Like our official definition of absolute discernibility in Section \ref{4defined}, it will be a disjunction of intrinsic and external discernibility, the latter understood as just above. Thus we say that two objects $o_1, o_2$ in the domain of quantification are `absolutely
discernible' if they are discerned by their monadic properties, or by their relations to
other objects, {\em distinct} from either of them.

 Formally, we define: $o_1, o_2$  are {\em
absolutely discernible} if for some $n$, there is:\\
\indent (i) an $n$-place predicate $J^n$;\\
\indent (ii) some argument-place within $J^n$, say the $i$th ($1 \leqslant i \leqslant n$), and \\
\indent (iii) a sequence of $n-1$ objects, each of them distinct from both $o_1$ and $o_2$,
though not necessarily from each other, call it $\langle o^1, \ldots o^{i-1}, o^{i+1}, \ldots o^n
\rangle$ such that:\\
either (a) $\langle o^1, \ldots o^{i-1}, o_1, o^{i+1}, \ldots o^n \rangle \in \mbox{ext}(J^n)$ and
$\langle o^1, \ldots o^{i-1}, o_2, o^{i+1}, \ldots o^n \rangle \notin \mbox{ext}(J^n)$; or (b) {\em
vice versa}.\\
 (For the case $n=1$, i.e. the case of monadic properties, the sequence in (iii)
is the empty sequence; and this covers the case of intrinsic discernibility.)

\indent So absolute discernibility, thus defined, is still a kind of discernibility, as defined by the negation of
the right hand side of (HB). That is, it implies discernibility. 

\begin{figure}[h!]
   \centering
   \includegraphics{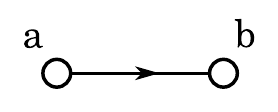}
   \caption{A structure in which $a$ and $b$ are not discerned intrinsically, nor by any third object, and yet are not related by any symmetry.}
   \label{fig: sym_ind6}
\end{figure}

But if we had opted for this semantic definition of absolute discernibility, then our two Theorems, in Sections \ref{parag1} and \ref{parag2} would both have \emph{failed}. (In saying this, 
we make the same stipulation that we did at the start of Section \ref{symmrevisit}: that `absolute indiscernibility' means just `not(absolute discernibility)'.)  We can see this failure with two counterexamples:
 
(1): Our first counterexample will falsify the rightward half of Theorem 2 in Section \ref{parag2}: that, in any finite structure, any two absolutely indiscernible objects are related by a symmetry.  According to the semantic definition just proposed, two objects are discerned absolutely if they are discerned either intrinsically---that is, by some monadic predicate---or externally---that is, by some relation to a \emph{third} object (or more).  So if a structure contains only \emph{two} intrinsically indiscernible objects, or contains two intrinsically indiscernible objects related in the same ways to all others in the structure, then the pair are absolutely indiscernible.  But the objects in the pair may well be related {\em to each other} asymmetrically, so that no symmetry relates them.  The simplest such example is shown in Figure \ref{fig: sym_ind6}.
 
According to the semantic definition under consideration, Figure \ref{fig: sym_ind6} exhibits a failure of both intrinsic and external discernibility (i.e.~it rules that $a$ and $b$ in Figure \ref{fig: sym_ind6} are absolutely indiscernible); while according to our syntactic definition, it does not. We admit that, in this case, the semantic definition better conforms with na\"ive intuition.  But our syntactic definition meshes (as well as it could do) with the topic of symmetries, whereas the semantic definition does not. Besides, we think the semantic definition better conforms with intuition only in its verdicts of \emph{indiscernibility} (i.e.~failures of what it calls absolute discernibility, as in Figure \ref{fig: sym_ind6}); not in its verdicts of \emph{discernibility}. This judgment is supported by the next counterexample.

(2): Our second counterexample will falsify the leftward half of Section \ref{parag2}'s Theorem 2: that any two objects related by a symmetry are absolutely indiscernible.  But recall that this leftward half of  Theorem 2 is the finite-domain case of Theorem 1 (from Section \ref{parag1}), that any symmetry leaves invariant the absolute indiscernibility classes.  So our second counterexample will also falsify Theorem 1.

\begin{figure}[h!]
   \centering
   \includegraphics{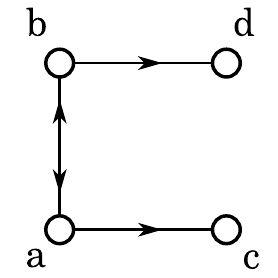}
   \caption{According to the semantic definition, in this structure $a$ and $b$ are absolutely discernible; but they are related by the symmetry $(ab)(cd)$.}
   \label{fig: sym_ind7}
\end{figure}

  This counterexample is provided by the left hand side of Figure \ref{fig: sym_ind5} (in Section \ref{paragButton}); thus cf.~Figure \ref{fig: sym_ind7}.

In this structure, the permutation $\pi:= (ab)(cd)$ is a symmetry, since it preserves the extension of the single relation $R$.  This symmetry swaps $a$ and $b$; so, according to our Theorems, $a$ and $b$ should be absolutely indiscernible.  But, according to the currently proposed semantic definition, $a$ and $b$ are in fact absolutely \emph{discernible}, since we have e.g.~$\langle a,c \rangle \in \mbox{ext}(R)$ but $\langle b,c \rangle \notin \mbox{ext}(R)$, which satisfies the conditions of the above semantic definition.

Here the discussion surrounding the counterexample to the converse of Theorem 1 (in Section \ref{paragButton}) recurs. The objects $a$ and $b$ are absolutely discerned by an object (namely $c$) which is {\em not itself an individual}; that is why $a$ and $b$ can be related by a symmetry.  In general, so long as $b$ is related to some other object (namely $d$) just as $a$ is related to $c$, and $c$ is taken to $d$ by some symmetry, then that symmetry can also take $a$ to $b$.  We might call the pair $a,c$ {\em correlated non-individuals}: correlated because of the existence of the appropriately related pair $b, d$. (Similarly, the pair $b, d$ are correlated non-individuals.) Note that $a$ and $c$ may be absolutely discerned, on our syntactic definition, from {\em each other}: they are iff $b$ and $d$ are.

We think the phenomenon of correlated non-individuals is the fly in the ointment for the semantic definition of absolute discernibility. For surely we should be persuaded by the external discernment of two objects (here, $a$ and $b$) by a third (or more), only if our grasp on the third object $c$ is suitably ``firm''.  If the discerning object is just as ``slippery" as the two we are trying to discern, then we get nowhere by appealing to it.  Agreed, it is too much to ask that the discerning object $c$ be specifiable, in Quine's (1976, p. 113) sense; (i.e.~uniquely satisfying some description, which nearly 
coincides with our sense of `individual': see Section \ref{4defined}'s Interlude). It should suffice that $c$ has no ``twin'' correlated to $b$ and $a$, in the same ways that $c$ is correlated to $a$ and $b$, respectively.  (Notice how symmetry has crept into our considerations!)

To sum up: these results suggest that the proposed semantic definition of absolute discernibility is unnatural from the point of view of symmetries. And from a syntactic point of view, our definition in Section \ref{4defined} is, we submit, the best we can do.
But if one still favours semantic definitions, one might seek an alternative semantic definition which meshes neatly with symmetries. In fact, there is one, though it makes this meshing a triviality.  We now turn to it.

\subsection{A definition that ``preserves'' the results}\label{vindicate}
Another natural way to define `absolute discernibility' in semantic terms is to say that two objects are absolutely discernible iff there is {\em not} a symmetry that maps one to the other. So, making the same stipulation as at the start of Section \ref{symmrevisit}---that `absolute indiscernibility' means just `not(absolute discernibility)'---two objects are absolutely indiscernible iff there is a symmetry that maps one to the other.

Clearly, this definition makes our two Theorems, in Sections \ref{parag1} and \ref{parag2}, trivially true. Besides, Theorem 2, in Section  \ref{parag2}, is now true for all structures, not just those with a finite domain.

Having previously adopted a syntactic definition of `absolute indiscernibility', it is perhaps natural to respond to these trivializations by quoting Russell's well-known jibe that this approach has `all the advantages of theft over honest toil'. But we take a more irenic view. After all, one can define a technical term that lacks a firm pre-existing meaning as one likes: and no one could claim that `absolute indiscernibility' has a firm pre-existing meaning! Thus our position is that whether one defines `absolute indiscernibility' syntactically, as we did, or by this semantic definition, there is `honest toil' to be done, showing the relations between the two definitions---and the honest toil pays off handsomely, in that the theorems show the definitions to mesh neatly. Indeed, the meshing is especially neat for finite domains: (a limitation that one would naturally expect, given the famous L\"{o}wenheim-Skolem limitations to the powers of first-order languages to express the size of domains). 

We also note that absolute discernibility, thus defined, may be taken as the disjunction of two notions of intrinsic discernibility and external discernibility: where again, these two notions are given reasonable semantic definitions. That is: suppose we now say that $a, b$ are intrinsically discernible iff there is some $n$-place predicate $J^n$ in the primitive vocabulary which has the ``constant'' $n$-tuple of $n$ copies of $a$, $\langle a,a, ..., a  \rangle$, as an instance, but not $\langle b,b, ..., b  \rangle$ (or vice versa). And suppose we also say that $a, b$ are externally discernible iff they are absolutely discernible, but not intrinsically discernible. (Though disjunctive, this definition of `externally discernible' is reasonably natural; though unlike ({\bf Ext}) in Section \ref{4defined}, it does not include its ``preceding kind'', i.e. intrinsic discernibility.) Then it now follows, by mere definition, that absolute discernibility  is the disjunction of  intrinsic discernibility and external discernibility.

\end{document}